\documentclass[prd,preprint,preprintnumbers,amsmath,amssymb]{revtex4}

\usepackage{bm,curves}
\usepackage{epsfig}
\usepackage[colorlinks = true, linkcolor = blue]{hyperref}
\usepackage{color}
\usepackage{dcolumn}

\newcommand{\nc}{\newcommand}   
\nc{\be}{\begin{equation}}
\nc{\ee}{\end{equation}}
\nc{\bea}{\begin{eqnarray}}
\nc{\eea}{\end{eqnarray}}
\nc{\F}{{\scriptscriptstyle F}}  
\nc{\xF}{{x_{\F}}}
\nc{\Fcc}{F_2^c}
\nc{\red}{\color{red}}

\def\IE{{\it i.e.,}}

\def\EA{{\it et al.}}

\def\Psl{\displaystyle{\not} P }
\def\psl{\displaystyle{\not} p }
\def\ksl{\displaystyle{\not} k }
\def\qsl{\displaystyle{\not} q }
\def\Dsl{\displaystyle{\not} \Delta }
\graphicspath{./IC-Fig/}

\begin{document}

\preprint{JLAB-THY-13-1813}

\title{Phenomenology of nonperturbative charm in the nucleon}

\author{T. J. Hobbs$^1$, J. T. Londergan$^1$, W. Melnitchouk$^2$}
\affiliation{
	$^1$Department of Physics and
	 Center for Exploration of Energy and Matter,
	 Indiana University, Bloomington, Indiana 47405, USA	\\
	$^2$\mbox{Jefferson Lab, 12000 Jefferson Avenue,
	 Newport News, Virginia 23606, USA}}

\date{\today}

\begin{abstract}
We perform a comprehensive analysis of the role of nonperturbative
(or intrinsic) charm in the nucleon, generated through Fock state
expansions of the nucleon wave function involving five-quark virtual
states represented by charmed mesons and baryons.  We consider
contributions from a variety of charmed meson--baryon states and find
surprisingly dominant effects from the $\bar{D}^{* 0}\, \Lambda_c^+$
configuration.  Particular attention is paid to the existence and
persistence of high-$x$ structure for intrinsic charm, and the $x$
dependence of the $c-\bar c$ asymmetry predicted in meson--baryon
models.  We discuss how studies of charmed baryons and mesons in
hadronic reactions can be used to constrain models, and outline
future measurements that could further illuminate the intrinsic
charm component of the nucleon.
\end{abstract}

\maketitle

\section{Introduction}
\label{sec:intro}

Charm in the nucleon can be produced in two fundamentally distinct
processes.  In the first, commonly referred to as ``extrinsic charm'',
charm-anticharm pairs arise through gluon radiation.  This perturbative
process involving gluon radiation is a significant feature of QCD
evolution.  Charm arising in this way is concentrated at very low
parton momentum fraction $x$, and with increasing $Q^2$ one expects
to see progressively more charm being produced.  Extrinsic charm
production ensures essentially identical distributions of charm and
anticharm \cite{Catani04}.

A second source of charm is called ``intrinsic charm.''  This source of
charm arises through nonperturbative fluctuations of the nucleon state
to four quark--one antiquark states.  Compared with extrinsic charm,
intrinsic charm has two rather striking features.  First, intrinsic
charm tends to be ``valence-like,'' \IE~it is produced at relatively
large $x$; second, it is not unusual for intrinsic charm and anticharm
to be unequal.

The first calculation of intrinsic charm was made by Brodsky, Hoyer,
Peterson and Sakai (BHPS) \cite{BHPS}.  They noticed that
the first direct measurements of charm production \cite{Drijard:1978gv,
Giboni:1979rm, Lockman:1979aj, Drijard:1979vd} were larger than could
be accounted for by contemporary calculations of extrinsic charm, and
were also more concentrated at relatively large values of Feynman $\xF$.
BHPS suggested adding in addition to extrinsic charm an ``intrinsic''
component arising from fluctuations of the nucleon to four quark--one
antiquark states.  By assuming that the mass of the charm quark in
the five-quark configuration was much greater than all other masses
and transverse momenta, the distribution of intrinsic charm in $x$
could be computed analytically, with only the overall normalization a
free parameter.  Although the BHPS {\it ansatz} is rather simplistic,
it does provide a convenient benchmark for comparing with estimates of
charm and anticharm probabilities in more sophisticated treatments.
We will review the BHPS formalism in more detail in Sec.~\ref{ssec:BHPS}.

In the intervening years there have also been a number of theoretical
calculations of intrinsic charm using various models for production
of four quark--one antiquark states \cite{BPS, HM83, HSV, VB, Vogt95,
ICHT, VB96, Ing97, MT97, SMT99, Nav96, Nav98, Dur01, Gon10, Pum05,
Pum07, Dulat13}.
In contrast to the BHPS philosophy, a more dynamical type of model
for producing intrinsic quark distributions is based on a two-step
approach, in which one considers quark-antiquark production through
nonperturbative fluctuations of a proton to a baryon plus meson state.
In this fluctuation the heavy quark and antiquark will appear in the
resulting meson and baryon states.
An example of intrinsic charm produced in this way would involve a
fluctuation such as
   $p \to \Lambda_c^+ + \bar{D}^0$,
where here the charm quark resides in the $\Lambda_c^+$ baryon
(assumed to be a valence $udc$ state) and the $\bar{c}$ in
the $\bar{D}^0$ (with valence composition $\bar{c} u$).
Such models are also referred to as ``meson--baryon'' models,
since one is computing Fock states where a baryon fluctuates to
another baryon plus a meson.
(These models are also sometimes referred to as ``meson-cloud''
or ``convolution'' models, motivated by the SU(2) fluctuation
of a proton into a nucleon plus a diffuse ``cloud'' of nearly
massless pions that exist primarily outside a compact baryon core.
Of course, the ``cloud'' analogy is less appropriate for
fluctuations to particles containing heavy quarks).
A pedagogical review of some of these models was given recently
by Pumplin \cite{Pum05}.

Since the initial experimental measurements of charm production that
inspired the BHPS work, substantial additional experimental data on
charm production in high-energy reactions have been collected.
Large-$x$ measurements of charm production were carried out by
the European Muon Collaboration (EMC) \cite{EMC} at CERN, and the
charm structure function $\Fcc$ was measured at small $x$ by the
H1 \cite{H1} and ZEUS \cite{ZEUS} Collaborations at HERA.
The HERA measurements probe primarily the region where the
perturbative extrinsic charm dominates, while the EMC experiment
was more sensitive to nonperturbative intrinsic charm.

Despite a number of dedicated theoretical analyses, the magnitude of
the intrinsic charm component of the nucleon, which was expected from
the early analyses to be of the order of 1\%, is still inconclusive.
In leptoproduction, the dominant mechanism for perturbative
generation of charm is the photon--gluon fusion process.
Hoffman and Moore \cite{HM83} considered ${\cal O}(\alpha_s)$
corrections to intrinsic charm distributions, as well as quark
and target mass contributions to charm cross sections, finding
an intrinsic charm component of order 0.3\% in the proton.
Harris, Smith and Vogt \cite{HSV} reanalyzed the Hoffman-Moore
treatment of the EMC data, finding that the EMC charm production
data at large energy transfers, $\nu \sim 170$~GeV, required an
intrinsic charm component normalized to $0.86 \pm 0.60\%$.
A subsequent study by Steffens \EA~\cite{SMT99} used
the BHPS and meson--baryon intrinsic charm models, together
with an interpolating scheme, to map smoothly onto massless QCD
evolution at large $Q^2$ and photon-gluon fusion at small $Q^2$.
While the analysis found it difficult to simultaneously fit the
entire data set in terms of a single intrinsic charm scenario,
it did indicate a slight preference for intrinsic charm in a
meson--baryon model at a level of about 0.4\%, although the
evidence was not conclusive.

The most detailed study of intrinsic charm within a global QCD
analysis of parton distribution functions (PDFs) was made by
Pumplin, Lai and Tung \cite{Pum07}, using models for intrinsic
charm from Ref.~\cite{Pum05}, and recently updated by
Dulat {\it et al.} \cite{Dulat13} to include next-to-next-to
leading (NNLO) order $\alpha_s$ corrections.
Depending on the intrinsic charm model used, they found that the
high-energy data could accommodate up to about 2\% of intrinsic
charm without serious disagreement with the high-energy data.
On the other hand, the analyses of Refs.~\cite{Pum07, Dulat13}
did not include the original EMC data \cite{EMC} that suggested
an enhancement in $\Fcc$ at large $x$.  In fact, since there is
some tension between the EMC and HERA data at small values of $x$,
typically global PDF analyses omit the EMC data and fit the more
extensive HERA data in terms of perturbative contributions alone
\cite{MSTW08, JR09}.

Clearly, additional data are needed for a more definitive determination
of the role of intrinsic charm in the nucleon, especially at large $x$,
and several new charm production experiments that have been proposed
recently may shed light on the dynamical origin of nonperturbative charm.
Studies of $DN$ and $J/\psi\, N$ interactions, charmed baryon
spectroscopy and charmed nuclei will be possible, for example,
at the J-PARC facility in Japan \cite{Kumano11},
with its primary proton and secondary pion beams, while proton and
antiproton beams can be used to study open charm production at
GSI-FAIR in Germany \cite{FAIR}.
Near threshold $J/\psi$ photoproduction at Jefferson Lab, following
its 12~GeV upgrade, would allow gluon generalized parton distributions
to be explored, in addition to the $A$ dependence of the
$J/\psi$-nucleon cross section \cite{PR12-07-106, Brambilla11}.
For the proposed AFTER@CERN fixed-target facility \cite{Brodsky12},
charmed meson production has been suggested as a means of tagging the
gluon distribution at large $x$, which requires understanding of
competing contributions from charm quarks at similar kinematics.
Finally, a future Electron-Ion Collider \cite{Deshpande05, Boer11,
Accardi12} would provide opportunities to access charm in the nucleon
in the large-$x$ region.

Anticipating these future experimental efforts, and in view of the
somewhat conflicting results from the previous phenomenological studies,
it is therefore timely to revisit the problem of intrinsic charm using
the latest available theoretical techniques and models for calculating
$\Fcc$.  In this paper we undertake a comprehensive and critical
analysis of nonperturbative models of charm in the nucleon, with the
aim of obtaining a more robust understanding of both the magnitude
and shape of the intrinsic charm distribution.
We begin by reviewing in Sec.~\ref{sec:5q} the simple five-quark models
of the nucleon introduced by BHPS \cite{BHPS} and Pumplin
\cite{Pum05}.  In Sec.~\ref{sec:MBM}, we discuss the more
dynamical meson--baryon models for heavy quark generation, which are
constrained by coupled-channel analyses of baryon--baryon scattering
\cite{Sib01, Hai07, Hai08, Hai11}.  These were initially used to
describe nucleon-nucleon scattering, then were applied to the
strange sector, and have since been extended to the regime of charm.
We summarize the assumptions made in the extensions into the strange
and charmed sectors, and outline the experimental data that fix the
coupling constants and cutoff parameters for such potentials.
We compare the meson--baryon model calculations with hadronic data on
production of charmed baryons from the R608 Collaboration at the ISR
\cite{Chauvat:1987kb}, which measured inclusive $\Lambda_c$ production
in $pp$ collisions.  We also examine charge asymmetries in $\Lambda_c$
production from proton beams which has been measured recently by the
SELEX Collaboration \cite{Garcia:2001xj}.

The basic ingredients of the meson--baryon models, namely, the
proton $\to$ meson $+$ baryon splitting functions and the charm quark
distributions in the charmed $D$ mesons and baryons, are derived in
Secs.~\ref{sec:mb} and \ref{sec:cinc}, respectively.
The resulting convolutions of these distributions are presented
in Sec.~\ref{sec:results}, where we compare various prescriptions
for the regularization of the momentum integrals, and different
approximations for the distributions in the charmed hadrons.
We also confront the models with data on the charm structure
function $\Fcc (x)$, in particular with the EMC data \cite{EMC}
which measured charm at large $x$, and in addition the very precise
HERA data from the H1 and ZEUS Collaborations \cite{H1, ZEUS},
which is sensitive to charm in the low-$x$ region.
Finally, in Sec.~\ref{sec:conc} we summarize our results,
and discuss possibilities for the measurements of charmed
observables at current and future facilities.
Additional technical details of the derivations of the
meson--baryon splitting functions and the charm quark distributions
in charmed hadrons are given in Appendix~\ref{sec:A_MBM} and
\ref{sec:B-MBM}, respectively, and simple parametrizations of the
calculated $c$ and $\bar c$ distributions in the nucleon are provided
in Appendix~\ref{sec:C-MBM}.

\section{Five-quark models of nucleon structure}
\label{sec:5q}

In this section we review models of intrinsic charm based on particular
five-quark Fock state components of the nucleon wave function.
We will focus on models that describe the process by which a nucleon
initially containing three light valence quarks transitions to a four
quark plus one antiquark state containing a charm-anticharm quark pair.

\subsection{The BHPS model}
\label{ssec:BHPS}

The simplest model for producing intrinsic charm was proposed over
30 years ago by BHPS \cite{BHPS}.
In the infinite momentum frame (IMF) the transition probability for a
proton with mass $M$ to make a transition $p \to uudc\bar{c}$ (or to
a five-quark state containing any heavy quark pair) involves an energy
denominator that can be expressed in terms of the masses $m_i$ and
momentum fractions $x_i$ of the constituents,
\be
P(p \to uudc\bar{c})
\sim \left[ M^2 - \sum_{i=1}^5 \frac{k_{\perp i}^2 + m_i^2}{x_i}
     \right]^{-2}.
\label{eq:BHPSeq}
\ee
Here, $k_{\perp i}$ is the transverse momentum of quark $i$, and the
heavy quarks in the $c \bar{c}$ pair are assigned indices $4$ and $5$.

For simplicity, the BHPS calculation assumed a point coupling for the
$c \bar{c}$ production vertex, and neglected the effect of transverse
momentum in the five-quark transition amplitudes.  With the additional
assumption that the charm mass is much greater than the nucleon and
light quark masses, the probability for producing a single charm quark
can be derived analytically,
\be
P(x_5) = \frac{N x_5^2}{2}
	 \left[ \frac{(1-x_5)}{3}\left( 1 + 10x_5 + x_5^2 \right)
		+ 2 x_5 (1+x_5) \ln(x_5)
	 \right],
\label{eq:charmprob}
\ee
with the normalization $N$ fixed by the overall charm quark
probability in the proton.

Using the analytic expression (\ref{eq:charmprob}) one obtains a
``valence-like'' charm quark distribution that is significant for
$0.1 \leq x_5 \leq 0.5$.  The valence-like shape arises from the
structure of the energy denominator in Eq.~(\ref{eq:BHPSeq}),
which for large quark masses $m_i$ favors configurations with
large momentum fractions $x_i$.  This feature is common to all
similar five-quark models and invariably results in a valence-like
heavy quark distribution.

It is also possible to compute the charm quark probability numerically
without assuming that the charm quark mass is much greater than all
other masses.  With realistic masses, the charm quark distribution turns
out to be similar to the analytic form of Eq.~(\ref{eq:charmprob}).
Note also that since the charm and anticharm probabilities enter
Eq.~(\ref{eq:BHPSeq}) symmetrically, the charm production mechanism
in this model will produce equal probabilities for $c$ and $\bar{c}$.
Although the BHPS model is rather simplistic, it nevertheless
provides a useful reference point of comparison to test intrinsic
charm and anticharm distributions obtained from other prescriptions.

\subsection{Scalar five-quark model}
\label{ssec:scalar5q}

In a detailed study of intrinsic heavy quark probabilities, Pumplin
\cite{Pum05} considered a series of models for the Fock space
wave function on the light-front for a proton to make a transition to
a four quark plus one antiquark system, with the heavy $q\bar{q}$ pair
composed of either charm or bottom quarks.
A simplified case was studied where a point scalar particle of mass
$m_0$ couples with strength $g$ to $N$ scalar particles with masses
$m_1, m_2, \ldots, m_N$.  The light-front Fock space probability
density $dP$ for such a process then takes the form
\cite{Pum05}
\bea
dP
&=& \frac{g^2}{(16\pi^2)^{N-1}(N-2)!}\,
    \prod_{j=1}^N dx_j\, \delta\left( 1-\sum_{j=1}^N x_j\right)
    \int_{s_0}^\infty ds\, \frac{(s-s_0)^{N-2}}{(s-m_0^2)^2}\,
    |F(s)|^2,
\label{eq:dPFock}
\eea
where $s_0 = \sum_{j=1}^N m_j^2/x_j$, and the form factor $F(s)$
serves to suppress contributions from high-mass states.
If one neglects the effects of transverse momentum and the factors
of $1/x_j$ in Eq.~(\ref{eq:dPFock}), and assumes a point form factor
$F(s) = 1$, then in the limit that the charm mass is much larger than
all other masses one recovers the distribution in the BHPS model
\cite{BHPS}.

To incorporate the effects of the finite size of the nucleon,
Pumplin considered both an exponential form factor,
\be
|F(s)|^2 = \exp \left[-(s-m_0^2)/\Lambda^2 \right],
\label{eq:expFF}
\ee
and a power-law suppression factor,
\be
|F(s)|^2 = {1 \over (s + \Lambda^2)^n}, 
\label{eq:powerFF}
\ee
with $\Lambda$ a cutoff mass regulator.
Fixing the overall normalization to be a constant, the resulting
shape of the charm quark momentum distribution with the power-law
suppression, for a reasonable choice of $n=4$, turns out to be softer
than the BHPS prediction for a range of cutoffs, $\Lambda = 2-10$~GeV
\cite{Pum05}.
For the exponential suppression, the shape depends somewhat more
strongly on the cutoff parameter, with the distribution being harder
than the BHPS result for smaller $\Lambda$ values and softer for
larger $\Lambda$.
All of the resulting charm distributions are valence-like, however,
with significant tails even beyond $x \approx 0.4$.

While the simple five-quark models give some qualitative insights
into the possible generation of intrinsic charm at large $x$,
they retain a high degree of dependence on the model parameters,
whose connection with the underlying QCD theory is not clear.
Furthermore, it is also not obvious how one could constrain
the parameters phenomenologically by comparing deep-inelastic
scattering with other observables, for instance.
In the next section we discuss an alternative approach which may
offer greater promise for relating intrinsic charm distributions
to inputs determined from independent reactions.

\section{Meson--baryon models for intrinsic charm}
\label{sec:MBM}

A hybrid class of intrinsic charm models that involves both quark
and hadron degrees of freedom, and makes some unique and testable
predictions for the $c$ and $\bar c$ distributions in the nucleon,
are meson--baryon models.  Such models attempt to quantify the
fluctuations of the nucleon to states with a virtual meson $M$
plus baryon $B$,
\be
|N\rangle\
=\ \sqrt{Z_2}\, \left| N \right.\rangle_0\
+\ \sum_{M,B} \int\! dy\, d^2\bm{k}_\perp\,
   \phi_{MB}(y,k^2_\perp)\,
   | M(y,\bm{k}_\perp); B(1-y,-\bm{k}_\perp) \rangle,
\label{eq:Fock}
\ee 
where $\left| N \right.\rangle_0$ is the ``bare'', three-quark
nucleon state, and $Z_2$ is the wave function renormalization.
The function $\phi_{MB}(y,k^2_\perp)$ gives the probability
amplitude for the physical nucleon to be in a state consisting
of a virtual meson $M$ with longitudinal momentum fraction $y$
and transverse momentum $\bm{k}_\perp$, and a baryon $B$ with
longitudinal momentum fraction $1-y$ and transverse momentum 
$-\bm{k}_\perp$.
The total invariant mass squared of the meson--baryon system
$s_{MB}$ can be written in the IMF as
\be
s_{MB}(y,k^2_\perp)
= \frac{k^2_\perp + m_M^2}{y} + \frac{k^2_\perp + M_B^2}{1-y},
\label{eq:CoM_En}
\ee
where $m_M$ and $M_B$ are the meson and baryon masses, respectively.
If the meson--baryon terms include states containing charm quarks,
the resulting probability distributions for anticharm and charm
quarks in the nucleon can be written in the form of convolutions,
\begin{subequations}
\label{eq:mesoncloud}
\bea
\bar{c}(x)
&=& \sum_{M,B}\,
    \Big[ \int_x^1 \frac{dy}{y}\,
	  f_{MB}(y)\, \bar{c}_M\Big(\frac{x}{y}\Big)
	+ \int_x^1 \frac{d\bar y}{\bar y}\,
	  f_{BM}(\bar{y})\, \bar{c}_B\Big(\frac{x}{\bar{y}}\Big)
   \Big],
\label{eq:mesoncloud_cb}			\\
c(x)
&=& \sum_{B,M}\,
    \Big[ \int_x^1 \frac{d\bar y}{\bar y}\,
	  f_{BM}(\bar y)\, c_B\Big(\frac{x}{\bar y}\Big)
	+ \int_x^1 \frac{dy}{y}\,
	  f_{MB}(y)\, c_M\Big(\frac{x}{y}\Big)
    \Big],
\label{eq:mesoncloud_c}
\eea
\end{subequations}%
where $\bar y \equiv 1-y$, and for ease of notation we have omitted
the dependence of the distributions on the scale $Q^2$.

In analogy with the quark-gluon splitting functions of perturbative
QCD, in Eqs.~(\ref{eq:mesoncloud}) $f_{MB}(y)$ represents the
splitting function for a nucleon to fluctuate to meson $M$ with
fraction $y$ of the proton's momentum, and a spectator baryon $B$.
The charm and anticharm distributions in the baryon $B$ are denoted
by $c_B(z)$ and $\bar{c}_B(z)$, respectively, and carry a fraction
$z=x/\bar y$ of the baryon's momentum.
Similarly, $f_{BM}(\bar y)$ represents the splitting function for a
nucleon fluctuating into a baryon $B$ with fraction $\bar y$ of the
proton's momentum, with a spectator meson $M$.  The quark distributions
inside the meson $M$ are denoted by $c_M(z)$ and $\bar{c}_M(z)$,
respectively.
If the charm quark resides exclusively in the baryon, with the
anticharm in the meson, Eqs.~(\ref{eq:mesoncloud}) simplify further
since
\bea
c_M(x)\ \to\ 0,\ \ \ 
\bar{c}_B(x)\ \to\ 0.
\label{eq:ccbar-simp}
\eea

The splitting functions in Eqs.~(\ref{eq:mesoncloud}) are related
to the probability amplitudes $\phi_{MB}$ by
\bea
f_{MB}(y)
&=& \int_0^\infty d^2\bm{k}_\perp \,|\phi_{MB}(y,k^2_\perp)|^2\
 =\ f_{BM}(\bar y),
\label{eq:fphi}
\eea
where the reciprocity relation in the second equality arises
from the conservation of three-momentum at the $MBN$ vertex 
\cite{Zoller92, MT93, Holtmann96, Speth98, Kum98}.
It can be shown to be satisfied explicitly in the infinite momentum
frame (or on the light-front), but is violated in covariant calculations
\cite{MT93, Holtmann96, Speth98, Kum98, MST94} in the presence of $MBN$
form factors (or other ultraviolet regulators) which do not exhibit
the $y \leftrightarrow \bar y$ symmetry of the amplitudes $\phi_{MB}$
\cite{Bur13}.

The convolution equations (\ref{eq:mesoncloud}) allow the symmetries
of the splitting functions to be represented in terms of moments of
the parton distributions $C^{(n)}$ and $\overline C^{(n)}$, defined as
\begin{subequations}
\label{eq:c_mom}
\bea
\overline C^{(n)}
&=& \int_0^1 dx\, x^n\, \bar c(x)\
 =\ \sum_{M,B}\, {\cal F}_{MB}^{(n)}\, \overline C_M^{(n)},	\\
C^{(n)}
&=& \int_0^1 dx\, x^n \,c(x)\
 =\ \sum_{B,M}\, {\cal F}_{BM}^{(n)}\, C_B^{(n)},
\eea
\end{subequations}%
where
\begin{subequations}
\label{eq:f_mom}
\bea
{\cal F}_{MB}^{(n)}
&=& \int_0^1 dy\, y^n\, f_{MB}(y),		\\
{\cal F}_{BM}^{(n)}
&=& \int_0^1 d\bar y\, \bar y^n\, f_{BM}(\bar y),
\eea
\end{subequations}%
are the $n$-th moments of the splitting functions.
The corresponding moments of the $\bar c$ and $c$ distributions
in the meson $M$ and baryon $B$ are denoted by $\overline C_M^{(n)}$
and $C_B^{(n)}$, respectively.
In particular, the lowest moment of the splitting functions
gives the average multiplicity of mesons $M$ and baryons $B$,
\bea
\langle n \rangle_{MB}
&\equiv& {\cal F}_{MB}^{(0)}\ =\ {\cal F}_{BM}^{(0)},
\label{eq:recip1}
\eea
which reflects global charge conservation, while conservation
of momentum implies that the momentum fractions
$\langle y \rangle_{MB} \equiv {\cal F}_{MB}^{(1)}$ and
$\langle y \rangle_{BM} \equiv {\cal F}_{BM}^{(1)}$ are
related by
\bea
\langle y \rangle_{BM}\ +\ \langle y \rangle_{MB}
&=& \langle n \rangle_{MB}.
\label{eq:conserv}
\eea

In contrast to the five-quark models discussed in Sec.~\ref{sec:5q},
in which the $x$ dependence of the $c$ and $\bar c$ distributions
was identical, in the MBM the distributions of heavy quarks and
antiquarks in the nucleon are generally expected to be different.
Indeed, since the $c$ in the baryon and $\bar c$ in the meson
reside in rather different local environments, an asymmetry
$c(x) \ne \bar{c}(x)$ is almost unavoidable.
Of course, since the proton has no net charm, the lowest moments
of $c$ and $\bar c$ must cancel; however, all higher moments will
be nonzero,
\be
C^{(0)} - \overline C^{(0)} = 0,\ \ \ \ \ \ \
C^{(n)} - \overline C^{(n)} \neq 0\ \ (n \geq 1).
\label{eq:c_moments}
\ee
Whenever Eq.~(\ref{eq:recip1}) is satisfied, this will guarantee that
the proton has no net charm.

Because quarks and antiquarks possess opposite intrinsic parities,
parity conservation will require that the quark wave functions respect
overall parity conservation.  For example if the initial proton state
is treated as three constituent $(uud)$ quarks in $S$-wave orbitals
and a $c \bar{c}$ pair is added, then if one of the charm quarks is
placed in an $S$ state the other needs to be in an odd-parity state.
In the MBM, this behavior is accommodated provided one uses physical
vertices that correctly account for the spin degrees of freedom of
the relevant fields. Models that treat quarks as scalar point-like
particles, for example, will generally not satisfy these constraints
\cite{Pum05, Pum07}.

\begin{table}[bt]
\caption{Lowest mass meson--baryon Fock states of the proton containing
	charm and anticharm quarks.  For each state the isospin $I$,
	spin $J$ and parity $P$ are listed for the meson and baryon,
	together with the masses.}
\centering
\begin{tabular}{l c l c}
\hline\hline
Baryon  & \ $I(J^{P})$\ \ \ \ \ & \ \ Meson & $I(J^{P})$	\\ 
						[0.5ex]\hline
$p\, (938)$            & $\frac{1}{2}\, (\frac{1}{2}^+)$
                                               & $J/\psi(3097)$\ \ \
	& $0\, (1^-)$		\\
$\Lambda_c^+(2286)$    & $0\, (\frac{1}{2}^+)$ & $\bar{D}^0(1865)$
	& $\frac{1}{2}\, (0^-)$	\\
                       &                       & $\bar{D}^{*0}(2007)$
	& $\frac{1}{2}\, (1^-)$	\\
$\Sigma_c^+(2455)$     & $1\, (\frac{1}{2}^+)$ & $\bar{D}^0(1865)$
	& $\frac{1}{2}\, (0^-)$	\\
                       &                       & $\bar{D}^{*0}(2007)$
	& $\frac{1}{2}\, (1^-)$	\\
$\Sigma_c^{++}(2455)$  & $1\, (\frac{1}{2}^+)$ & $D^-(1870)$
	& $\frac{1}{2}\, (0^-)$	\\
                       &                       & $D^{*-}(2010)$
	& $\frac{1}{2}\, (1^-)$	\\
$\Sigma_c^{*+}(2520)$  & $1\, (\frac{3}{2}^+)$ & $\bar{D}^0(1865)$
	& $\frac{1}{2}\, (0^-)$	\\
                       &                       & $\bar{D}^{*0}(2007)$
	& $\frac{1}{2}\, (1^-)$	\\
$\Sigma_c^{*++}(2520)$ & $1\, (\frac{3}{2}^+)$ & $D^-(1870)$
	& $\frac{1}{2}\, (0^-)$	\\
                       &                       & $D^{*-}(2010)$
	& $\frac{1}{2}\, (1^-)$	\\ [1ex]
\hline
\end{tabular}
\label{table:mass_spect}
\end{table}

In the present analysis, we consider various meson--baryon states
containing charm quarks that could contribute to the intrinsic
charm in the proton, as summarized in Table~\ref{table:mass_spect}.
These include the SU(4) octet isoscalar $\Lambda_c$ and isovector
$\Sigma_c$ baryons, and the decuplet $\Sigma_c^*$, while for the
mesons, the pseudoscalar $D$ and vector $D^*$ mesons are included.
In addition, the state involving a proton fluctuation to
$p + J/\psi$, where both the $c$ and $\bar{c}$ reside in
the $J/\psi$, was considered in Ref.~\cite{Pum05}.
Although this has a combined mass which is actually lower
than all the other charmed meson--baryon configurations,
its contribution should be strongly suppressed by the OZI rule.

In the following section the splitting functions $f_{MB}(y)$
for the various configurations in Table~\ref{table:mass_spect}
will be presented, with details of the derivations given in
Appendix~\ref{sec:A_MBM}.
To constrain the model parameters in the calculations,
namely, the hadronic couplings and form factor cutoffs,
we use phenomenological input from $DN$ and $\bar{D} N$
scattering analyses \cite{Sib01, Hai07, Hai08, Hai11},
together with inclusive charmed baryon production data in
$pp$ collisions.

\section{Charmed meson--baryon splitting functions}
\label{sec:mb}

For a given meson--baryon state $MB$, the splitting function can
be evaluated as in Eq.~(\ref{eq:fphi}) in terms of an integral over
the transverse momentum of the square of the probability amplitude
$\phi_{BM}(y,k^2_\perp)$ which was defined in Eq.~(\ref{eq:Fock}).
To compute the probability amplitudes we use time-ordered perturbation
theory, characterized by a particular choice of forward-moving
kinematics in the IMF.  Details of the framework and complete
derivations follow in Appendix~\ref{sec:A_MBM}.

\begin{figure}[t]
\includegraphics[width=10cm]{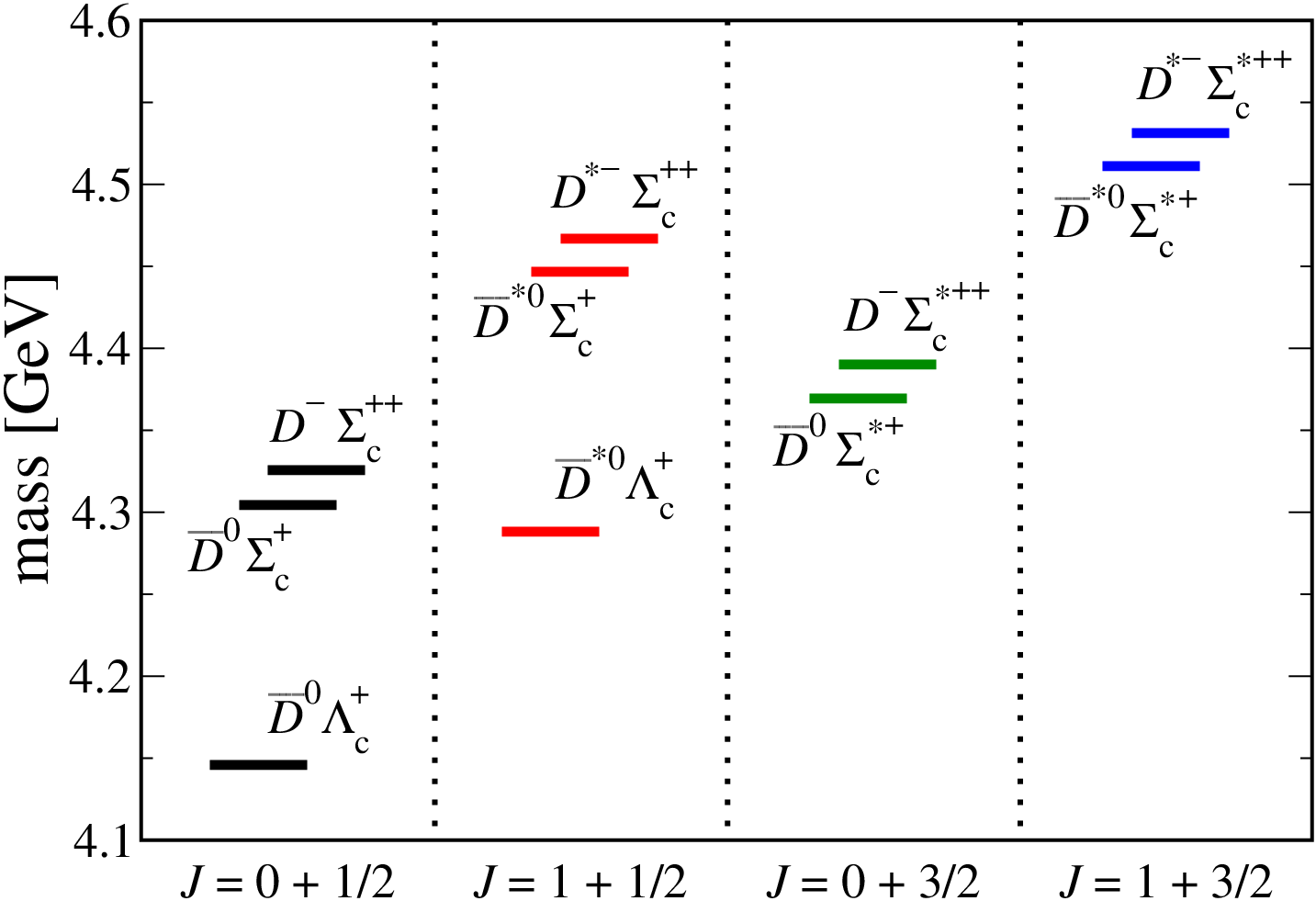}
\caption{(color online)
	Spin configurations ($J=$ meson + baryon spin) and masses
	for the spectrum of charmed hadron configurations included
	in the present MBM calculation.}
\label{fig:Spectrum}
\end{figure}

We consider splitting functions for $p \to MB$ fluctuations for the
spin transitions
\mbox{
$\bm{\frac{1}{2}} \to
	 \bm{0} \bigoplus \bm{\frac{1}{2}}$},
	$\bm{1} \bigoplus \bm{\frac{1}{2}}$,
	$\bm{0} \bigoplus \bm{\frac{3}{2}}$,
and	$\bm{1} \bigoplus \bm{\frac{3}{2}}$,
with the specific states listed in Table~\ref{table:mass_spect}
and illustrated in Fig.~\ref{fig:Spectrum} (with the exception of
the $p\, J/\psi$ state).
Couplings to these states are then determined from the lowest-order
effective hadronic Lagrangian for each transition, given in
Appendix~\ref{sec:A_MBM}.

In the framework of the MBM, the splitting function for the
fluctuation of the proton to a spin-0 charmed meson
	$D = \bar{D}^0$ or $D^-$
and a spin-1/2 charmed baryon
	$B = \Lambda_c^+$, $\Sigma_c^+$ or $\Sigma_c^{++}$
is given by
\begin{eqnarray}
f_{DB}(y)
&=& T_B
    \frac{g^2}{16\pi^2} \int{ dk_\perp^2 \over y (1-y) }
    { |F(s)|^2 \over (s - M^2)^2 }
    \left[ \frac{k_\perp^2 + (M_B - (1 -y) M)^2}{1-y} \right],
\label{eq:DLsplit}
\end{eqnarray}
where for ease of notation we have used for the coupling constant
  $g \to g_{DBN}$ and for the $DB$ invariant mass
  $s \to s_{DB}$.
The isospin transition factor $T_B$ is given by
\begin{equation}
T_B = 1 + \delta_{t_B,+1}
\label{eq:TB}
\end{equation}
where the third component of the isospin of the charmed baryon is
$t_B =  0$ for $B = \Lambda_c^+$ and $\Sigma_c^+$, and
$t_B = +1$ for $B = \Sigma_c^{++}$.
The states described by the splitting function $f_{DB}(y)$
include the lowest-mass configuration
	$\bar{D}^0 \Lambda_c^+$,
as well as the isovector charmed baryon states
	$\bar{D}^0 \Sigma_c^+$ and
	$D^- \Sigma_c^{++}$.

At large transverse momenta, the invariant mass $s \sim k_\perp^2$,
making the integral in Eq.~(\ref{eq:DLsplit}) logarithmically divergent.
A simple way to regulate this divergence is with a form factor $F(s)$,
which acts to suppress the ultraviolet contributions.
A convenient way to parametrize this regularizing form factor is
using an exponential function of $s$,
\begin{equation}
F(s) = \exp[-(s-M^2)/\Lambda^2],
\label{eq:FF_exp}
\end{equation}
which has the merit of possessing simple normalization properties
on-shell, although multipole, or power-law, functional forms as
in Eq.~(\ref{eq:powerFF}) would also suffice.
The splitting functions also depend upon the magnitude of the coupling
$g$ to each meson--baryon state, which we take from baryon--baryon
scattering models extended to the charm sector.  For simplicity,
we assume a universal exponential form factor for all couplings,
where the scale factor $\Lambda$ for that form factor is varied
to fit charmed baryon production in hadronic interactions,
as discussed below.
(Note, however, that the exponential form is used to define
the form factor $F(s)$ in Eq.~(\ref{eq:FF_exp}), in contrast
to the square of the form factor, as in Eq.~(\ref{eq:expFF});
the two can of course be related by a simple rescaling of the
cutoff mass $\Lambda^2 \to 2 \Lambda^2$).

For the dissociation of the proton to a charmed vector meson
$D^* = \bar{D}^{*0}$ or $D^{*-}$ and spin-1/2 charmed baryon, the
corresponding splitting function is given by a sum of vector ($G_v$),
tensor ($G_t$) and vector-tensor interference ($G_{vt}$) terms,
\bea 
f_{D^* B}(y)
&=& T_B \frac{1}{16\pi^2} \int{ dk_\perp^2 \over y (1-y) }
    { |F(s)|^2 \over (s - M^2)^2 }			\nonumber\\
& & \times
    \left[
	g^2\, G_v(y,k_\perp^2)\
     +\ {g f \over M}\, G_{vt}(y,k^2_\perp)\
     +\ \frac{f^2}{M^2}\, G_t(y,k^2_\perp)
    \right],
\label{eq:spin1-fMB}
\eea 
where 
\begin{subequations}
\label{eq:vectABC-app}%
\bea
G_v(y,k_\perp)
&=& - 6 M M_B\
 +\ \frac{4(P \cdot k) (p \cdot k)}{m_D^2}\
 +\ 2 P \cdot p,					\\
G_{vt}(y,k_\perp)
&=& 4(M + M_B)(P \cdot p - M M_B)			\nonumber\\
& &
 -\ \frac{2}{m_D^2}
    \left[ M_B (P \cdot k)^2
	 - (M + M_B)(P \cdot k)(p \cdot k)
	 + M (p \cdot k)^2
    \right],						\\
G_t(y,k_\perp)
&=& -(P \cdot p)^2\
 +\ (M + M_B)^2\, P \cdot p\
 -\ M M_B (M^2 + M_B^2 + M M_B)				\nonumber\\
& &
 +\ \frac{1}{2m_D^2}
    \Big[ (P \cdot p - M M_B) [(P-p) \cdot k]^2
	- 2 (M_B^2 P\cdot k - M^2 p\cdot k) [(P-p) \cdot k]	
\nonumber\\
& & \hspace*{1cm}
	+ 2 (P \cdot k) (p \cdot k) (2P \cdot k - M_B^2 - M^2)
    \Big],
\eea
\end{subequations}%
where $p$ is the four-momentum of the baryon,
and the inner products $P \cdot p$, $P \cdot k$ and $p \cdot k$
can be computed from Eq.~(\ref{eq:kin-TOPT}).
The splitting function $f_{D^* B}(y)$ describes transitions
to the states
        $\bar{D}^{*0} \Lambda_c^+$,
        $\bar{D}^{*0} \Sigma_c^+$ and
        $D^{*-} \Sigma_c^{++}$,
and the isospin transition factor $T_B$ is as in Eq.~(\ref{eq:TB}).
Again, for ease of notation, we have used the shorthand notation
for the couplings
  $g \to g_{D^* B N}$ and
  $f \to f_{D^* B N}$, with
  $s \to s_{D^* B}$.

For completeness, we also include fluctuations to spin-3/2 charmed
baryons $B^* = \Sigma_c^{* +}$ and $\Sigma_c^{* ++}$.  For the
dissociations of a proton to states with a spin-0 $D$ meson, 
	$\bar{D}^0 \Sigma_c^{* +}$ and
	$D^- \Sigma_c^{* ++}$,
the splitting function is given by
\bea
f_{D B^*}(y)
&=& T_{B^*}
    \frac{g^2}{16\pi^2} \int{ dk_\perp^2 \over y (1-y) }
    { |F(s)|^2 \over (s - M^2)^2 }			\nonumber\\
& & \times
    { \left[ k_\perp^2 + (M_{B^*} - (1-y) M)^2 \right]
      \left[ k_\perp^2 + (M_{B^*} + (1-y) M)^2 \right]^2
    \over 6 M_{B^*}^2 (1-y)^3},
\label{eq:DS*-split}
\eea
with 
  $g \to g_{D B^* N}$ and
  $s \to s_{D B^*}$.
The isospin transition factor $T_{B^*}$ here is similar to that
in Eq.~(\ref{eq:TB}), but with the third component of the charmed
baryon isospin
$t_{B^*} =  0$ for $B^* = \Sigma_c^{* +}$ and
$t_{B^*} = +1$ for $B^* = \Sigma_c^{* ++}$.

Finally, for the fluctuations to states with $D^*$ mesons
and spin-3/2 baryons $B^*$, the splitting function is
\bea 
f_{D^* B^*}(y)
&=& T_{B^*}
    \frac{g^2}{m_{D^*}^2 16\pi^2}
    \int{ dk_\perp^2 \over y(1-y) }
    { |F(s)|^2 \over (s - M^2)^2 }			\nonumber\\
& &
 -\ \left[
    \frac{4 M M_{B^*}}{3}
    \Big( 2 M_{B^*}^2 + M M_{B^*} + 2 M^2 \Big)
     - \frac{4 M M_{B^*}}{3 m_{D^*}^2} ((P-p) \cdot k)^2
    \right.						\nonumber\\
& &
 -\ \frac{4}{3m_{D^*}^2}
    \Big( M_{B^*}^2 (P \cdot k)^2 + M^2 (p \cdot k)^2 \Big)
     + \frac{4 P \cdot p}{3}
	\Big( 2 M_{B^*}^2 + 4 M M_{B^*} + M^2 \Big)	\nonumber\\
& & \left.
 +\ \frac{4 P \cdot p}{3m_{D^*}^2}(p \cdot k)^2
	 \Big( 1-\frac{M^2}{M_{B^*}^2} \Big)
    - 4 (P \cdot p)^2
        \left(
	  1 - \frac{2 (P \cdot k)(p \cdot k)}{3m_{D^*}^2 M_{B^*}^2}
	    - \frac{P \cdot p}{3 M_{B^*}^2}
	\right)
    \right],						\nonumber\\
& &
\label{eq:RS-fMB}
\eea
with
  $g \to g_{D^* B^* N}$,
  $s \to s_{D^* B^*}$,
and the inner products in Eq.~(\ref{eq:RS-fMB}) obtained
from Eq.~(\ref{eq:kin-TOPT}) with the replacements
$D \to D^*$ and $B_c \to \Sigma_c^*$.
The details of the derivations of all of the functions
$f_{DB}(y)$, $f_{D^* B}(y)$, $f_{DB^*}(y)$ and $f_{D^* B^*}(y)$
can be found in Appendix~\ref{sec:A_MBM}.

\subsection{Constraints from inclusive charmed hadron production}
\label{ssec:HSM}

To calculate the contributions of the various charmed mesons and baryons
listed in Table~\ref{table:mass_spect} requires the couplings of these
states to the proton.  In this analysis we take the coupling constants
from boson-exchange models that were originally applied to pion-nucleon
interactions \cite{Machleidt87}, and later generalized to $KN$
scattering \cite{HDHPS95}.  In the extension to the strange sector,
the relevant couplings are taken from non-strange analyses with SU(3)
arguments used to incorporate the corresponding strange particles.
The off-shell behavior of the amplitudes is typically regulated by
a multipole form factor of the type
\begin{equation}
F(t) = \left( \frac{\Lambda^2 + m_M^2}{\Lambda^2 - t} \right)^n,
\label{eq:t-chan_FF}
\end{equation}
where $t$ is the usual Mandelstam variable for the squared momentum
transfer of the exchanged meson with mass $m_M$.  A monopole form factor
($n=1$) is generally used for low-spin states, while for higher-spin
states a dipole form factor ($n=2$) is typically employed to damp the
higher powers of momentum that enter into the transition amplitudes.

The extension of meson--baryon couplings from the non-strange to the
strange sector has generally been quite successful phenomenologically.
Continuing this program further to the charm sector, Haidenbauer
{\it et al.} \cite{Hai07, Hai08, Hai11} used SU(4) symmetry arguments
to describe exclusive charmed hadron production in $\bar{D}N$ and $DN$
scattering within a one-boson-exchange framework.
We fix the couplings for the spin-1/2 charmed baryons $\Lambda_c$ and
$\Sigma_c$ to those found in Ref.~\cite{Hai11}, as summarized in
Table~\ref{table:couplings}.  For the couplings to spin-3/2 states
$\Sigma_c^*$, we take the couplings from those obtained for the
analogous strange states by Holzenkamp \EA~\cite{Holzenkamp89}.
The signs of the couplings are related to the value of the $\pi NN$
coupling, for which we use $g_{\pi NN}/\sqrt{4\pi} = -3.795$.

\begin{table}[t]
\caption{Charm-sector coupling constants, deduced from $DN$
	and $\bar D N$ scattering analyses \cite{Hai11} for
	the spin-1/2 charmed baryons $\Lambda_c$ and $\Sigma_c$,
	and by extending the SU(3) sector analysis of
	Ref.~\cite{Holzenkamp89} for the spin-3/2 $\Sigma_c^*$
	states.}
\centering
\begin{tabular}{l c c}				\hline\hline
Vertex 				& $g_{MBN}/\sqrt{4\pi}$\ \ \ \ \
				& $f_{MBN}/\sqrt{4\pi}$	\\
[0.5ex] \hline
$p \to \bar{D}^0 \Lambda_c^+$ 	& 3.943    & ---      \\
$p \to \bar{D}^{*0}\Lambda_c^+$	& 1.590    & 5.183    \\
$p \to \bar{D}^0 \Sigma_c^+,\
	D^- \Sigma_c^{++}$ 	& 0.759    & ---      \\
$p \to \bar{D}^{*0}\Sigma_c^+,\
	D^{*-} \Sigma_c^{++}$ 	& 0.918    & $-2.222$ \\
$p \to \bar{D}^0 \Sigma_c^{*+},\
	D^- \Sigma_c^{*++}$	& $-0.193$ & ---      \\
$p \to \bar{D}^{*0}\Sigma_c^{*+},\
	D^{*-} \Sigma_c^{*++}$	& $-1.846$ & ---      \\
[1ex] \hline
\end{tabular}
\label{table:couplings}
\end{table}

The remaining parameters of the model are the form factor cutoffs,
$\Lambda$, which could in principle be constrained for the various
meson--baryon vertices by data from exclusive or inclusive charmed
baryon production.  In practice, such data are rather limited,
and the most direct constraints come from inclusive $\Lambda_c$
production in proton--proton scattering, $p p \to \Lambda_c X$,
measured by the R680 Collaboration at the ISR \cite{Chauvat:1987kb}.
Since it is currently not possible to constrain the cutoffs for
the individual charmed meson--baryon configurations, we assume a
universal exponential form factor cutoff as in Eq.~(\ref{eq:FF_exp})
for all the fluctuations listed in Table~\ref{table:couplings},
and tune $\Lambda$ to best reproduce the shape and normalization
of the inclusive $\Lambda_c$ production cross section data.
This will place an upper bound on $\Lambda$ and the magnitude
of the charmed meson--baryon contribution in the MBM.

Within the same one-boson exchange framework as adopted in the $DN$
and $\bar{D} N$ scattering analyses \cite{Hai07, Hai08, Hai11},
the contribution from charmed meson exchange to the differential
cross section for inclusive baryon production in $pp$ scattering
can be written \cite{Holtmann96}
\begin{equation}
E \frac{d^3\sigma}{d^3\bm{p}}\
=\ \frac{\bar{y}}{\pi}
   \frac{d^2\sigma}{d\bar{y}\, dk_\perp^2}\
=\ \frac{\bar{y}}{\pi} \sum_M
   \left| \phi_{BM}(\bar{y},k_\perp^2) \right|^2\,
   \sigma_{\rm tot}^{Mp}(sy),
\label{eq:diff3_CS}
\end{equation}
where $E$ is the energy of the proton beam, and the sum over
$M$ includes incoherent contributions from processes involving
the exchange of meson $M$ leading to a final baryon $B$.
The total meson--proton cross section $\sigma_{\rm tot}^{Mp}$
here is evaluated at the meson energy $s y$, with $s$ being
the total $pp$ invariant mass squared.
For the case of $\Lambda_c^+$ production, the sum is
restricted to the $\bar{D}^0$ and $\bar{D}^{*0}$ mesons.
The $k_\perp^2$-integrated cross section for $\Lambda_c^+$
production is then given by
\begin{equation}
\frac{d\sigma}{d\bar{y}}
= \sum_{M=D,D^*}
  f_{\Lambda_c^+ M}(\bar{y})\,
  \sigma_{\rm tot}^{Mp}(sy).
\label{eq:diff_CS}
\end{equation}
Note that in Ref.~\cite{Cazaroto13} this cross section is defined
with an additional factor $(\pi/\bar y)$ on the right hand side.
For the total charmed meson--proton cross section
$\sigma_{\rm tot}^{Mp}$ we take a constant value, as suggested
by the analysis of pion-nucleon scattering \cite{Holtmann96},
where $\sigma_{\rm tot}^{\pi p} \approx \sigma_{\rm tot}^{\rho p}$.
Adopting a similar approach to the strange and charmed meson cross
sections, we have
\begin{equation}
\sigma_{\rm tot}^{Dp}\
\approx\ \sigma_{\rm tot}^{D^*p}\
\approx\ \sigma_{\rm tot}^{\bar{K}p}\
\approx\ (20 \pm 10) \ \mathrm{mb},
\label{eq:ISR_Mp}
\end{equation}
where the value of the $\bar{K}p$ total cross section is taken
from Ref.~\cite{Beringer:1900zz}, and we assign a conservative
50\% uncertainty on the central value.

\begin{figure}[t]
\includegraphics[width=8.0cm]{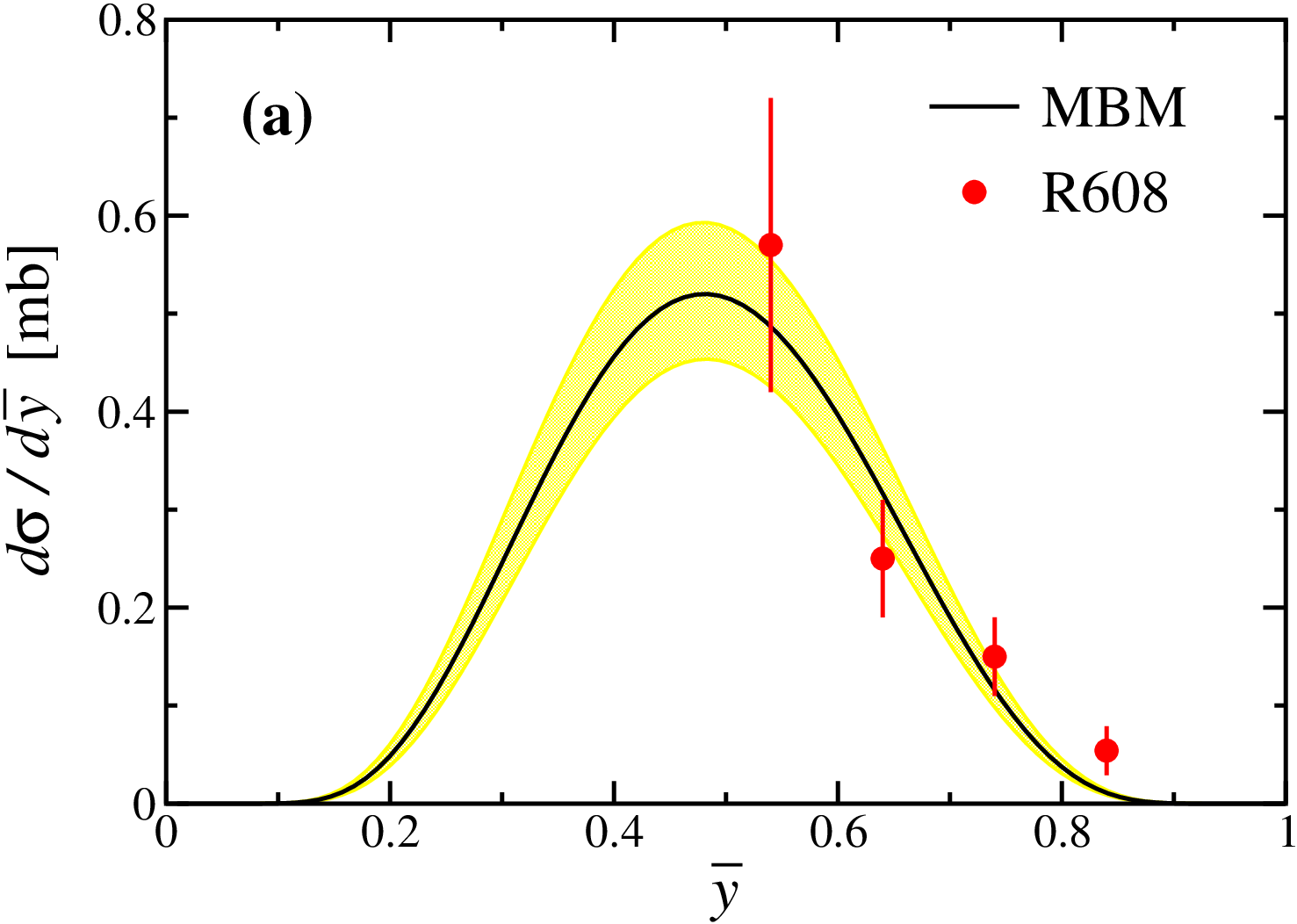}
\includegraphics[width=8.2cm]{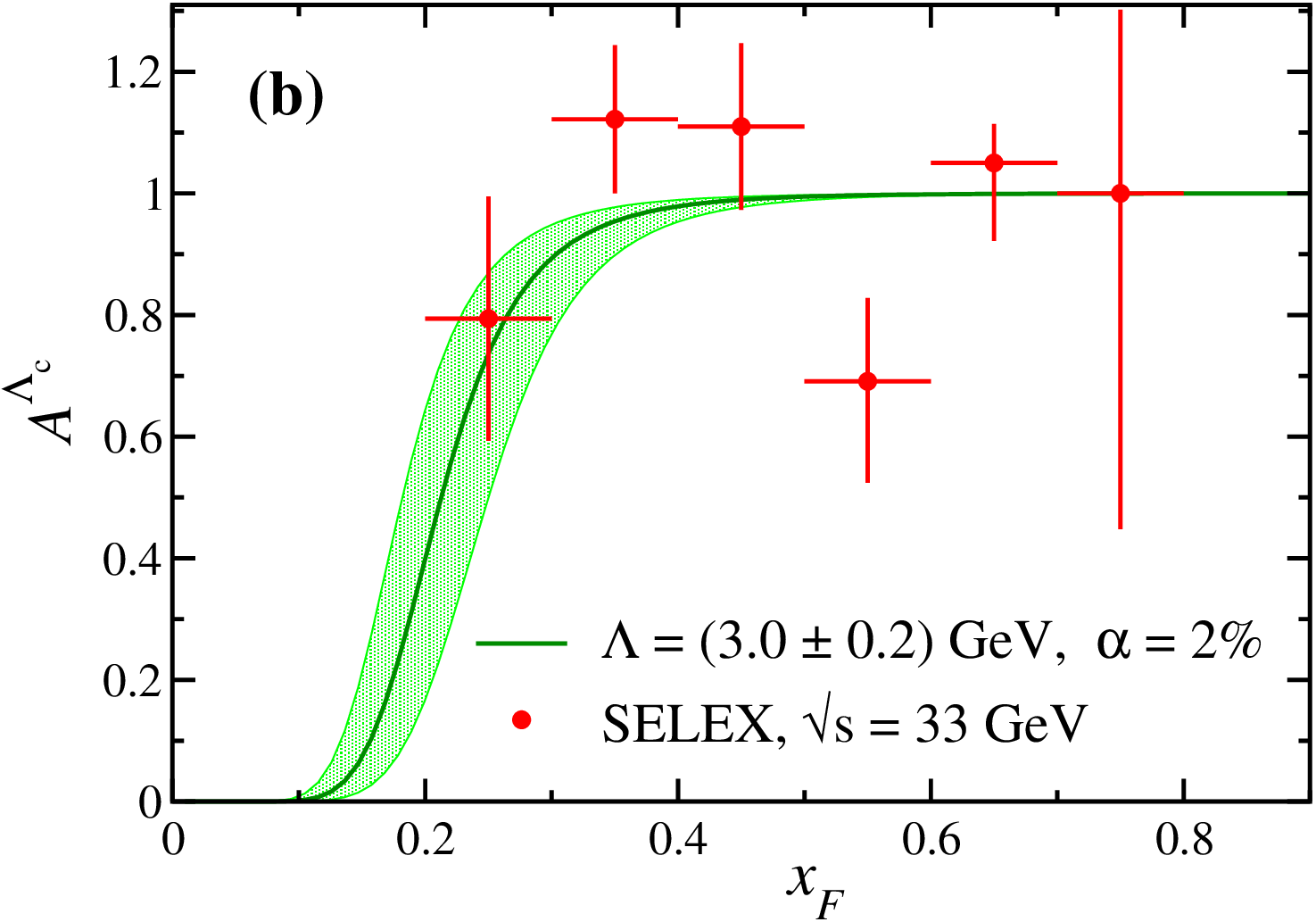}
\caption{(color online)
	{\bf (a)}
	Differential cross section $d\sigma/d\bar y$ for the inclusive
	charm production reaction $pp \to \Lambda_c^+ X$ as a function
	of the momentum fraction $\bar y$ carried by the $\Lambda_c^+$.
	The MBM cross section (solid) is computed using the central value
	for the $Dp$ cross section $\sigma^{Dp}_{\rm tot} = 20$~mb,
	and the resulting error band (shaded) represents the
	purely statistical uncertainty.  The data (red circles)	are
	from the R608 collaboration at the ISR \cite{Chauvat:1987kb}.
	{\bf (b)}
	Charge asymmetry $A^{\Lambda_c}$ for
	$\Lambda_c^+/\bar\Lambda_c^-$ production in the MBM (solid),
	using the $\xF$ dependence of the $\bar\Lambda_c$ cross section
	in Eq.~(\ref{eq:SELEX-iii}), compared with data from the
	SELEX Collaboration \cite{Garcia:2001xj}.}
\label{fig:R608_CS}
\end{figure}

Using Eqs.~(\ref{eq:diff_CS}) and (\ref{eq:ISR_Mp}), the calculated
cross section is shown in Fig.~\ref{fig:R608_CS}(a) at the kinematics
of the $\Lambda_c^+$ production data from the R608 Collaboration
\cite{Chauvat:1987kb} at the ISR.  The kinematical coverage of the
ISR data was restricted to $k_\perp \le 1.1$~GeV, which we impose
in the computed cross section.  For the central value of the $Dp$
total cross section, $\sigma_{\rm tot}^{Dp} = 20$~mb, the best fit
value of the cutoff parameter is found to be
	$\Lambda = (2.89 \pm 0.04)$~GeV,
which gives a good fit to both the overall normalization and the
shape of the inclusive $\Lambda_c^+$ data.  Including the uncertainty
in the $Dp$ cross section from Eq.~(\ref{eq:ISR_Mp}), the cutoff becomes
	$\Lambda = (3.0 \pm 0.2)$~GeV.

Note that in calculating the inclusive $\Lambda_c^+$ production
cross section we have included the possibility that higher-mass
baryons such as $\Lambda_c^*$ and $\Sigma_c^*$ are produced
and subsequently decay to a $\Lambda_c^+$, using the relevant
branching ratios for the decays.  With the coupling constants
from Table~\ref{table:couplings}, we find that the dominant
contribution to inclusive $\Lambda_c$ production arises from
the state $\bar{D}^{*0} \Lambda_c^+$.  This is in contrast to
earlier analyses, where the largest contribution was assumed
to be from the lowest-energy $\bar{D}^0 \Lambda_c^+$ state.
As we discuss below, this will have significant ramifications
for intrinsic charm production in electromagnetic reactions.

Note also that the original data from Ref.~\cite{Chauvat:1987kb}
were recorded in terms of the variable
    \mbox{$\xF = 2 p_{\Lambda}^0 / \sqrt{s}$},
which in general differs from the momentum fraction $\bar y$
that scales the calculations of the MBM.
It can be shown, however, that at high energies
(\IE~$s \gg M_B^2$, $k_\perp^2$), one has $\bar y \to \xF$.

More recently the SELEX Collaboration at Fermilab \cite{Garcia:2001xj}
has produced data on the charge asymmetry for inclusive $\Lambda_c^+$
and $\bar\Lambda_c^-$ production in the scattering of 540~GeV protons
from copper and carbon targets,
\be
A^{\Lambda_c}(\xF)
= { \sigma^{\Lambda_c}(x_F) - \sigma^{\bar{\Lambda}_c}(x_F)
    \over
    \sigma^{\Lambda_c}(x_F) + \sigma^{\bar{\Lambda}_c}(x_F) },
\label{eq:SELEX-i}
\ee
where $\sigma^{\Lambda_c}(x_F) \equiv d\sigma^{\Lambda_c}/d\xF$.
While the contribution to the production of $\Lambda_c^+$ can be
calculated in the MBM, the computation of the asymmetry $A^{\Lambda_c}$
requires in addition an estimate of the $\bar{\Lambda}_c$ cross
section.  Following Ref.~\cite{Garcia:2001xj} we approximate this
using a simple monomial parametrization.
Furthermore, we assume that the $\Lambda_c^+$ cross sections can be
written as the sum of valence and sea components, with the generation
of the former described by the nonperturbative MBM and dominating at
intermediate and high values of $\xF$, and the latter concentrated
at small $\xF$,
\be
\frac{d\sigma^{\Lambda_c}}{d\xF}
= \frac{d\sigma^{\Lambda_c}_{\rm (val)}}{d\xF}
+ \frac{d\sigma^{\Lambda_c}_{\rm (sea)}}{d\xF},
\label{eq:SELEX-ii}
\ee
where
\begin{subequations}
\label{eq:SELEX-iii}
\begin{eqnarray}
\frac{d\sigma^{\Lambda_c}_{\rm (val)}}{d\xF}
&\approx& \sigma_0 \sum_M f_{\Lambda_c M}(\xF),
\label{eq:SELEX-iiia}					\\
\frac{d\sigma^{\Lambda_c}_{\rm (sea)}}{d\xF}
&\equiv& \frac{d\sigma^{\bar{\Lambda}_c}}{d\xF}\
 \approx\ \bar\sigma_0 (1-\xF)^{\bar{n}}.
\label{eq:SELEX-iiib}
\end{eqnarray}%
\end{subequations}%
In Eq.~(\ref{eq:SELEX-iiia}) the factor $\sigma_0$ corresponds to
the total meson--proton cross section in Eq.~(\ref{eq:diff_CS}),
which we take to be independent of the flavor and spin of the
meson, as in Eq.~(\ref{eq:ISR_Mp}), while $\bar\sigma_0$ is a
normalization parameter for the corresponding $\bar{\Lambda}_c$
production cross section.
Using Eqs.~(\ref{eq:SELEX-iii}), the asymmetry in Eq.~(\ref{eq:SELEX-i})
can then be written
\be
A_{\Lambda_c}(\xF)
= \frac{ \sum_M f_{\Lambda_c M}(\xF)}
       { \sum_M f_{\Lambda_c M}(\xF)
	 + 2 \alpha (1-\xF)^{\bar{n}} },
\label{eq:SELEX-iv}
\ee
where
    $\alpha = \bar\sigma_0 / \sigma_0$
is the ratio of the sea to valence contributions to the
$\Lambda_c$ cross sections.
For $\bar{\Lambda}_c$ production induced by $\Sigma^-$
beams, the SELEX Collaboration found for the exponent
    $\bar{n} \approx 6.8$,
which we assume also for the $x_F$ dependence of the proton
induced cross section in Eq.~(\ref{eq:SELEX-iiib}).
Using the MBM cutoff parameter $\Lambda = (3.0 \pm 0.2)$~GeV,
a good fit to the SELEX charge asymmetry data can then be
obtained with $\alpha \approx 2.0\%$, as illustrated in
Fig.~\ref{fig:R608_CS}(b).
We should note, however, that $\Lambda_c$ charge asymmetry
data are rather sensitive to the form of the $\bar{\Lambda}_c$
cross section, so that agreement with the SELEX data should
not be considered as a stringent test of the MBM; rather,
with an appropriate choice of parameter $\alpha$ the model
is able to accommodate the empirical results.

Having constrained the scale parameter for the meson--baryon
form factor by the inclusive $\Lambda_c$ production data,
and with the coupling constants for the various meson--baryon
states given in Table~\ref{table:couplings}, we are now able to
compute the meson--baryon splitting functions in Eq.~(\ref{eq:fphi}),
which we consider in the following section.

\subsection{Phenomenology of charmed meson--baryon splitting functions}
\label{ssec:splitfin}

\begin{figure}[t]
\includegraphics[width=11cm]{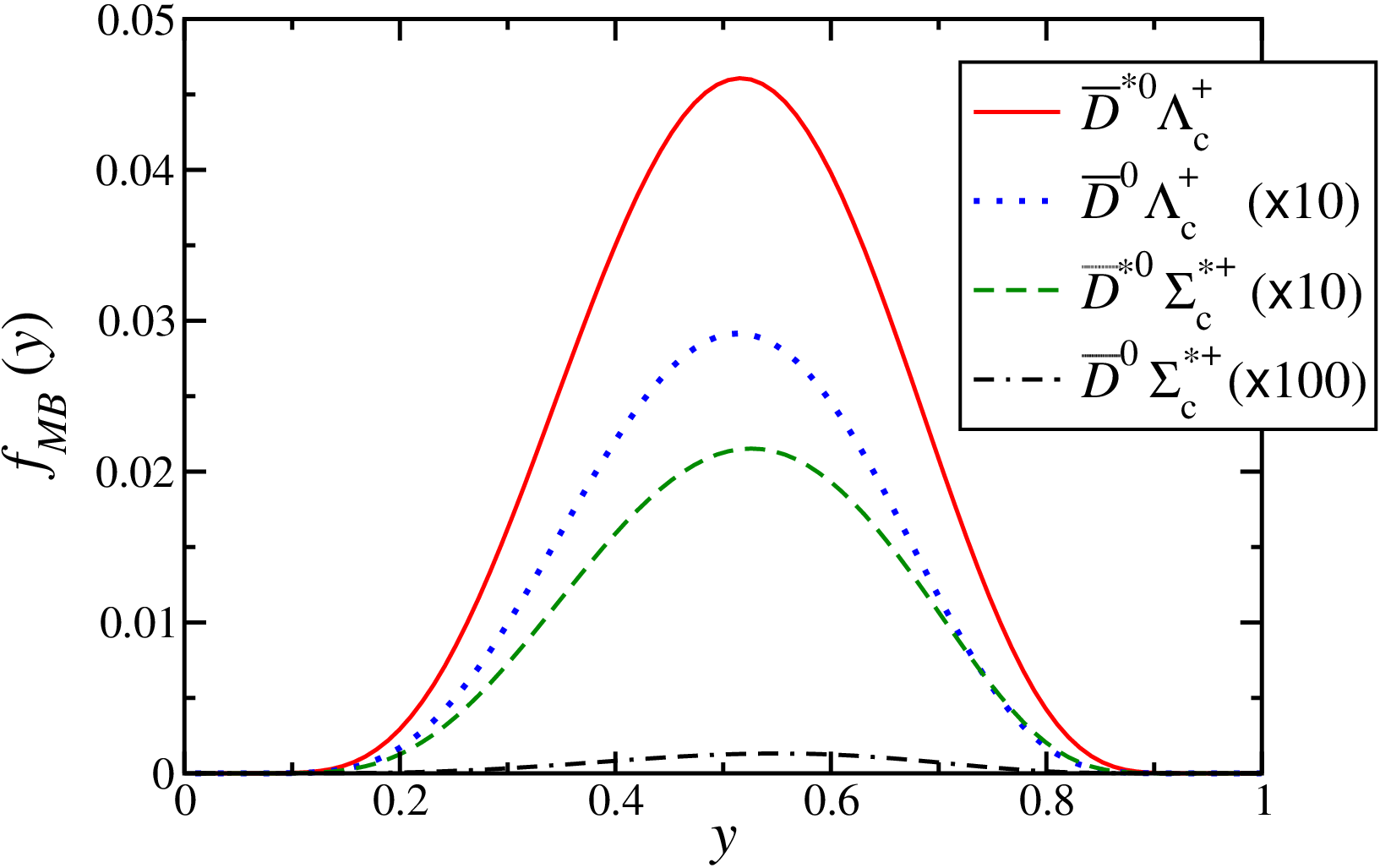}
\caption{(color online)
	Splitting functions for the four basic dissociations
	of a proton into charmed meson--baryon states, for the
	spin-1 meson + spin-1/2 baryon state
	$\bar{D}^{*0} \Lambda_c^+$
	  (red solid),
	spin-0 meson + spin-1/2 baryon state
	$\bar{D}^0 \Lambda_c^+$
	  (scaled $\times 10$, blue dotted),
	spin-1 meson + spin-3/2 baryon state
	$\bar{D}^{*0} \Sigma_c^{*+}$
	  (scaled $\times 10$, green dashed), and
	spin-0 meson + spin-3/2 baryon state
	$\bar{D}^0 \Sigma_c^{*+}$
	  (scaled $\times 100$, black dot-dashed).
	A universal exponential cutoff mass $\Lambda = 3$~GeV is
	used with the couplings from Table~\ref{table:couplings}.}
\label{fig:f_MB}
\end{figure}

The complete set of the four basic splitting functions representing
the dissociation of a proton to charmed meson--baryon states
$p \to D B$ (pseudoscalar meson + octet baryon),
    $D^* B$ (vector meson + octet baryon),
    $D B^*$ (pseudoscalar meson + decuplet baryon) and
  $D^* B^*$ (vector meson + decuplet baryon)
is illustrated in Fig.~\ref{fig:f_MB}.  The functions are shown
for the neutral $\bar{D}^0$ and $\bar{D}^{*0}$ mesons, and all
the other charge states in Table~\ref{table:couplings} can be
obtained using appropriate Clebsch-Gordan coefficients.
For the best fit value of the universal cutoff parameter
$\Lambda = 3$~GeV from the inclusive $\Lambda_c^+$ production
data, the $\bar{D}^{*0} \Lambda_c^+$ contribution is found to be
dominant, an order of magnitude larger than the corresponding
$\bar{D}^0 \Lambda_c^+$ and $\bar{D}^{*0} \Sigma_c^{*+}$ contributions.
The $\bar{D}^0 \Sigma_c^{*+}$ contribution is two orders of magnitude
smaller still, and effectively plays no role in the phenomenology.
To a good approximation, therefore, one can represent the total charm
distribution in the proton by the single $\bar{D}^{*0} \Lambda_c^+$ state.

Interestingly, the shapes of the various charmed meson--baryon
distributions $f_{MB}(y)$ are rather similar, peaking just above
$y = 1/2$.  This is in contrast to the distributions in the light
flavor sector, where the corresponding $\pi N$ splitting function
is considerably more skewed in~$y$~\cite{MT93, Holtmann96}.
The skewedness arises from the large difference in mass between
the pion and nucleon in the dissociation, whereas the masses of
both the charmed meson and baryon are of the order $\sim 2$~GeV.
This is also one reason why the lowest mass $\pi N$ configuration
is the dominant one in the pion sector, unlike the lowest mass
charmed state $\bar{D}^0 \Lambda_c^+$, which as Fig.~\ref{fig:f_MB}
indicates gives a significantly smaller contribution than the
$\bar{D}^{*0} \Lambda_c^+$.
The dominance of the SU(2) flavor sector by the $\pi N$ state is
ensured by the relatively large energy gap between higher mass
configurations involving $\rho$ mesons or $\Delta$ baryons,
whereas no significant energy gap exists between the various states
in the charm sector.

\begin{figure}[t]
\includegraphics[width=11cm]{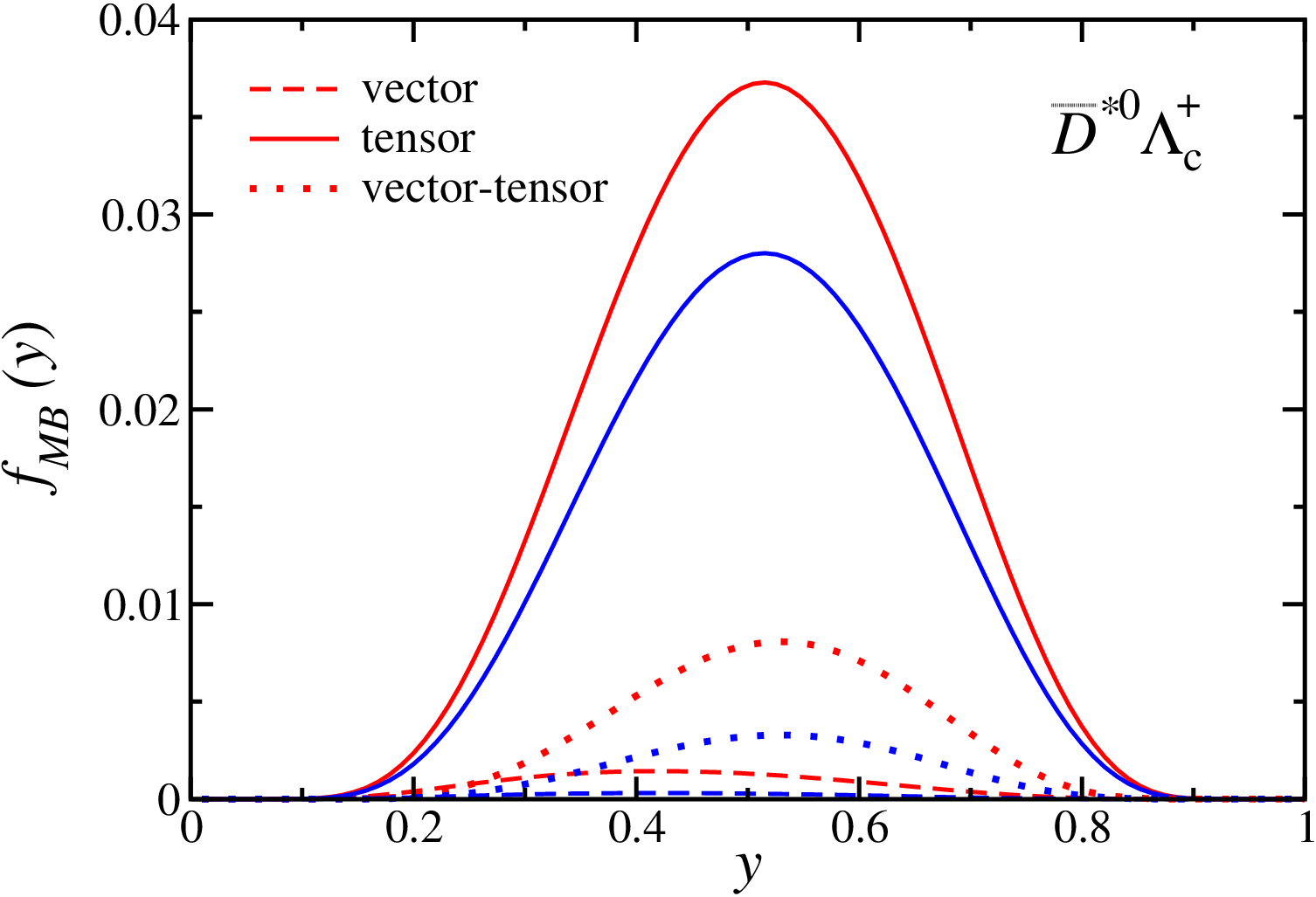}
\caption{(color online)
	Contributions to the $p \to \bar{D}^{*0} \Lambda_c^+$
	splitting function from the vector (dashed), tensor (solid)
	and vector-tensor interference (dotted) terms in
	Eq.~(\ref{eq:spin1-fMB}).
	The curves are computed for a cutoff mass $\Lambda = 3$~GeV
	with charm coupling constants from SU(4) symmetry (red)
	and using the corresponding SU(2) values for $\rho NN$ (blue).}
\label{fig:fD*_terms}
\end{figure}

To explore further the origin of the dominance of the
$\bar{D}^{*0} \Lambda_c^+$ contribution, we note the relatively
strong coupling to the vector meson state, particularly for the
tensor coupling term, as seen in Table~\ref{table:couplings},
where the tensor to vector coupling ratio is
    $f_{D^* \Lambda_c N} / g_{D^* \Lambda_c N} = 3.26$
\cite{Hai07, Hai08, Hai11}.
This is analogous to the large tensor coupling for the $\rho$
meson in one-boson exchange models of the $NN$ interaction
\cite{Machleidt87}, where in the Bonn-J{\"u}lich model,
for instance, one has an even larger tensor/vector ratio,
    $f_{\rho NN} / g_{\rho NN} = 6.1$
\cite{Hohler75}.
The individual vector, tensor, and vector-tensor interference
contributions to the $f_{\bar{D}^{*0} \Lambda_c^+}$ splitting
function are shown in Fig.~\ref{fig:fD*_terms}, for the universal
$\Lambda = 3$~GeV exponential cutoff as in Fig.~\ref{fig:f_MB}.
Using the charm couplings from Table~\ref{table:couplings},
the tensor contribution clearly dominates over the vector term.
This feature is preserved even if one uses the SU(2) couplings from
the $\rho$ exchange in the $NN$ analysis instead of the SU(4) couplings
\cite{Hai07, Hai08, Hai11} (but with the same charm hadron masses).
In particular, since the SU(4) vector coupling
    $g_{D^* \Lambda_c N}^2/4\pi = 2.53$
is around 5 times larger than that found for the $\rho$ from
$NN$ analyses,
    $g_{\rho NN}^2/4\pi = 0.55$
\cite{Hohler75}, the vector contribution to the charm splitting
function is significantly larger than for the SU(2) coupling case.
This is compensated somewhat by the $\sim 2$ times smaller SU(4)
tensor/vector ratio, making the total contribution to the charmed
vector meson splitting function $f_{\bar{D}^{*0} \Lambda_c^+}(y)$
similar.

\begin{figure}[t]
\includegraphics[width=8.1cm]{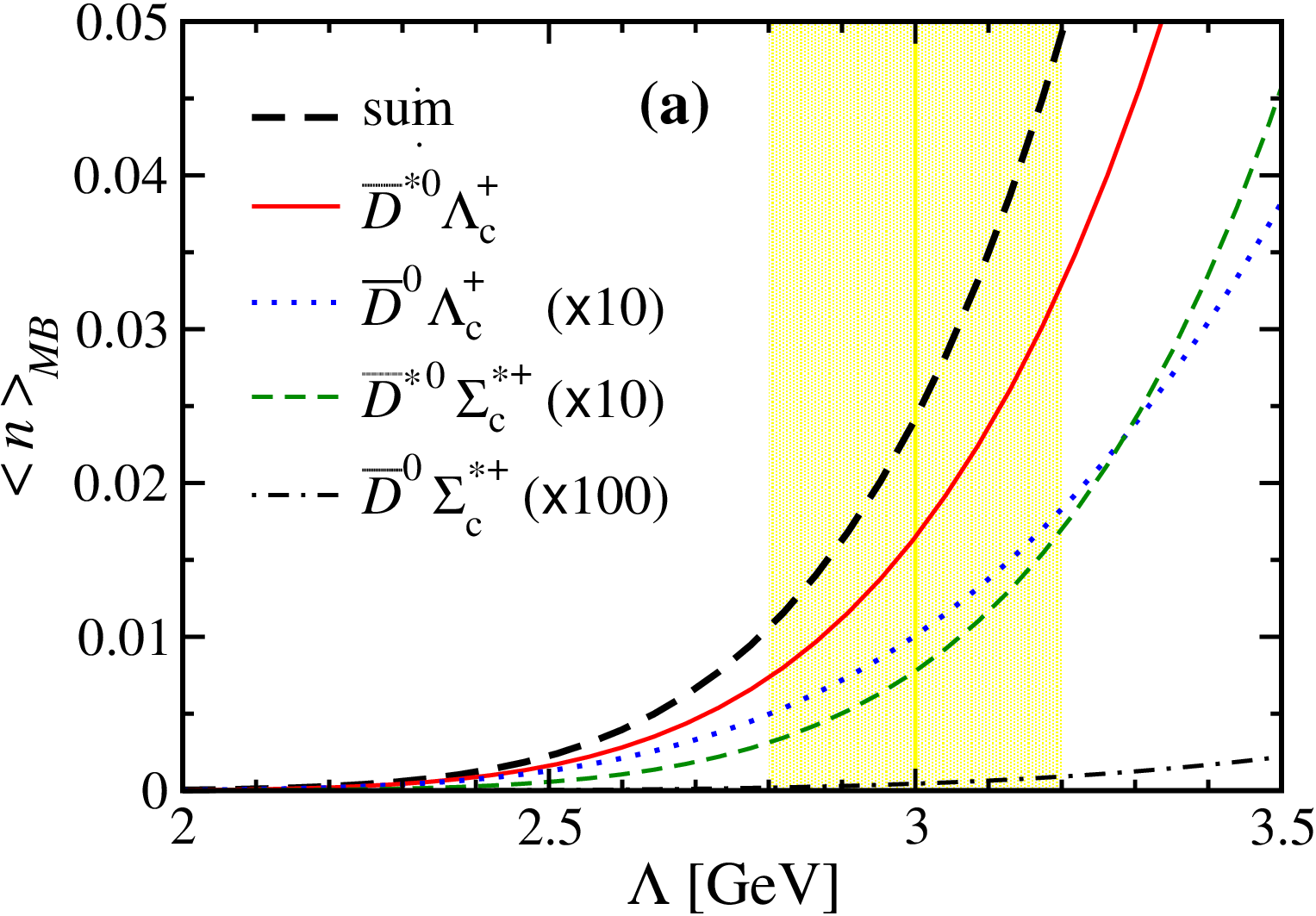}
\includegraphics[width=7.9cm]{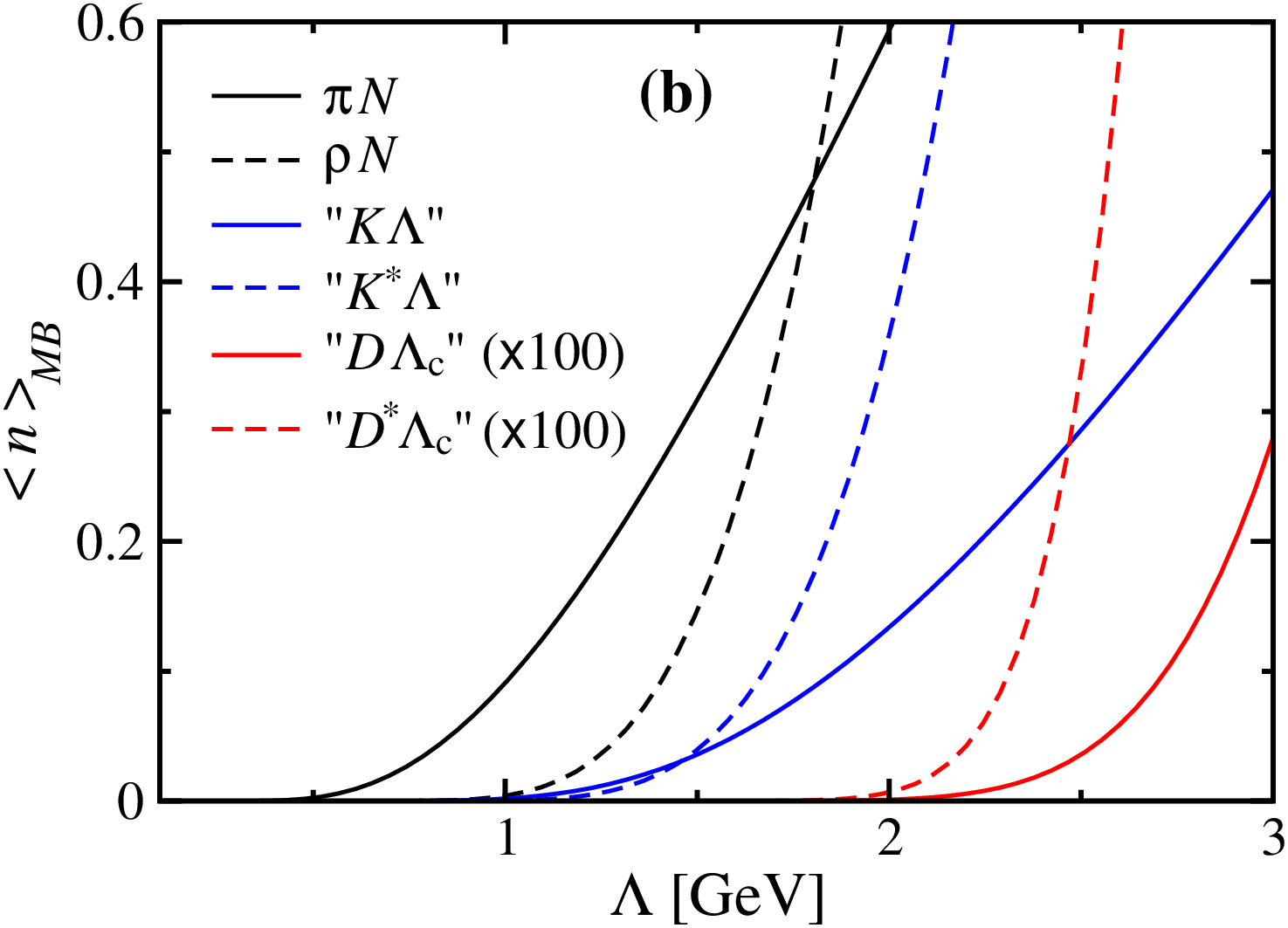}
\caption{(color online)
	{\bf (a)}
	Normalizations $\langle n \rangle_{MB}$ of the charmed
	meson--baryon splitting functions as a function	of the
	form factor cutoff $\Lambda$, for the states
	$\bar{D}^{*0} \Lambda_c^+$
	  (red solid),
        $\bar{D}^0 \Lambda_c^+$
	  (scaled $\times 10$, blue dotted),
        $\bar{D}^{*0} \Sigma_c^{*+}$
	  (scaled $\times 10$, green dashed), and
	$\bar{D}^0 \Sigma_c^{*+}$
	  (scaled $\times 100$, black dot-dashed),
	as well as the sum of all contributions (black dashed).
	The (yellow) shaded band represents the uncertainty on the cutoff
	obtained from fits to inclusive $\Lambda_c^+$ production data.
	{\bf (b)}
	Normalizations of the splitting functions to pseudoscalar
	(solid) and vector (dashed) mesons computed with
	SU(2) sector ($\pi N$ and $\rho N$) masses (black),
	SU(3) masses, denoted by ``$K \Lambda$'' and
	``$K^* \Lambda$'' (blue), and
	SU(4) masses, denoted by ``$D \Lambda_c$'' and
	``$D^* \Lambda_c$'' (scaled $\times 100$, red),
	all for the same SU(2) couplings.}
\label{fig:f_MB_int}
\end{figure}

At the effective Lagrangian level, the large tensor contribution
is associated with the additional momentum dependence induced
by the derivative coupling in the tensor interaction, which is
a general feature of couplings to states with higher spin
[see Eq.~(\ref{eq:L-S1})].  This additional momentum dependence
can have a significant impact on the relative importance of various
charmed meson--baryon transitions, as is evident from the form of
the splitting function in Eq.~(\ref{eq:spin1-fMB}).
The effect of the momentum dependence of the meson--baryon vertices
on the splitting functions can be illustrated even more dramatically
by considering the normalizations
    $\langle n \rangle_{MB} = \int dy\, f_{MB}(y)$
as a function of the cutoff $\Lambda$.
These are displayed in Fig.~\ref{fig:f_MB_int}(a) for the four
charmed states shown in Fig.~\ref{fig:f_MB}, together with the
sum over all contributions.
At the best fit value of $\Lambda \sim 3$~GeV, the lowest mass
vector state $\bar{D}^{*0} \Lambda_c^+$ makes up around 70\%
of the total charm normalization of
    $\langle n \rangle_{MB}^{\rm (charm)} = 2.40\%$.
Including the uncertainty on the cutoff (indicated by the shaded
band), the total charm normalization ranges from $\approx 1.04\%$
to $\approx 4.87\%$.

While the variation of the charm splitting functions with the choice
of SU(4) or SU(2) couplings was found in Fig.~\ref{fig:fD*_terms}
to be relatively mild, a significant effect is seen for the
dependence of the splitting functions on the hadron masses.
In Fig.~\ref{fig:f_MB_int}(b) the normalizations
$\langle n \rangle_{MB}$ of splitting functions to pseudoscalar
and vector mesons are illustrated for the light-quark, strange
and charmed sectors, using the SU(2) coupling constants for
$\pi NN$ and $\rho NN$ listed above.
The curves labeled ``$K \Lambda$'' and ``$K^* \Lambda$'' are
obtained from the $\pi N$ and $\rho N$ splitting functions by
replacing the pion and recoil baryon masses with the appropriate
kaon and hyperon masses, and those labeled ``$D \Lambda_c$''
and those labeled ``$D^* \Lambda_c$'' are obtained by using the
corresponding charmed meson and baryon masses.
For small values of the cutoff, the normalizations of the
pseudoscalar meson--baryon states is larger than for the vector
meson states, but with increasing $\Lambda$ the contributions
from the vector meson states eventually dominate.
With increasing hadron masses the cross-over point between
the pseudoscalar and vector meson states occurs at progressively
smaller $\Lambda$ values.  Neglecting differences between the
coupling constants (which are small if quark model symmetries
are assumed), the size of hadronic masses relative to the cutoff
scale $\Lambda$ is the main determinant of the balance between
the pseudoscalar and vector states for a given flavor sector.

The best fit value $\Lambda = 3$~GeV for the charmed splitting
functions corresponds to a region where the vector meson term
clearly dominates over the pseudoscalar meson contribution.
Had we found a significantly softer cutoff $\Lambda \sim 1$~GeV,
the pseudoscalar contribution would have dominated, although for
such values the total charm contribution would be negligible.
Since only a single charmed baryon production cross section was
available to constrain the charm splitting functions, only a
single parameter $\Lambda$ could be determined.
The existence of data for various charmed channels, on the other
hand, would allow the cutoffs to be determined for individual
meson--baryon states.  This could in principle lead to hard form
factor cutoffs for some states and soft cutoffs for others,
which would affect the degree to which the charmed vector meson
states dominate the splitting functions.
The results of our MBM calculations imply that the production of
charmed mesons in $pp$ reactions would occur almost entirely through
$D^*$ mesons, with subsequent decays of $D^*$ to $D$ mesons.

\section{Charm distributions in charmed hadrons}
\label{sec:cinc}

Within the two-step, meson--baryon convolution picture in 
Eqs.~(\ref{eq:mesoncloud}), the charm and anticharm distributions in
the nucleon require knowledge of the $c$ and $\bar c$ distributions
in the charmed baryon and meson.
In applications of the MBM to the light quark sector, where the
$\bar u$ and $\bar d$ distributions arise from the pionic component
of the nucleon wave function, one can use data from measurements of
the pion structure function \cite{Conway89, Wij05}.  There is even
some empirical information on the kaon structure function from the
Drell-Yan reaction \cite{Aitkenhead80}, which has been utilized
in calculations of strange quark distributions in the nucleon
\cite{Signal87, Malheiro97}.  In contrast, nothing at all is known
about the partonic structure of charmed hadrons, so that in practice
these need to be modelled.

In the literature estimates of the distributions of heavy quarks in
heavy hadrons have been made using the heavy quark limit \cite{MT97},
and within a scalar constituent quark model \cite{Pum05}.
Here we combine several features of these approaches in constructing
a relativistic quark model, with the correct spin degrees of freedom,
that parallels the splitting function calculations of Sec.~\ref{sec:mb}.
We apply the time-ordered perturbation theory framework in the IMF
at the parton level, defining the IMF momentum fraction
$\hat{y} = \hat{k}_L/P_L$ to be the ratio of the longitudinal
momentum of the constituent quark or antiquark ($\hat{k}_L$) to
that of the parent charmed meson or baryon ($P_L$).
Convolution with the leading twist point-like structure of constituent
quarks gives distributions as functions of the quark-level Bjorken
limit variable, denoted here by $z$ to prevent confusion with quark
distributions in the proton.
In the following we summarize the results for the $\bar c$ distributions
in $D$ and $D^*$ mesons, and the $c$ distributions in the $\Lambda_c$
and $\Sigma_c^*$ baryons; technical details of the calculations are
presented in Appendix~\ref{sec:B-MBM}.

\subsection{Anticharm in charmed mesons}
\label{ssec:cbar_in_D}

The distribution of a relativistic $\bar c$ quark in a pseudoscalar
$D$ meson, with a spectator $u$ or $d$ quark, can be computed in
analogy with the $p \to D \Lambda_c$ splitting function in
Eq.~(\ref{eq:DLsplit}).  Using a pseudoscalar meson--quark--antiquark
vertex parametrized by the structure $\gamma_5 G(\hat{s})$, where
$G(\hat{s})$ is the $D$-$\bar c$-$q$ vertex function ($q=u, d$),
the distribution of anticharm quarks in the $D$ meson is given by
\begin{eqnarray}
\bar{c}_D(z)
&=& \frac{N_D}{16\pi^2} \int_0^\infty 
    \frac{d\hat{k}_\perp^2}{[z (1-z)]^2}
    \frac{|G(\hat{s})|^2}{(\hat{s} - m_D^2)^2}
    \Big[ \hat{k}^2_\perp + (z\, m_q + (1-z)\, m_{\bar{c}})^2 \Big],
\label{eq:cbar_Dbar}
\end{eqnarray}
where the integration is over the transverse momentum $\hat{k}_\perp^2$
of the interacting heavy quark, and $z$ is the Bjorken scaling variable
of the heavy quark inside the charmed hadron.
The total invariant mass squared of the $\bar{c} q$ pair is defined
[see also Eq.~(\ref{eq:CoM_En})] for the corresponding invariant mass
of the meson--baryon system) as
\be
\hat{s}(z,\hat{k}^2_\perp)
= \frac{m_{\bar c}^2 + \hat{k}^2_\perp}{z}
+ \frac{m_q^2 + \hat{k}^2_\perp}{1-z},
\label{eq:s-hat}
\ee
where $m_{\bar{c}}$ is the constituent anticharm quark mass,
$m_q$ is the mass of the (light) spectator quark, and $N_D$
is the overall normalization factor determined by the valence
normalization condition,
\be
\int_0^1 dz\, \bar{c}_D(z) = 1.
\label{eq:cbarDNorm}
\ee
For point particles, the ultraviolet behavior of the $\hat{k}_\perp^2$
integration would be logarithmically divergent for the $\bar{c}_D(z)$
distribution in Eq.~(\ref{eq:cbar_Dbar}).
The divergence can be regulated by defining the vertex function
$G(\hat{s})$ to suppress contributions from large parton momenta.
Following Sec.~\ref{sec:mb}, we can use, for example, an exponential
functional dependence on $\hat{s}$,
\bea
G(\hat{s}) &=& \exp\left[-(\hat{s}-m_D^2)/\hat{\Lambda}^2\right],
\label{eq:G_exp}
\eea
with $\hat{\Lambda}$ serving the role of a corresponding momentum
cutoff on the partonic quark-antiquark system.

At low momenta, on the other hand, a mass singularity can arise
in the energy denominator $(\hat{s} - m_D^2)^{-2}$ in the
infrared limit ($\hat{k}_\perp^2 \to 0$) for physical quark masses
$m_q$ and $m_{\bar c}$.  A simple solution adopted by Pumplin
\cite{Pum05} was to assume an artificially large effective
mass for the anticharm quark, $m_{\bar c}^{\rm eff}$, and a large
constituent quark mass for the spectator $u$ or $d$ quark,
$m_q^{\rm eff}$, such that
\bea
m_{\bar c}^{\rm eff} + m_q^{\rm eff} &>& m_D.
\label{eq:mceff}
\eea
In our numerical analysis we fix the effective charm mass to be
$m_{\bar c}^{\rm eff} = 1.75$~GeV and the light constituent quark
mass $m_q^{\rm eff} = M/3 = 0.31$~GeV, similar to that used in
Ref.~\cite{Pum05}, which is sufficient to remove the
propagator singularity.

An alternative method to avoid the pole is to utilize a form factor
that simulates confinement by directly cancelling the singular
denominator, similar to that advocated in Ref.~\cite{MST94}.
A form that satisfies this is
\bea
G(\hat{s}) &=& (\hat{s} - m_D^2)\,
	       \exp\left[-(\hat{s}-m_D^2)/\hat{\Lambda}^2\right].
\label{eq:FF_con}
\eea
An attractive feature of this form of the vertex function is that it
permits any values of the quark masses to be used, allowing the partons
to be confined without the need for {\it ad hoc} constraints to avoid
singularities through judicious choice of effective quark masses.

For the $\bar c$ distribution in a vector $D^*$ meson, there
exist in principle both the Dirac and Pauli couplings of the
$D^*$ to a quark and antiquark, as for the $D^* B$ splitting
function in Eq.~(\ref{eq:spin1-fMB}).  To reduce the number
of free parameters in the calculation, we make the simplifying
assumption that the $D^*$--quark--antiquark coupling is governed
by a purely vector interaction, $\gamma_\alpha\, G(\hat{s})$.
In this case the distribution of a $\bar c$ quark in the $D^*$
meson is given by
\bea 
\bar{c}_{D^*}(z)
&=& \frac{N_{D^*}}{16\pi^2} \int_0^\infty 
    \frac{d\hat{k}_\perp^2}{[z (1-z)]^2}
    \frac{|G(\hat{s})|^2}{(\hat{s} - m_{D^*}^2)^2}
    \Biggl[
       \left( \frac{\hat{k}^2_\perp + m_q^2}{m_{D^*}^2} + (1-z)^2 \right)
       \left( \hat{k}^2_\perp + m_{\bar c}^2 + z^2\, m_{D^*}^2    \right)
\nonumber\\
& & \hspace*{4cm}
     +\ \hat{k}^2_\perp\
     +\ \left( z\, m_q + (1-z)\, m_{\bar c} \right)^2\
     +\ 4z(1-z)\, m_q m_{\bar c}
    \Biggr], 
\label{eq:cbar_D-st}
\eea 
where the normalization factor $N_{D^*}$ is again determined
by the valence quark number conservation condition in
Eq.~(\ref{eq:cbarDNorm}).
As for the $\bar c$ distribution in the $D$ meson, for point
interactions the integral in Eq.~(\ref{eq:cbar_D-st}) would be
divergent, in this case linearly in $\hat{k}_\perp^2$.
Vertex form factors $G(\hat{s})$ such as in Eqs.~(\ref{eq:G_exp})
or (\ref{eq:FF_con}) would again act to regularize this
divergence.

\begin{figure}[t]
\includegraphics[width=8cm]{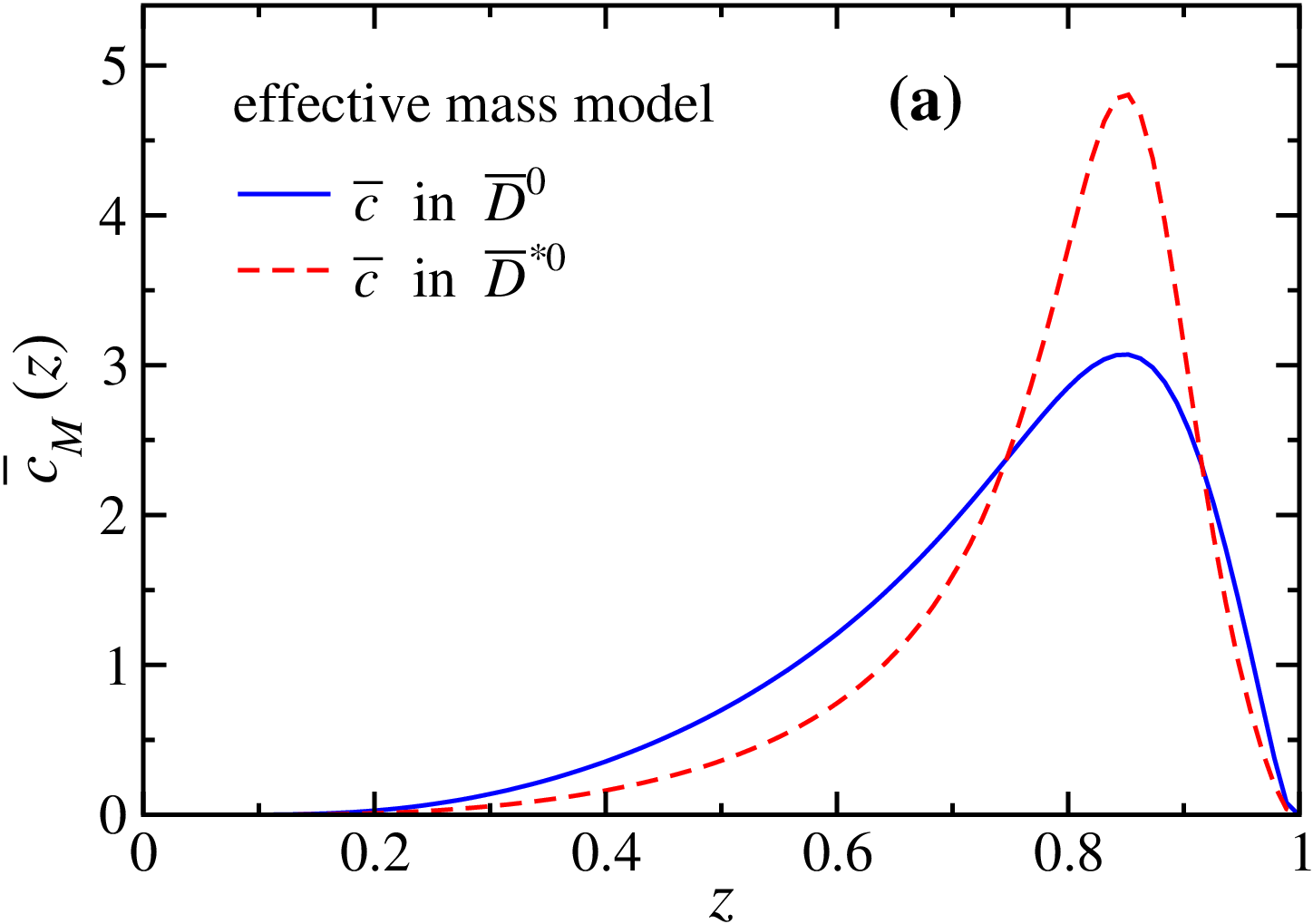}
\includegraphics[width=8cm]{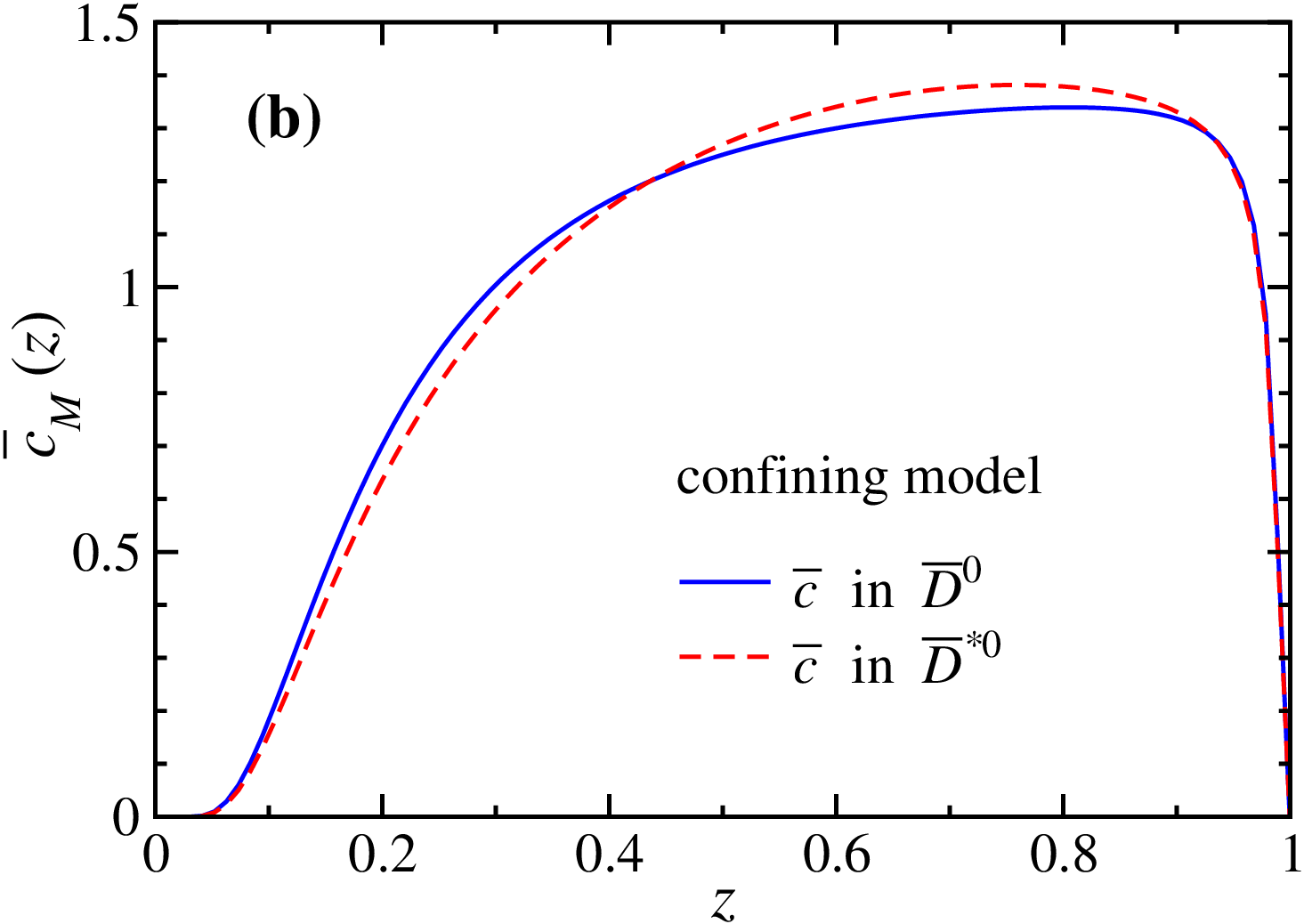}
\caption{(color online)
	Anticharm quark distributions in charmed $D$ (solid)
	and $D^*$ (dashed) mesons, within the
	{\bf (a)} effective mass model, with the vertex form
	factor in Eqs.~(\ref{eq:G_exp}) and (\ref{eq:mceff}),
	and
	{\bf (b)} confining model, with the form factor	in
	Eq.~(\ref{eq:FF_con}).}
\label{fig:QDF_mc}
\end{figure}

The results for the $\bar{c}$ distributions in the $D$ and $D^*$ mesons
are illustrated in Fig.~\ref{fig:QDF_mc} for both types of vertex
functions $G(\hat{s})$.  In the absence of empirical constraints on PDFs
in charmed mesons, the partonic cutoff $\hat\Lambda$ is a free parameter.
However, since for heavy quarks the typical masses of the intermediate
states ($D B$ or $\bar c q$) are comparable, to a first approximation we
can fix $\hat\Lambda$ to the meson--baryon cutoff, $\hat\Lambda=\Lambda$.
In the effective mass model, Eqs.~(\ref{eq:G_exp}) and (\ref{eq:mceff}),
the peak of the anticharm distribution in $z$ reflects the fraction of
the meson mass carried by the $\bar{c}$ quark.  For both the $D$ and
$D^*$ mesons, the $\bar{c}$ distribution peaks at $z \sim 0.85$, with
the latter being slightly narrower.  The distributions in the confining
model, Eq.~(\ref{eq:FF_con}), also peak at similarly large momentum
fractions, but are significantly broader.

A numerical feature of the effective mass model is the presence of the
energy denominator $\propto (\hat{s} - m_D^2)^{-2}$, which largely
determines the qualitative shapes of the $\bar c$ distributions.
For specific mass choices, $\hat{s} - m_D^2$ is minimized at a unique
value of $z$, resulting in the strongly-peaked shapes observed in
the effective mass model.  In the confining model, on the other hand,
the energy denominator responsible for this $z$ dependence is suppressed
directly such that the resulting distribution shapes no longer possess
pronounced maxima.  In the effective charm model, however, the closer
the energy denominator approaches its pole value, the more ``singular''
the behavior at the distribution maximum; as such, if we fix the charm
and spectator masses according to Eq.~(\ref{eq:mceff}), the energy
denominator approaches the zero pole for heavier hadron masses,
producing the more sharply peaked distributions seen in 
Fig.~\ref{fig:QDF_mc}(a).

\subsection{Charm in charmed baryons}
\label{ssec:c_in_B}

The calculation of the charm quark distributions in charmed baryons
proceeds in similar fashion to that for the $D$ and $D^*$ mesons,
but is more involved since the spectator system consists of two
(or more) particles.  In practice, however, one can simplify the
calculation by treating the spectator $qq$ system as an effective
``diquark'' with a fixed mass $m_{qq}$.
For spin-1/2 charmed baryons, in general the spectator diquark state
can have either spin~0 or spin~1, with corresponding scalar and
pseudovector vertex functions describing the momentum dependence.
The spin of the spectator diquark can affect the spin and flavor
dependence of the associated parton distribution; for example,
the suppression of the $d/u$ ratio in the proton at large $x$ is
usually attributed to a higher energy of the spin-1 diquark in
the proton compared with the spin-0 diquark \cite{MT97, Close88}.
Since here we are concerned with the total effect on the charm quark
distribution, rather than the flavor dependence, it will be sufficient
to consider only the leading contribution arising from the scalar
spectators, for which we take an effective mass of $m_{qq} = 1$~GeV.

The charm quark distribution in a spin-1/2 charmed baryon $B$
($B = \Lambda_c$ or $\Sigma_c$) with a scalar $qq$ spectator
is then given by
\bea
c_B(z) &=& \frac{N_B}{16\pi^2} \int_0^{\infty} 
  \frac{d\hat{k}_\perp^2}{z^2 (1-z)}
  \frac{|G(\hat{s})|^2}{(\hat{s} - M_B^2)^2}
  \Big[ \hat{k}^2_\perp + (m_c + z M_B)^2 \Big],
\label{eq:c_in_Lambda}
\eea
where for the charm quark mass we take $m_c = m_{\bar{c}}$,
and the invariant mass squared of the quark--diquark system here
is defined as
\be
\hat{s}(z,\hat{k}^2_\perp)
= \frac{m_c^2 + \hat{k}^2_\perp}{z}
+ \frac{m_{qq}^2 + \hat{k}^2_\perp}{1-z}.
\label{eq:s-hatB}
\ee
The functional form of the $B$-$c$-$qq$ vertex function $G(\hat{s})$
is taken to be the same as for the $D$-$\bar c$-$q$ function in the
models of Eqs.~(\ref{eq:G_exp}) and (\ref{eq:FF_con}), and the
normalization constant $N_B$ determined from an analogous valence
charm quark number condition to that in Eq.~(\ref{eq:cbarDNorm}),
\be
\int_0^1 dz\, c_B(z) = 1.
\label{eq:cLambdacNorm}
\ee
Note that for the $\Sigma_c^{++}$ baryon, the $uu$ spectator diquark
has spin 1, so that the calculation of its $c$ quark distribution
here is approximated by neglecting the diquark's spin structure.
While it is straightforward to include both spin-0 and spin-1 diquark
contributions, in analogy with the spin structures discussed in
Sec.~\ref{ssec:cbar_in_D}, since the overall contribution from the
dissociation of the proton to $D^- \Sigma_c^{++}$ is at least an
order of magnitude smaller than for $\bar{D}^{*0} \Lambda_c^+$,
this will have negligible effect on the numerical results.

For the spin-3/2 $\Sigma_c^*$ baryons, the charm quark is always
accompanied by a spin-1 diquark.  Incorporating the fully
relativistic Rarita-Schwinger structure for the spin-3/2 state
(see Appendix~\ref{sec:B-MBM}), the charm quark distribution
in the $\Sigma_c^*$ is given by
\bea
c_{B^*}(z)
&=& \frac{N_{B^*}}{12\pi^2 m_{qq}^2} \int_0^\infty
    \frac{d\hat{k}_\perp^2}{z (1-z)}
    \frac{|G(\hat{s})|^2}{(\hat{s} - M_{B^*}^2)^2}
    \Biggl( (\hat{k} \cdot \hat{\Delta})(P \cdot \hat{\Delta})
	  + 2 m_c M_{B^*} \hat{\Delta}^2
\nonumber\\
&+& \frac{1}{m_{qq}^2}
    \Bigl[
      m_c M_{B^*} (\hat{p} \cdot \hat{\Delta})^2
    - (\hat{p} \cdot \hat{\Delta})
      \Bigl( (\hat{p} \cdot \hat{\Delta})(\hat{k} \cdot \hat{p})
	   - (P \cdot \hat{\Delta})(\hat{k} \cdot \hat{p})
	   - (\hat{k} \cdot \hat{\Delta})(P \cdot \hat{p})
      \Bigr)
\nonumber\\ 
&-&   (P \cdot \hat{p})(\hat{k} \cdot \hat{p}) \hat{\Delta}^2
     - \frac{P \cdot \hat{k}}{M_{B^*}^2} 
       \Bigr(
	 (P \cdot \hat{p})^2 \hat{\Delta}^2
	- m_{qq}^2 (P \cdot \hat{\Delta})^2
	- 2 (P \cdot \hat{\Delta})
	    (\hat{p} \cdot \hat{\Delta})
	    (P \cdot \hat{p})
       \Bigr)
    \Bigr] 
    \Biggr),
\nonumber\\
\label{eq:c_Sigma-st}
\eea 
in which $\hat{p}$ is the momentum of the spectator diquark $qq$,
and $\hat{\Delta} \equiv P - \hat{k}$ [see Eq.~(\ref{eq:kin-TOPT})].

\begin{figure}[t]
\includegraphics[width=8cm]{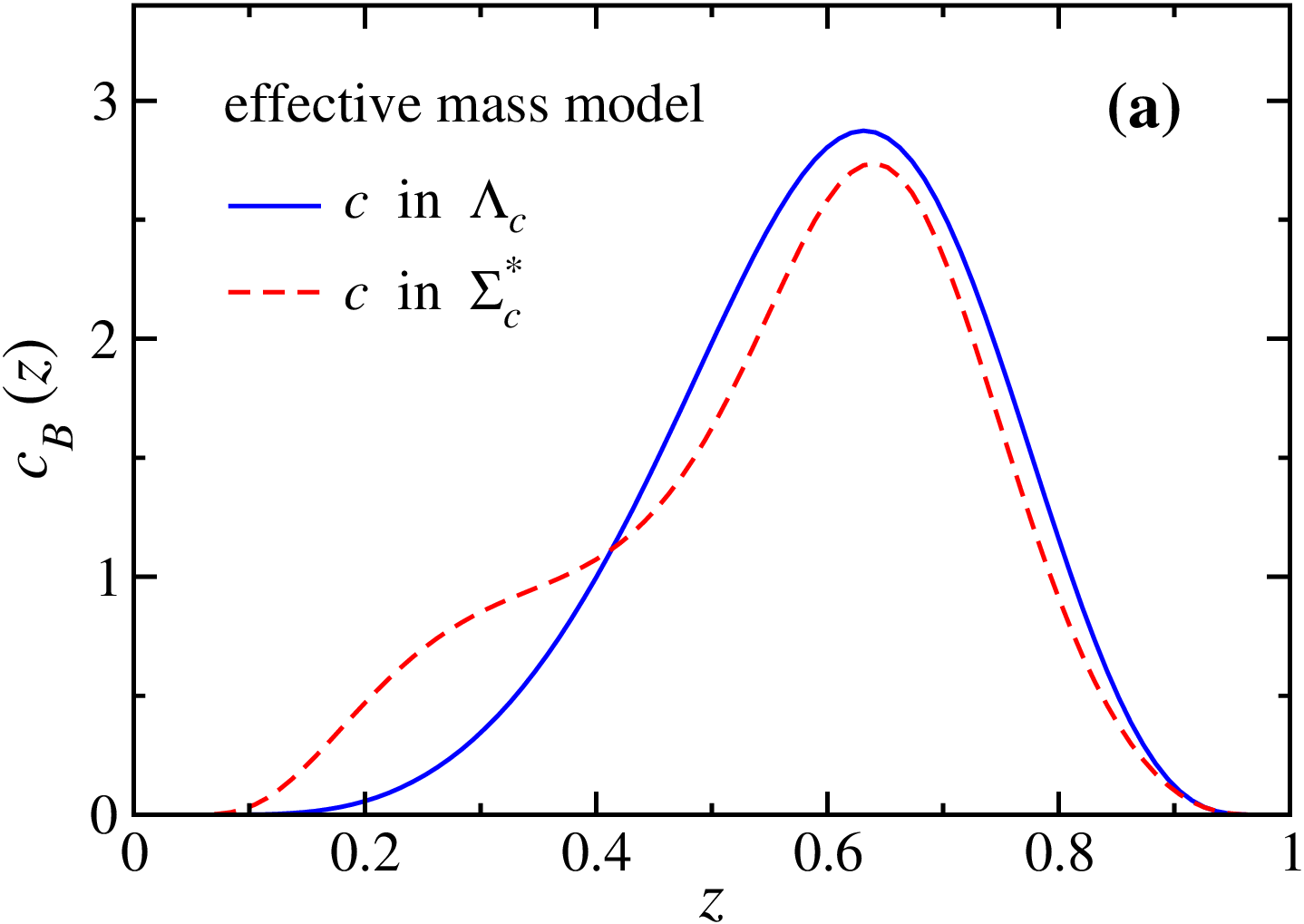}
\includegraphics[width=8cm]{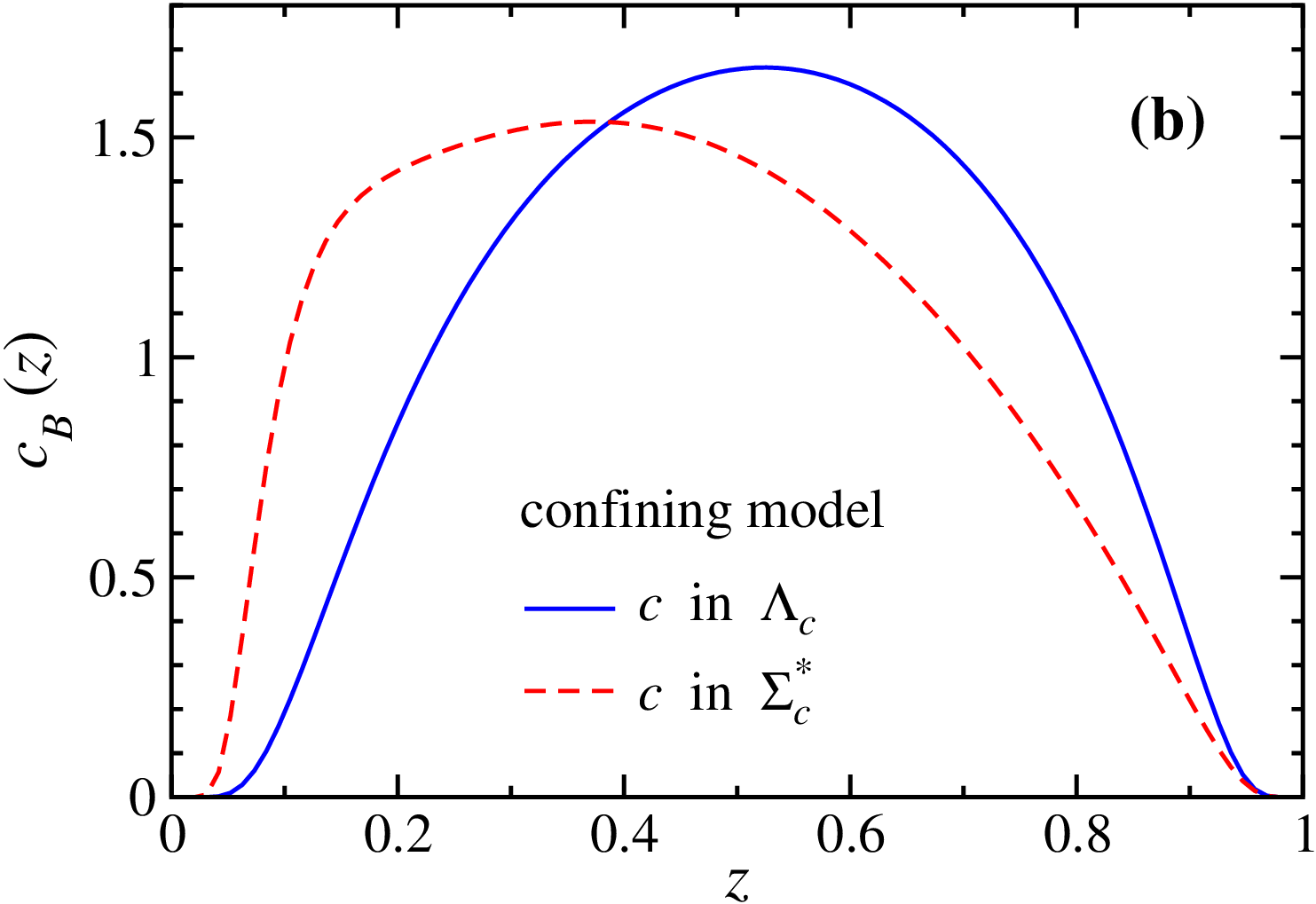}
\caption{(color online)
	Charm distributions in the charmed $\Lambda_c$ (solid)
	and $\Sigma_c^*$ (dashed) baryons, within
	{\bf (a)} the effective mass model, and
	{\bf (b)} the confining model.}
\label{fig:QDF}
\end{figure}

The resulting $c$ quark distributions in the spin-1/2 and
spin-3/2 charmed baryons are illustrated in Fig.~\ref{fig:QDF},
using the same numerical values for the masses and cutoffs
as in the $\bar c$ calculation in the charmed mesons above.
Compared with the $\bar d$ distributions in $D$ and $D^*$, the
$c$ quark PDFs are peaked at somewhat smaller values of $z$.
In the effective mass model for the $B$-$c$-$qq$ vertex function,
both the $c$ distributions in $\Lambda_c$ and in $\Sigma_c^*$
are maximal at $z \approx 0.6-0.65$, with a relatively narrow
distribution in $z$.  The bulge in the $c$ distribution in the
$\Sigma_c^*$ baryon is associated with the more complicated
spin algebra compared with the $\Lambda_c$.
The $c$ distributions with the confining model vertex function
are once again somewhat broader, peaking at smaller $z$ values,
$z \approx 0.55$ for the $\Lambda_c$ baryon and $z \approx 0.4$ for
the $\Sigma_c^*$, with the latter having a sharp drop off at $z \to 0$.
The broader distributions here are generated by the suppression of the
energy denominator $(\hat{s} - M_B^2)^{-2}$, which is mostly responsible
for the strongly-peaked distributions in the effective mass model.
In all cases the distributions have been normalized to respect the
valence quark number sum rule, as in Eq.~(\ref{eq:cLambdacNorm}).

Having assembled the various ingredients for the calculation
of the convolution expressions in Eqs.~(\ref{eq:mesoncloud}),
in the next section we combine these inputs to compute the
$c$ and $\bar c$ distributions in the nucleon.

\section{Intrinsic charm in the nucleon}
\label{sec:results}

Combining the distributions of $c$ and $\bar c$ quarks in the
charmed mesons and baryons discussed in the previous section
with the splitting functions summarized in Sec.~\ref{sec:mb},
here we present the resulting $c$ and $\bar c$ distributions in
the physical nucleon.  We consider in Sec.~\ref{ssec:c_in_MBM}
contributions from the various meson--baryon configurations in
the MBM, and the dependence of the results on the models for the
charm distributions inside the charm hadrons.  We compare our
results with other prescriptions for intrinsic charm distributions
in Sec.~\ref{ssec:approx}, and with measurements of the charm
structure function in Sec.~\ref{ssec:EMC}.

\subsection{Intrinsic charm in the MBM}
\label{ssec:c_in_MBM}

\begin{figure}[t]
\includegraphics[width=8cm]{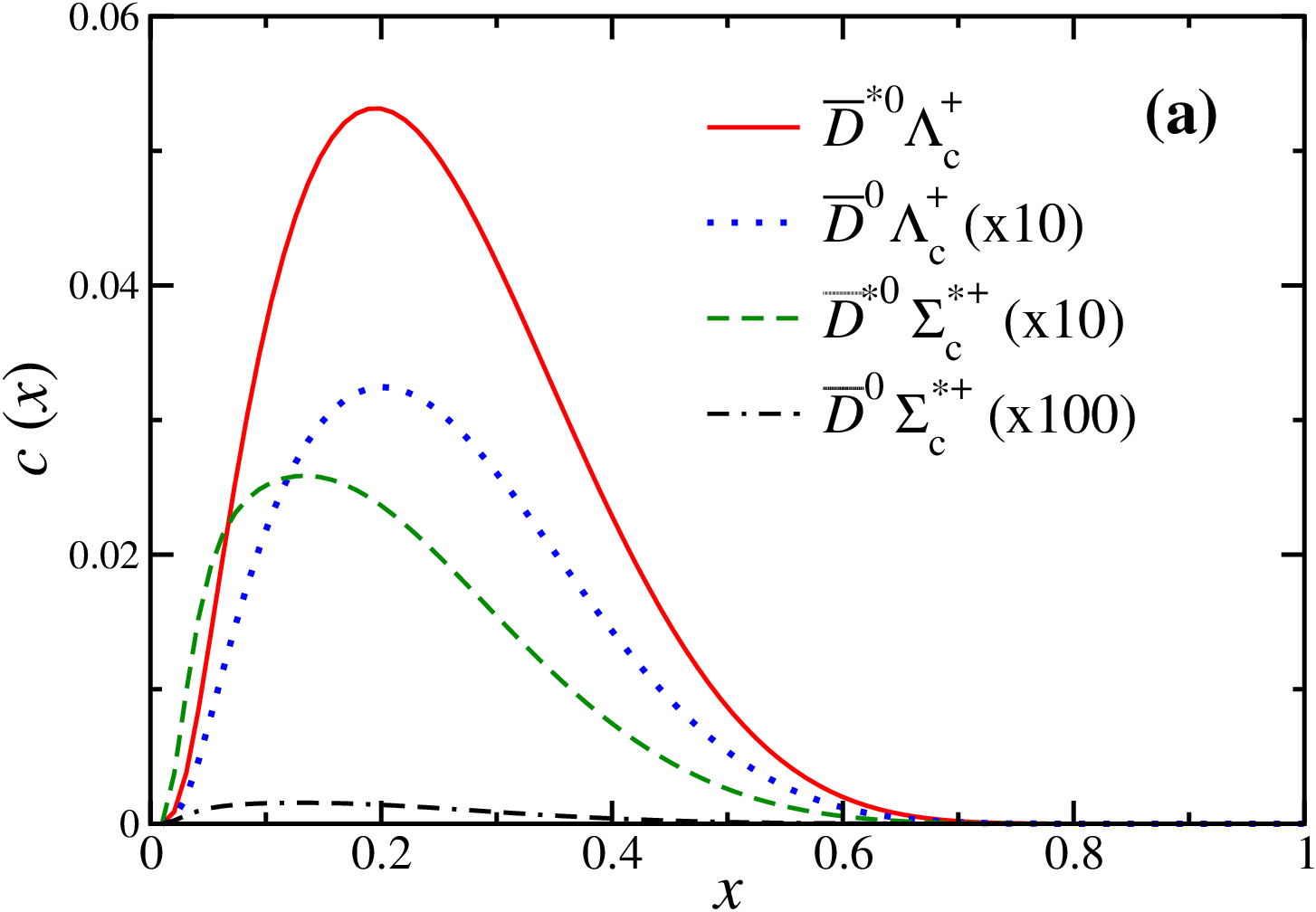}\ \
\includegraphics[width=8cm]{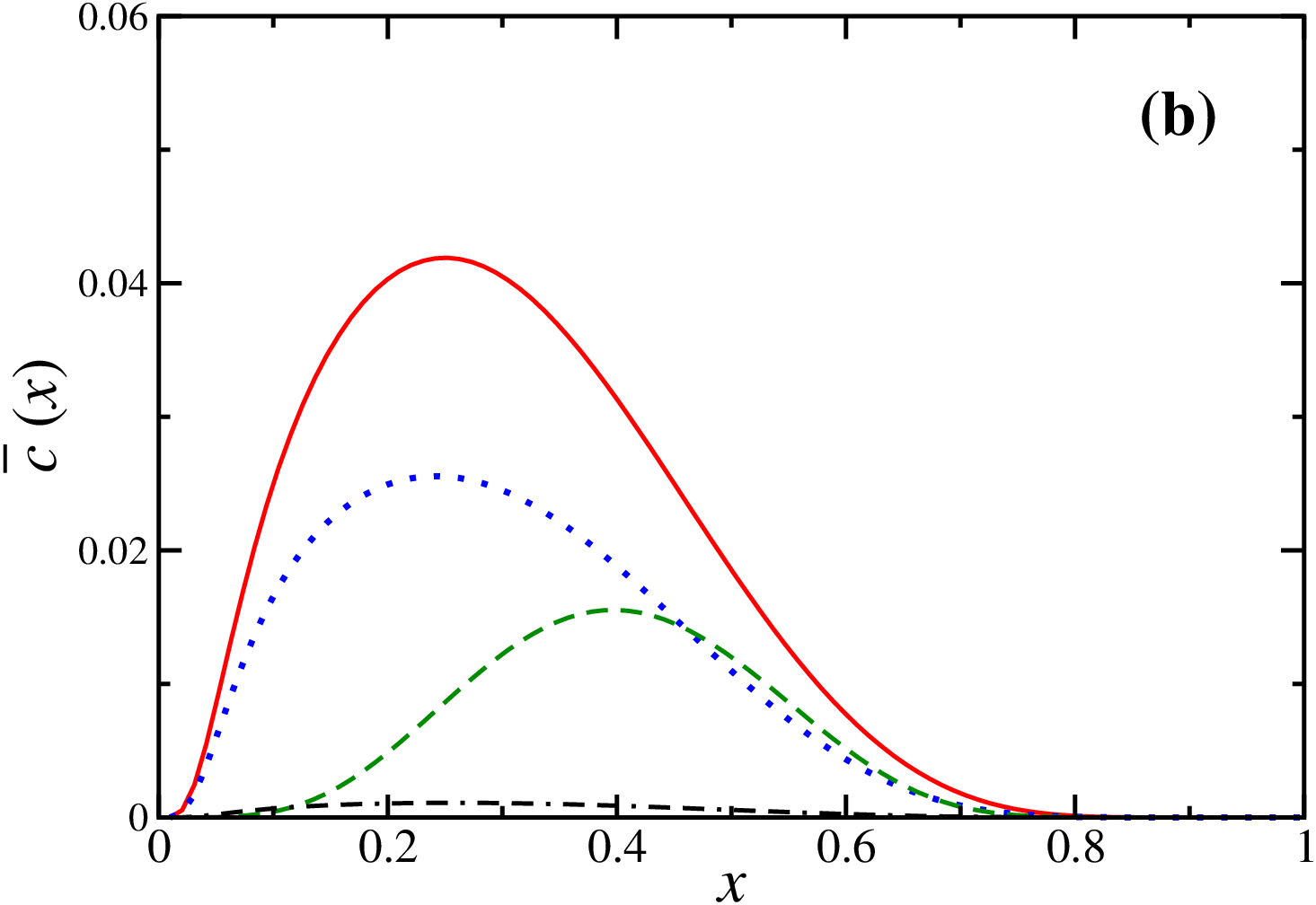}
\caption{(color online)
	{\bf (a)} Charm and {\bf (b)} anticharm quark distributions
	in the nucleon in the MBM, with contributions from the
	meson--baryon configurations as in Fig.~\ref{fig:f_MB}:
        $\bar{D}^{*0} \Lambda_c^+$ (red solid),
        $\bar{D}^0 \Lambda_c^+$ (scaled $\times 10$, blue dotted),
        $\bar{D}^{*0} \Sigma_c^{*+}$ (scaled $\times 10$, green dashed),
	and
	$\bar{D}^0\Sigma_c^{*+}$ (scaled $\times 100$, black dot-dashed).}
\label{fig:PS_mc}
\end{figure}

The contributions to the charm and anticharm quark distributions
in the nucleon from various meson--baryon states are presented
in Fig.~\ref{fig:PS_mc}, using the confining model for the PDFs
in the charmed hadrons (\ref{eq:FF_con}) with a mass parameter
$\Lambda = 3$~GeV.  The contributions correspond to the same
configurations as in Fig.~\ref{fig:f_MB}, namely, the dominant
  $\bar{D}^{*0} \Lambda_c^+$ state, the
  $\bar{D}^0 \Lambda_c^+$ and
  $\bar{D}^{*0} \Sigma_c^{*+}$ states, as well as the (negligible)
  $\bar{D}^0 \Sigma_c^{*+}$ contribution.
As expected from the magnitudes of the splitting functions in
Fig.~\ref{fig:f_MB}, the $\bar{D}^{*} \Lambda_c^+$ state produces
the dominant meson--baryon contribution to charm.

\begin{figure}[t]
\includegraphics[width=7.8cm]{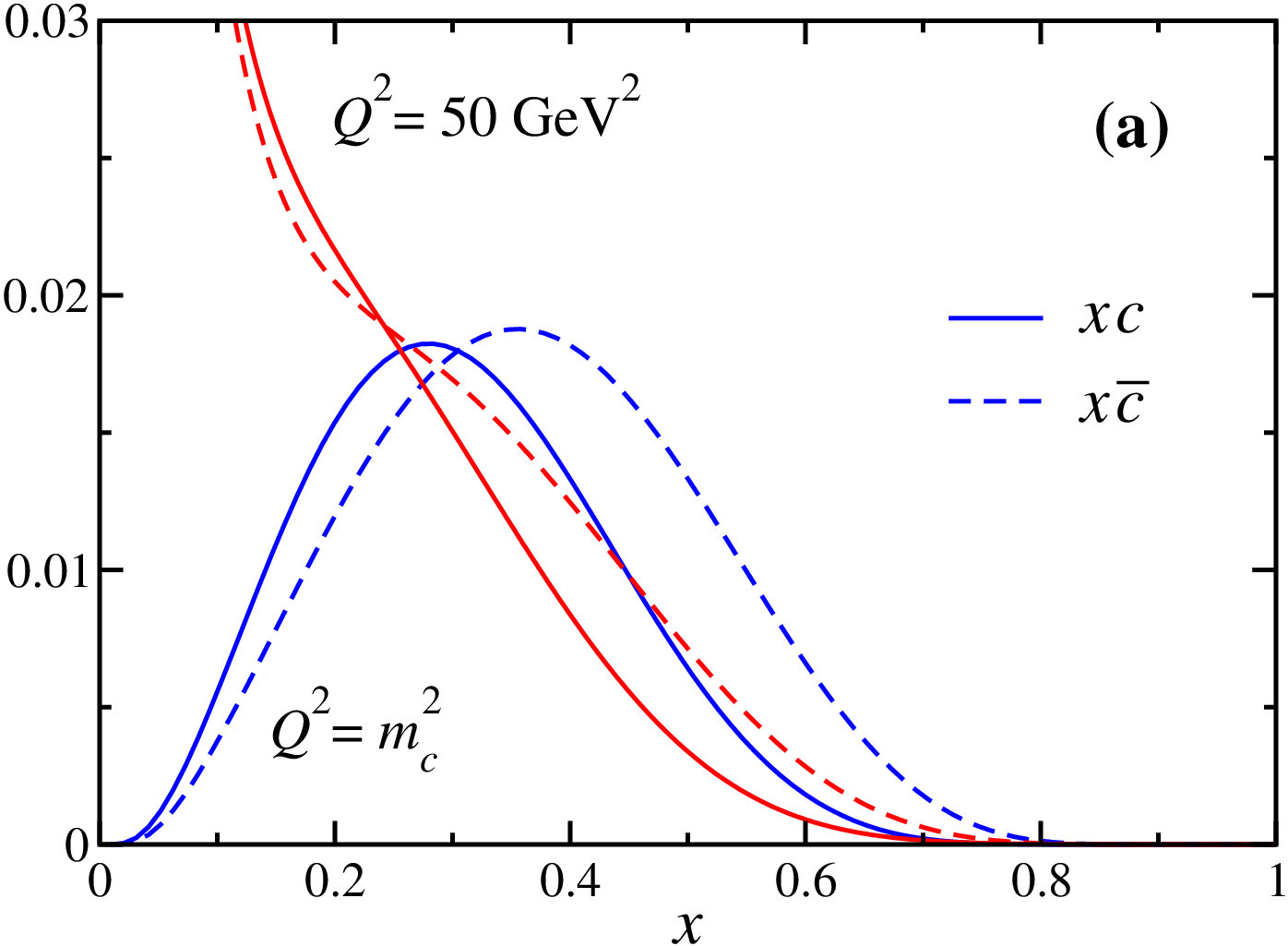}\ \
\includegraphics[width=8.3cm]{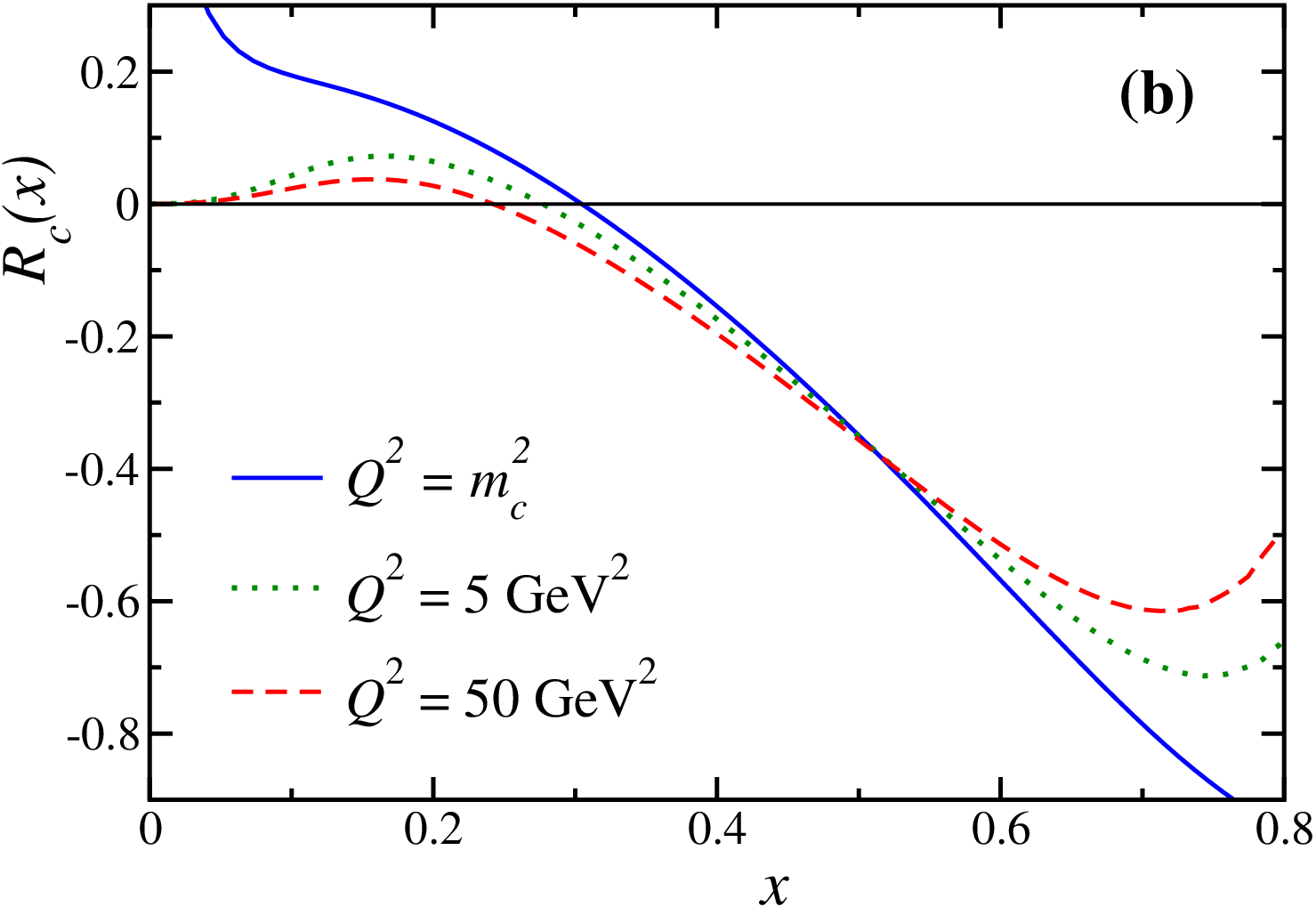}
\caption{(color online)
	{\bf (a)}
	Total $xc$ (solid lines) and $x\bar{c}$ (dashed lines)
	distributions in the MBM with the confining model for
	the PDFs in the charmed hadrons, Eq.~(\ref{eq:FF_con}), at
	$Q^2 = m_c^2$ (blue) and evolved to
	$Q^2 = 50$~GeV$^2$ (red).
	{\bf (b)}
	Corresponding charm--anticharm asymmetry
	$R_c(x) = (c(x)-\bar{c}(x))/(c(x)+\bar{c}(x))$ at
	$Q^2 = m_c^2$ (solid),
	$Q^2 = 5$~GeV$^2$ (dotted), and
	$Q^2 = 50$~GeV$^2$ (dashed).}
\label{fig:c-c+}
\end{figure}

Summing over all the contributions listed in Fig.~\ref{fig:Spectrum},
the total $xc$ and $x\bar{c}$ distributions are shown in 
Fig.~\ref{fig:c-c+}, at the input scale $Q^2 = m_c^2$ and evolved
to $Q^2=50$~GeV$^2$ (which is typical for charm structure function
measurements).  Note that the dominant $\bar{D}^{*0} \Lambda_c^+$
contribution accounts for nearly 70\% of the total.
A unique feature of the MBM which is evident in Fig.~\ref{fig:c-c+}
is the fact that $\bar{c}$ distribution is clearly harder than the $c$
distribution.  This is true for every meson--baryon configuration in
the MBM, and simply reflects the fact that the charm quark represents
a larger fraction of the total mass of the meson than of the baryon.
Since the peak in the charm distribution in a hadron is related to the
fraction of the hadron mass carried by the charm quark, the resulting
distribution of $\bar{c}$ in the $\bar{D}$ meson will typically be
harder than that for the $c$ in the $\Lambda_c$.
While it is possible to make the intrinsic $c$ distribution as hard
as the $\bar{c}$ distribution in a convolution model, this requires
rather unnatural parton distributions inside the baryon and meson
states.

To quantify the magnitude of the nonperturbative charm in the nucleon,
we can compute the total proton momentum carried by charm and anticharm
quarks,
\bea
P_c &=& C^{(1)} + \overline{C}^{(1)},
\label{eq:cmom_tot}
\eea
where the moments $C^{(1)}$ and $\overline{C}^{(1)}$ are defined
in Eqs.~(\ref{eq:c_mom}).  For the confining model distributions
in the charmed mesons and baryons, the momentum fraction at the
input model scale (which is naturally of the order of the charm
quark mass) is found to be
  $P_c = 1.34^{\, +1.35}_{\, -0.75}\, \%$
for the cutoff mass parameter $\Lambda = (3.0 \pm 0.2)$~GeV
obtained from the inclusive $\Lambda_c$ production data,
Sec.~\ref{ssec:HSM}.  Again we note that these first moments
differ numerically from the charm multiplicities mentioned in
Sec.~\ref{ssec:splitfin}, where we found
$\langle n \rangle_{MB}^{\rm (charm)}
= 2.40^{\, +2.47}_{\, -1.36}\, \%$.
The strong dependence of the total momentum on $\Lambda$
stems from the strong dependence of the dominant meson--baryon
splitting function on the hadronic form factor, as seen in
Fig.~\ref{fig:f_MB_int}(a).
In the BHPS model, in contrast, when the charm quark is normalized
to 1\% probability in the nucleon, charm quarks carry a momentum
fraction $P_c = 0.57$\%.  For our best fit form factor cutoff mass
$\Lambda = 3$~GeV, therefore, nonperturbative charm quarks carry
about twice the momentum as in the BHPS model.

Valence quark normalization requires that the first moment of
$c-\bar{c}$ vanishes, as in Eq.~(\ref{eq:c_moments}), which follows
for any splitting function that obeys the reciprocity relation,
Eq.~(\ref{eq:fphi}).  Higher moments, on the other hand, are not
required to vanish.  In fact, the magnitude of the $c-\bar{c}$
asymmetry can be quantified in terms of the difference of the
second moments (momentum carried by charm and anticharm quarks),
\bea
\Delta P_c &=& C^{(1)} - \overline{C}^{(1)}.
\label{eq:cmom_dif}
\eea
At the model scale, $Q^2 = m_c^2$, we find
  $\Delta P_c = -(0.13^{\, +0.14}_{\, -0.08})\, \%$
for $\Lambda = (3.0 \pm 0.2)$~GeV.
The overall negative values of $\Delta P_c$ reflect the fact
that in the MBM the $\bar c$ distribution is harder than the $c$.

The momentum imbalance of anticharm quarks compared to charm
can be estimated from the ratio of the difference $\Delta P_c$
to the sum $P_c$, for which we find $\Delta P_c/P_c \approx -10\%$.
As a function of $x$, however, the imbalance is not uniformly
distributed.  Defining the ratio
\bea
R_c(x) &=& \frac{c(x) - \bar{c}(x)}{c(x) + \bar{c}(x)},
\label{eq:Rx}
\eea
from Fig.~\ref{fig:c-c+}(b) we observe that the relative asymmetry
can exceed 50\% at large values of $x$ ($x \gtrsim 0.5$).
The $Q^2$ dependence of the ratio indicates relatively mild
effects over the large range considered (up to $Q^2=50$~GeV$^2$),
with the slope of the asymmetry becoming slightly more shallow
with increasing $Q^2$.
Note that the ratio $R_c$ is nonzero at $x=0$ at the model scale,
but the effects of perturbative evolution force $R_c(x=0)$ to vanish
at large $Q^2$ values due to the growth of the denominator
$c+\bar c$.

\subsection{Comparison with other models}
\label{ssec:approx}

While some of the features of the nonperturbative $c$ and $\bar c$
distributions in the MBM are relatively robust, such as the generally
harder $x$ dependence compared with the perturbatively generated
distributions and the presence of a $c-\bar{c}$ asymmetry, the
detailed $x$ dependence does depend on the specifics of the model.
To estimate the model dependence of the calculated $c$ and $\bar c$
PDFs, we compare the results obtained in the previous section, using
the splitting functions from Sec.~\ref{sec:MBM} and the confining
model for the PDFs in the charmed hadrons in Sec.~\ref{sec:cinc}, with
distributions computed under different assumptions and approximations.

\begin{figure}[t]
\includegraphics[width=8cm]{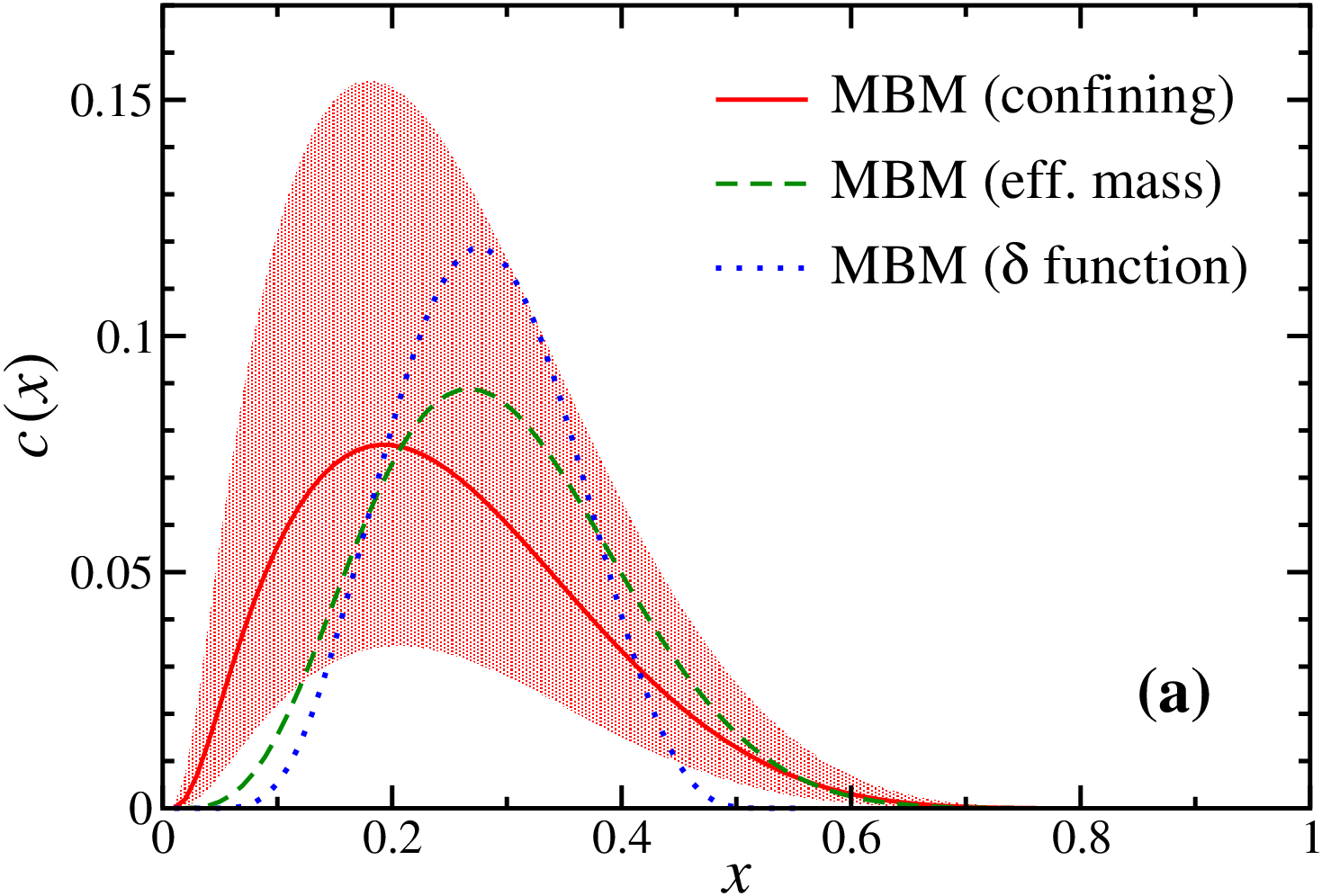}
\includegraphics[width=8cm]{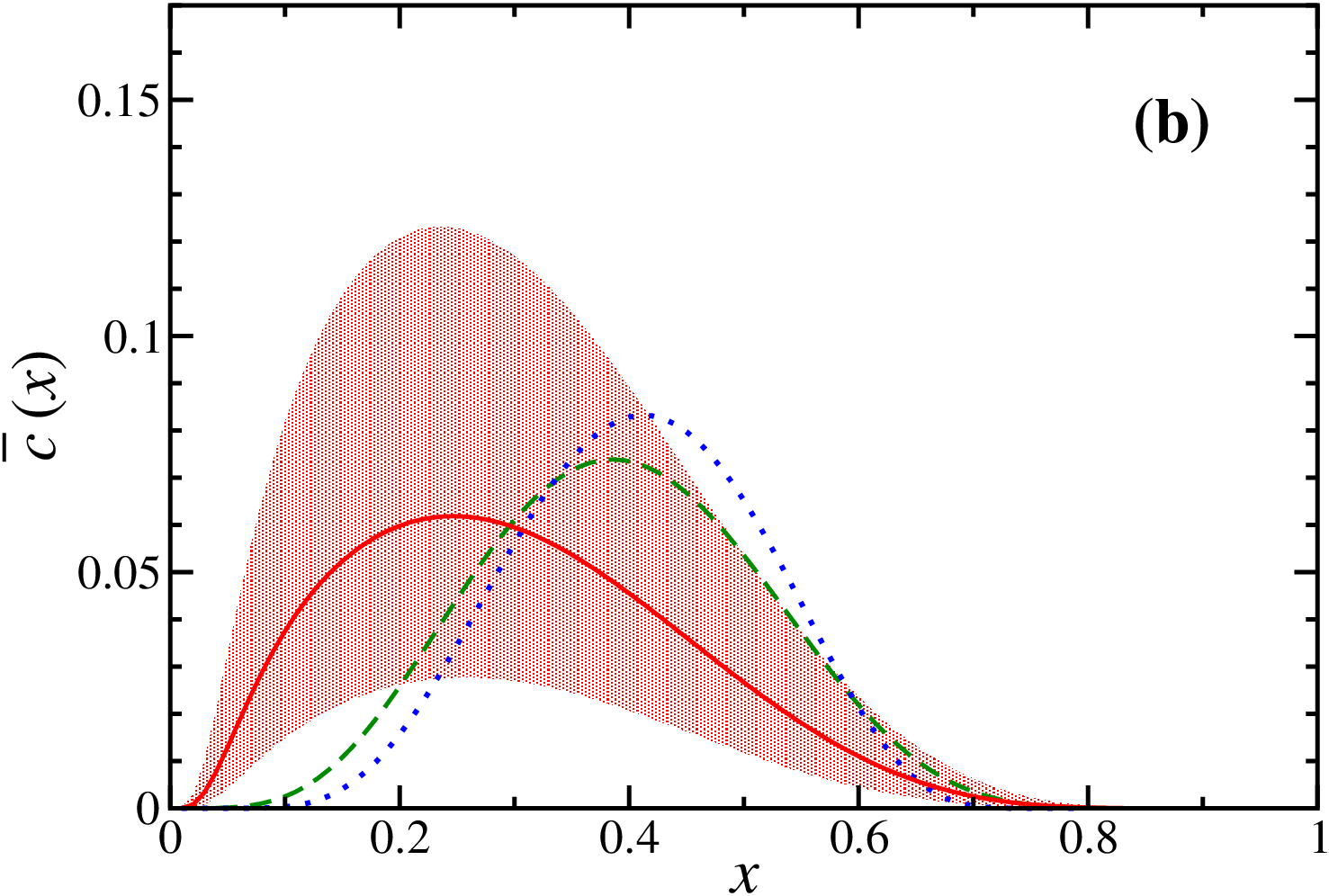}
\caption{(color online)
	Model dependence of the charm distributions in the nucleon
	for ({\bf a}) $c(x)$ and ({\bf b}) $\bar{c}(x)$, for the
	MBM with the confining model for the PDFs in charmed hadrons
	(red solid), the effective mass model (green dashed),
	and the $\delta$ function model (blue dotted).
	All distributions use a common value for the cutoff mass of
	$\Lambda = (3.0 \pm 0.2)$~GeV, with the uncertainty band
	shown for the confining model.}
\label{fig:cnaive2}
\end{figure}

Within the same MBM framework, if one uses the effective mass model
for the charm PDFs in the charmed mesons and baryons, the resulting
$c$ and $\bar c$ distributions in the nucleon are slightly harder,
especially for the $\bar c$, as Fig.~\ref{fig:cnaive2} illustrates.
This generally follows from the shape of the $\bar{c}_M$ distribution
in the confining and effective mass models in Fig.~\ref{fig:QDF_mc},
where the latter is more strongly peaked at large values of the
parton momentum fraction.
The corresponding value of the total nucleon momentum carried
by charm and anticharm quarks in the effective mass model is
  $P_c = 1.67^{\, +1.70}_{\, -0.94}\, \%$
for cutoff masses $\Lambda = (3.0 \pm 0.2)$~GeV,
and
  $\Delta P_c = -(0.24^{\, +0.28}_{\, -0.14})\, \%$
for the momentum asymmetry, with the resulting momentum imbalance
$\Delta P_c/P_c \approx -15\%$.  The $c-\bar{c}$ asymmetry in this
model is therefore more pronounced than in the confining model.

In a more simplified approach, the $c$ and $\bar c$ distributions
inside the charmed hadrons were approximated in Ref.~\cite{MT97}
by $\delta$ functions centered at the $x$ values corresponding to
the fraction of the hadron mass carried by the constituent charm
or anticharm quark,
\be 
c_B(x) = \delta(x - x_B) \hskip 0.3cm {\rm and} \hskip 0.3 cm 
\overline{c}_M(x) = \delta(x - x_M).
\label{eq:cdelta}
\ee
From Eqs.~(\ref{eq:mesoncloud}), the charm and anticharm distributions
in the nuclear are then given directly as sums over the various
meson--baryon splitting functions,
\begin{subequations}
\label{eq:cnaive}
\bea
c(x)
&=& \sum_{B,M}\, \frac{1}{x_B}\, f_{BM}\left(\frac{x}{x_B}\right), \\
\bar{c}(x)
&=& \sum_{M,B}\, \frac{1}{x_M}\, f_{MB}\left(\frac{x}{x_M}\right). 
\eea
\end{subequations}
Since the masses of the charm quark [$m_c = {\cal O}(1.5~{\rm GeV})$]
and the $D$ mesons [$m_D = {\cal O}(\mbox{1.8--2}~{\rm GeV})$] are
similar, as a first approximation one can take $x_M \approx 1$.
Similarly, for the fractional mass of the $c$ quark in the charmed
baryon, the approximation $x_B \approx 2/3$ was utilized \cite{MT97}.
In a somewhat more sophisticated approach, one can choose $x_M$
and $x_B$ to minimize the $\hat{s}$-dependent energy denominator,
which depends on the combination
  $m_c^2/x + m_{\rm spec}^2/(1-x)$,
where the spectator mass $m_{\rm spec}$ corresponds to the light quark
mass $m_{u,d}$ for a meson, and to an effective diquark mass $m_{qq}$
for a baryon.  Choosing $m_c = 1.3$~GeV, $m_{u,d} = 0.313$~GeV and
$m_{qq} = 1$~GeV, one has
\be
x_B\, =\, \frac{m_c}{m_c + m_{qq}} \approx 0.57,\ \ \ \ \ \
x_M\, =\, \frac{m_{\bar c}}{m_{\bar c} + m_{u,d}} \approx 0.81.
\label{eq:mB}
\ee
For the best fit form factor cutoff mass $\Lambda = 3$~GeV, the
momentum carried by charm in this $\delta$ function approximation
model is $P_c = 1.66$\%, which is slightly greater than in the
MBM confining or effective mass models.

In Fig.~\ref{fig:cnaive2} we compare the $c$ and $\bar{c}$
distributions in the MBM obtained using the confining model
PDFs in the charmed hadrons with those computed from the
effective mass model and $\delta$ function approximations,
with a common cutoff mass $\Lambda = (3.0 \pm 0.2)$~GeV.
The MBM confining model distributions are generally softer
than those in the effective mass and $\delta$ function models,
with the confining model giving a slightly broader shape, and
the $\delta$ function model having the narrowest distribution.
Within the uncertainty bands of the parameters (for clarity
we only shown the uncertainty band for the confining model),
the distributions are compatible with each other.
In all three models the anticharm distributions are clearly
harder than the charm, so that the qualitative features of
the ratio $R_c$ in Fig.~\ref{fig:c-c+} are largely retained.
Interestingly, the $\delta$ function model gives an $x$ dependence
for the charm PDFs that closely resembled the shape of the
effective mass model distributions for $\Lambda = 3$~GeV.
This feature may be exploited in simplified calculations that seek
only approximate feature of nonperturbative charm distributions.

A similar model to the MBM was constructed by Pumplin
\cite{Pum05}, based on the couplings of scalar mesons
and baryons.  The dominant meson-baryon contribution was assumed
to be from the $\bar{D}^0 \Lambda_c^+$ state, and described by 
Eq.~(\ref{eq:dPFock}) with $N=2$ and an exponential form factor
(\ref{eq:expFF}) with $\Lambda_p = 4$~GeV.  The $c$ PDF in the
$\Lambda_c$ was taken from Eq.~(\ref{eq:dPFock}) with $N=3$
and a dipole form factor with $\Lambda_{\Lambda_c} = 2$~GeV,
while the $\bar{c}$ PDF in $\bar{D}$ was obtained assuming
$N=2$ and a dipole form factor with $\Lambda_D = 2$~GeV.
The magnitude of the resulting charm distributions in the
nucleon was normalized so that the momentum carried by charm
and anticharm quarks was identical, with their sum equal to the
momentum carried by charm in the BHPS model with 1\% probability,
$P_c = 0.57\%$.  A feature of this model is that the requirement
that the momentum carried by $c$ and $\bar{c}$ quarks to be
identical implies a nonzero net charm in the proton,
$C^{(0)} - \bar{C}^{(0)} \neq 0$.

\begin{figure}[t]
\includegraphics[width=8cm]{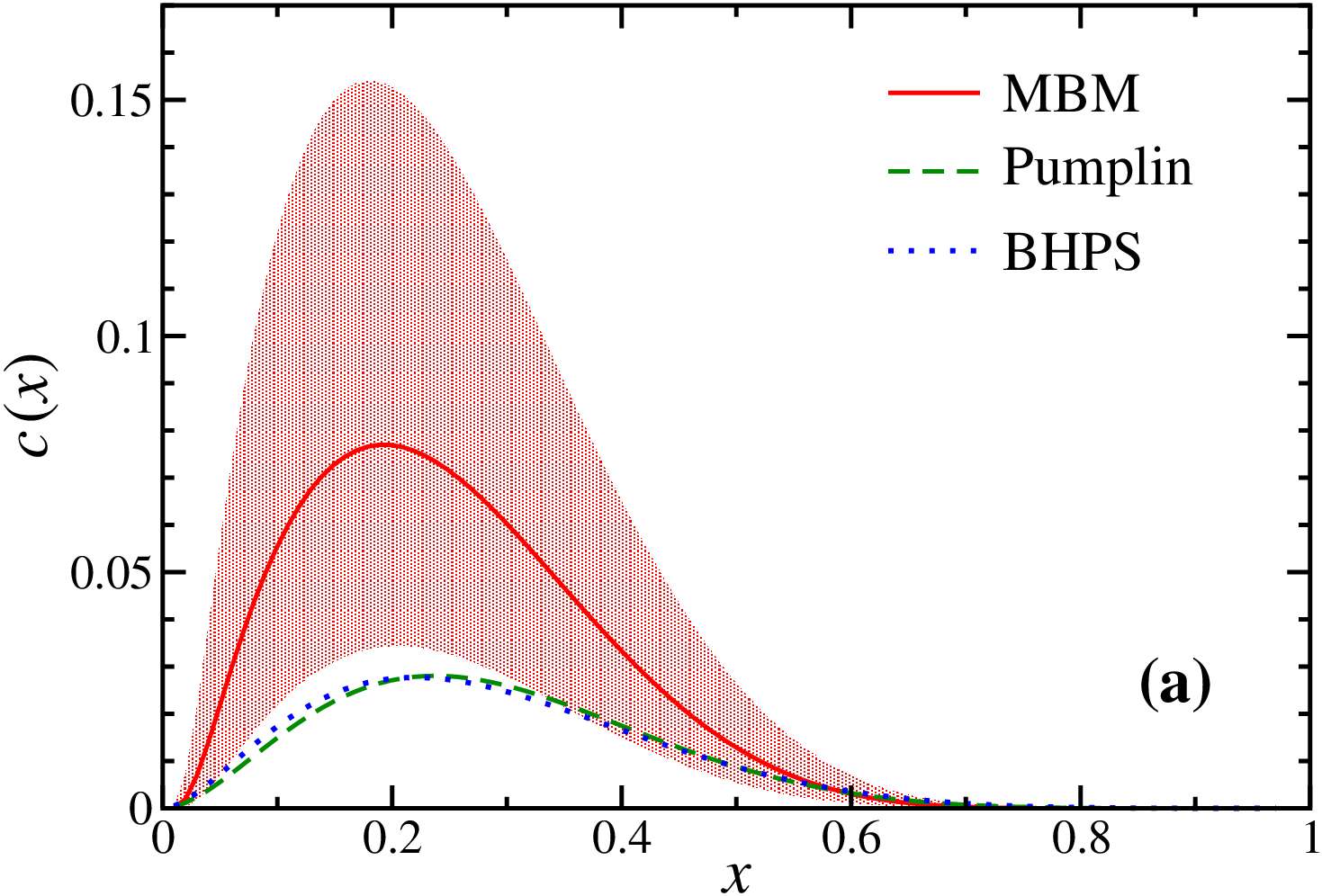}
\includegraphics[width=8cm]{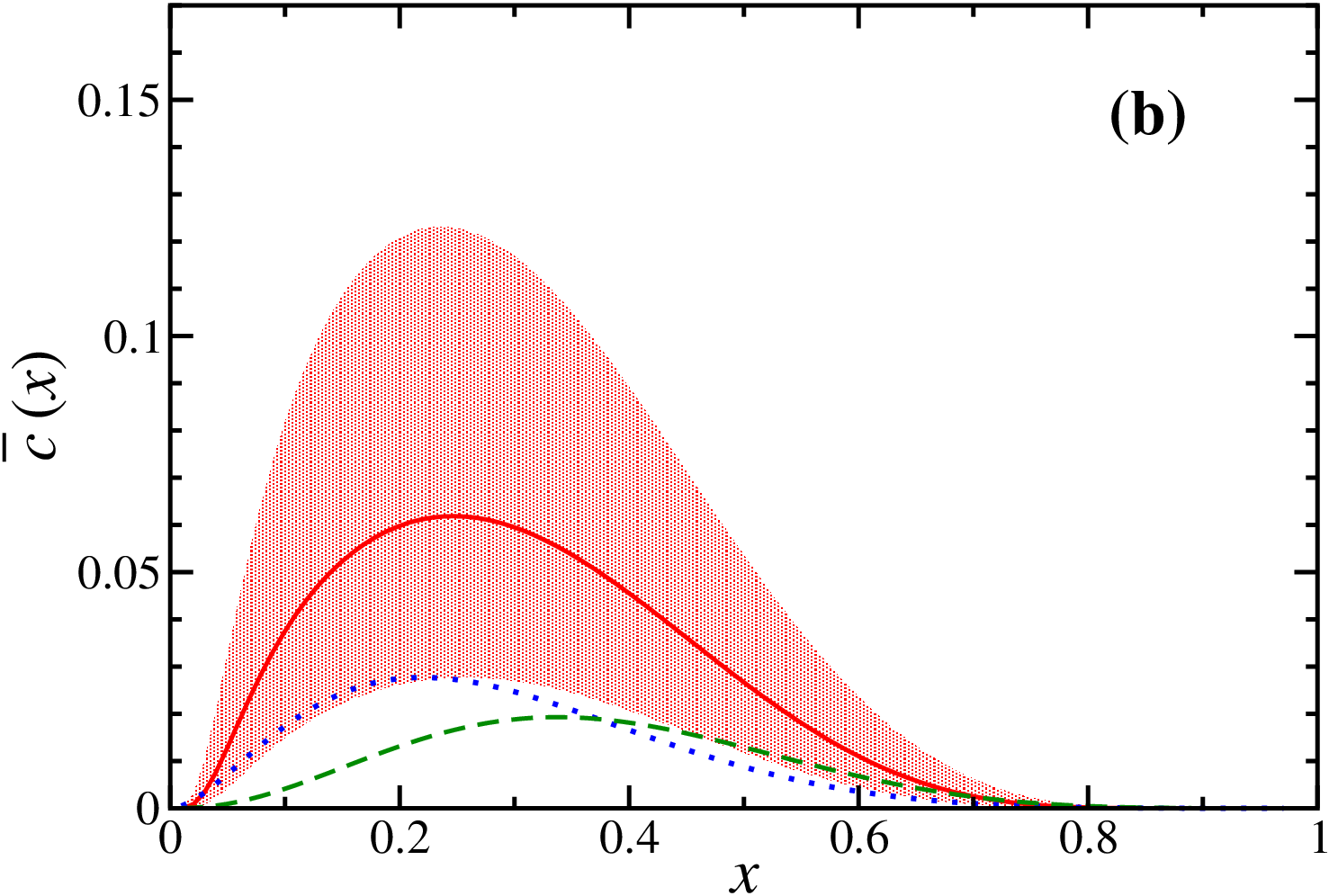}
\caption{(color online)
	Model dependence of the
	({\bf a}) $c(x)$ and
	({\bf b}) $\bar{c}(x)$ distributions in the nucleon,
	for the	MBM with the confining model for the PDFs in charmed
	hadrons	with cutoff mass $\Lambda = (3.0 \pm 0.2)$~GeV
	(red solid and shaded band), the Pumplin scalar meson cloud
	model (green dashed), and the BHPS intrinsic charm model
	(blue dotted).}
\label{fig:cnaive}
\end{figure}

The resulting charm distribution in the nucleon in the scalar meson
cloud model \cite{Pum05} is found to be similar in shape
to that in the MBM with the confining vertex function, as 
Fig.~\ref{fig:cnaive} illustrates.  The distribution is also
very similar to that in the simplified BHPS five-quark model
\cite{BHPS} discussed in Sec.~\ref{ssec:BHPS}
(which is to be expected perhaps given that the scalar model
is normalized to the charm in the BHPS model).
The $\bar{c}$ distribution in the scalar model \cite{Pum05}
is again harder than the $c$, while in the BHPS model the charm and
anticharm distributions are assumed identical.
Compared with the distributions in the confining MBM, the results
for the lower limit of the form factor cutoff $\Lambda = 2.8$~GeV,
are rather similar to the results of the other models shown
in Fig.~\ref{fig:cnaive}.  However, for the central value
$\Lambda = 3$~GeV the MBM model gives a substantially larger
intrinsic charm content.
A similar result holds for the anticharm distributions,
except that the MBM results with confining form factor are
slightly harder than the BHPS results.

Having explored the model dependence of the total intrinsic
$c$ and $\bar{c}$ distributions in the nucleon, we can now
directly confront the results with measurements of the charm
structure function, $\Fcc$.  This will provide additional
constraints on the model parameters, complementing those of
the inclusive $\Lambda_c$ production in $pp$ scattering
discussed in Sec.~\ref{ssec:HSM}.

\subsection{Charm structure function}
\label{ssec:EMC}

The calculations of intrinsic charm in this analysis are normalized
to inclusive charm production data in $pp$ collisions, as discussed
in Sec.~\ref{ssec:HSM}.  The results can also be confronted with data
on the charm structure function $\Fcc$ obtained from measurements of
charm production cross sections in deep-inelastic lepton scattering.
In comparing with experimental measurements of $\Fcc$, in addition
to intrinsic charm arising from nonperturbative fluctuations of the
nucleon into states with 5 or more quarks, one must also consider the
``extrinsic'' charm arising from gluon radiation to $c\bar c$ pairs,
which is described by perturbative QCD evolution.

To lowest order in the strong coupling constant $\alpha_s$,
the charm structure function is straightforwardly related to
the $c$ and $\bar c$ parton distributions in the nucleon,
\be
\Fcc (x,Q^2) = \frac{4x}{9}\left[ c(x,Q^2) + \bar{c}(x,Q^2) \right].
\label{eq:F2c}
\ee
When combining the two charm contributions, it is necessary to assign
a scale $Q_0^2$ at which the nonperturbative charm is generated,
and then evolve this, together with the perturbative component,
to a common $Q^2$ for comparison with experiment.
While the absolute scale of the intrinsic contribution is a
characteristic ingredient of the model (in our case, the MBM),
it is customary to set this to the effective charm quark mass,
$Q_0^2 = m_c^2 = 1.69$~GeV$^2$.

\begin{figure}[t]
\includegraphics[width=8cm]{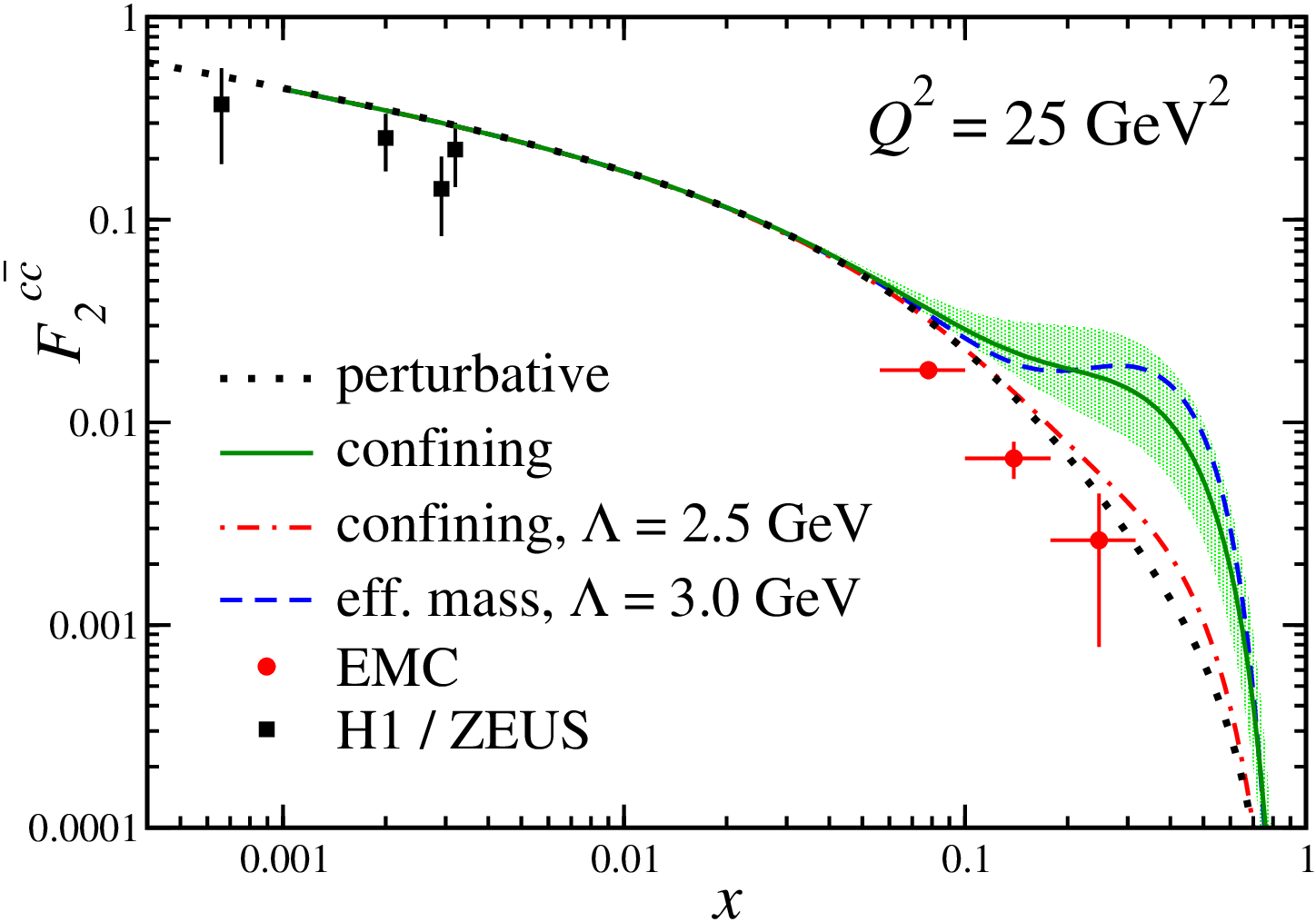}\ \
\includegraphics[width=8cm]{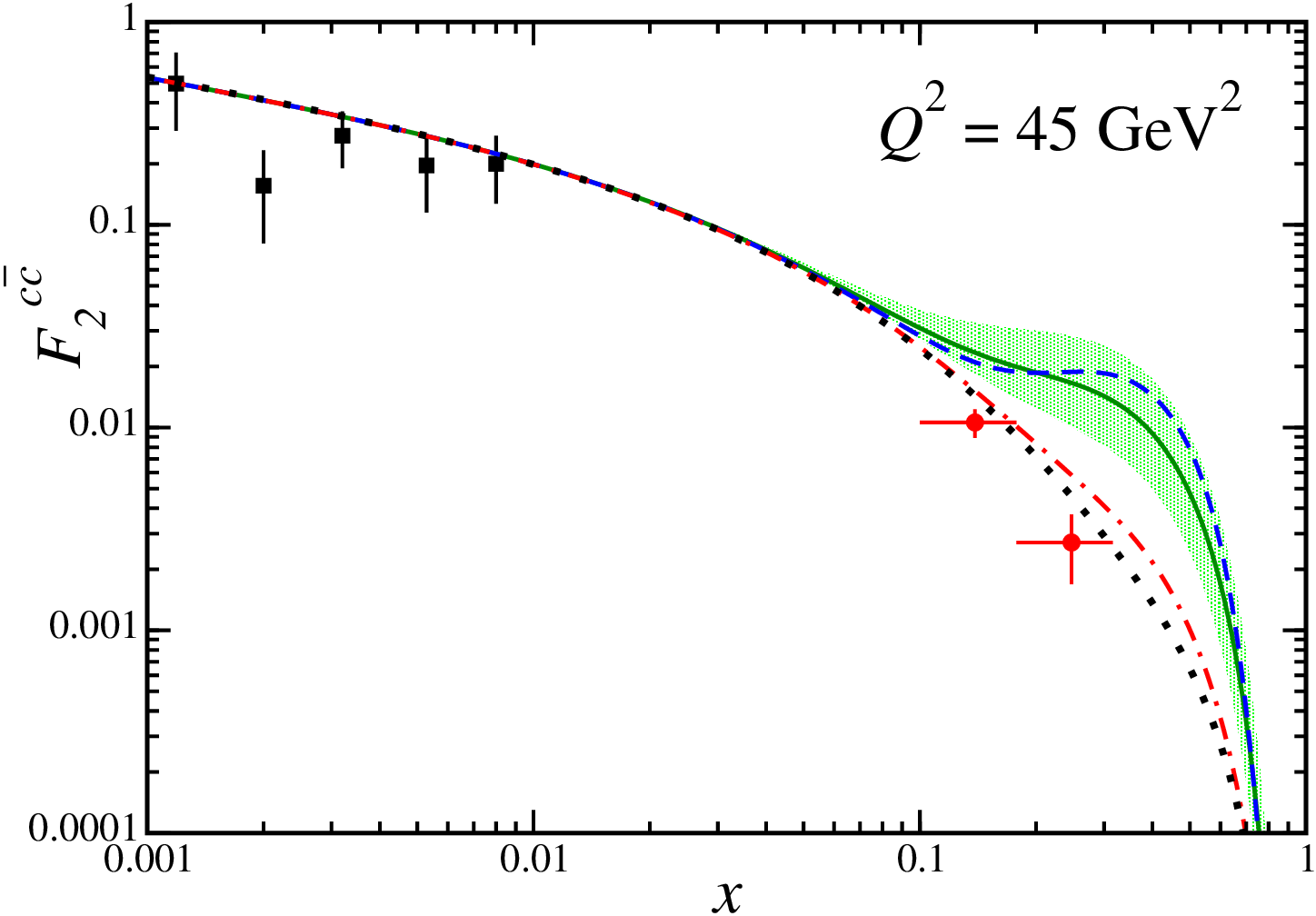}\vspace*{1cm}
\includegraphics[width=8cm]{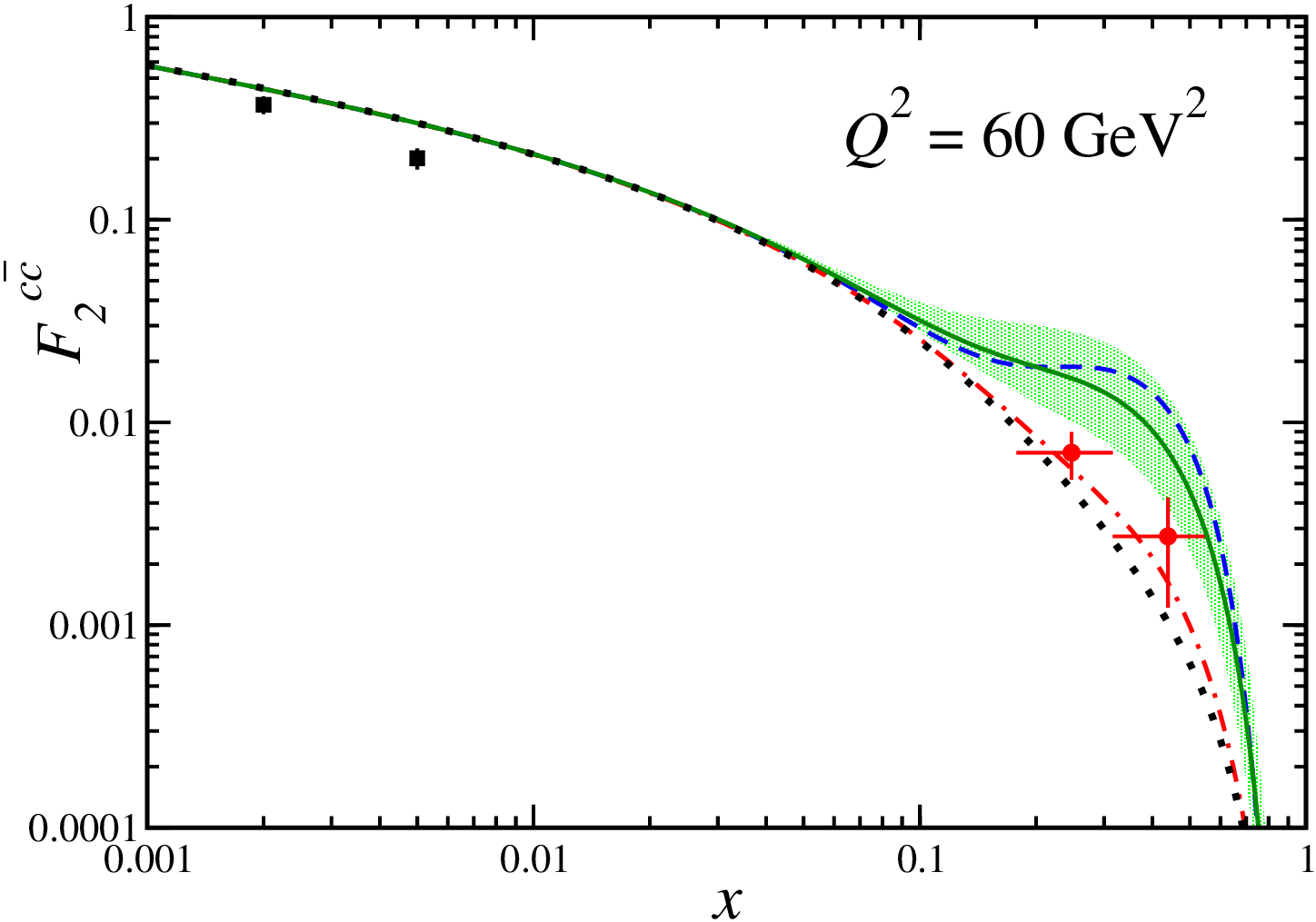}
\caption{(color online)
	Charm structure function $\Fcc$ at
	$Q^2 = 25$~GeV$^2$ (top),
	45~GeV$^2$ (middle), and
	60~GeV$^2$ (bottom).
	The perturbative QCD calculation (black dotted line)
	is compared with nonperturbative charm contributions
	in the MBM using the
	confining model with cutoff $\Lambda=(3.0 \pm 0.2)$~GeV
	  (green solid line and shaded band),
	confining model with $\Lambda=2.5$~GeV
	  (red dot-dashed line), and
	effective mass model with $\Lambda=3.0$~GeV
	  (blue dashed line).
	The data are from H1 and ZEUS (black squares)
	and EMC (red circles).}
\label{fig:F2c-MBM_DC-EMC}
\end{figure}

The calculated charm structure function $\Fcc (x,Q^2)$ is shown
in Fig.~\ref{fig:F2c-MBM_DC-EMC} for three different average $Q^2$
values ranging from 25~GeV$^2$ to 60~GeV$^2$, compared with data from
the H1 \cite{H1} and ZEUS \cite{ZEUS} Collaborations at HERA and
with higher-$x$ data from EMC \cite{EMC}.
The extrinsic charm distribution is obtained from the CTEQ6.5
parametrization \cite{Tung07}, where it is generated
perturbatively through QCD evolution.
The nonperturbative contribution to $\Fcc$ is computed from 
Eq.~(\ref{eq:F2c}) using the MBM at the scale $Q_0^2$ and evolved
to higher $Q^2$ using the next-to-leading order evolution code
from Ref.~\cite{Miyama96}.
Since the experimental $Q^2$ values are large compared to $m_c^2$,
standard massless QCD evolution, in the form of the Variable Flavor
Number Scheme, can be used.

At the lower two $Q^2$ values the extrinsic charm distributions
give a reasonable fit to the data, both in the low-$x$ and high-$x$
regions, although at $Q^2 = 60$~GeV$^2$ the perturbative results
generally underestimate the EMC data at high $x$.
The addition of the nonperturbative contribution raises the total
$\Fcc$, depending on the amount of intrinsic charm assumed in the
model.  For the MBM with the confining model vertex functions,
using the cutoff $\Lambda = (3.0 \pm 0.2)$~GeV obtained from the
fit to the inclusive $\Lambda_c^+$ production data in $pp$ scattering
generally overestimates the EMC $\Fcc$ data at large $x$ at the
lowest $Q^2$ points.  At the highest $Q^2$ value the calculated
structure function is marginally consistent with the data at the
lower edge of the error band.
Using instead the MBM with the effective mass model vertex
functions and the same cutoff $\Lambda = 3.0$~GeV, the peak
at large $x$ is more pronounced, and thus overestimates the
EMC data more somewhat more.
Lowering the cutoff to $\Lambda = 2.5$~GeV for the confining model,
the resulting $\Fcc$ is in better agreement with the data,
producing a smaller overestimate of the lower-$Q^2$ data points
and resulting in a better fit to the $Q^2 = 60$~GeV$^2$ data. 
Note that for such a small cutoff the average charm normalization
$\langle n \rangle_{MB}^{\rm (charm)} \lesssim 0.5\%$, which would
significantly underestimate the inclusive $\Lambda_c^+$ production
data (see Fig.~\ref{fig:R608_CS}).

In an earlier study \cite{Pum07}, Pumplin, Lai and Tung carried
out a global fit to high-energy data with phenomenological PDFs
including an intrinsic charm term at the starting scale $Q_0 = m_c$.
Using several different phenomenological forms for the intrinsic
charm (including the BHPS model, Eq.~(\ref{eq:charmprob}),
the scalar MBM discussed in Sec.~\ref{ssec:approx}, and a
``sea-like'' charm model), the magnitude of the charm contribution
was varied until a substantial increase in the $\chi^2$ was found
with the set of global high-energy data.
The analysis found that the global fits could accommodate charm
momentum fractions of $P_c \approx 2\%$ in the BHPS and scalar MBM
models and $P_c \approx 2.5\%$ in the sea-like model at the 90\%
confidence level, which are significantly larger than the constraints
from the EMC $\Fcc$ data, and at the upper boundary of the range
allowed by the ISR R608 $\Lambda_c^+$ production data.
The more recent update \cite{Dulat13} that includes NNLO corrections
finds $P_c \leq 1.5\%$ for the sea-like model and $P_c \leq 2.5\%$
for the BHPS model at the scale $Q_0$.

We should note, however, that while the analysis in Ref.~\cite{Pum07}
fitted the precision low-$x$ charm structure function data from
H1 and ZEUS, it {\em did not} include the EMC $\Fcc$ data at large $x$,
which it is likely would have resulted in tighter constraints on the
magnitude of the intrinsic charm.
In a future study \cite{Hob13}, we plan to perform a dedicated global
analysis of all high-energy data, including the EMC measurements of
$\Fcc$, to determine the constraints on the magnitude and shape of
the nonperturbative charm component of the nucleon.
Additional data on $\Fcc$ at large $x$ would of course be very
valuable in providing additional information on this question.

\section{Conclusion}
\label{sec:conc}

In this work we have presented a comprehensive analysis of intrinsic
charm in the nucleon using a phenomenological model formulated in terms
of effective meson--baryon degrees of freedom, with couplings taken
from $DN$ and $\bar{D} N$ scattering studies \cite{Hai07, Hai08, Hai11}.
Within the MBM framework, the $c$ and $\bar{c}$ distributions in the
nucleon are expressed as convolutions of $N \to$ charmed meson $+$
baryon splitting functions and charm PDFs in the mesons and baryons.
We have included in the calculation all
of the low-lying pseudoscalar $D$ and vector $D^*$ mesons, together
with the spin-1/2 $\Lambda_c$, $\Sigma_c$ and spin-3/2 $\Sigma_c^*$
baryons that couple to the proton, in contrast to some previous
analyses that neglected the spin structure \cite{Pum05}
and assumed dominance by the lowest mass $\bar{D}^0\, \Lambda_c^+$
state \cite{MT97, Nav96}.

The splitting functions are determined essentially in terms of a
single parameter, the momentum cutoff that regulates the ultraviolet
behavior of the nucleon--meson--baryon form factor.
The form factor is constrained by fitting to data on inclusive
$\Lambda_c^+$ production cross sections in $pp$ collisions from the
ISR \cite{Chauvat:1987kb}, which gives an exponential cutoff mass of
$\Lambda = (3.0 \pm 0.2)$~GeV.  The resulting distributions are also
consistent with the more recent SELEX data \cite{Garcia:2001xj} on
$\Lambda_c^+/\bar{\Lambda}_c^-$ asymmetries, although here there is
additional sensitivity to the $\bar{\Lambda}_c^-$ production mechanism.

The distributions of $c$ and $\bar c$ quarks in the charmed baryons
and mesons are computed from a relativistic quark--spectator model,
in which the momentum distributions of the quarks are parametrized
through phenomenological quark--spectator--nucleon vertex functions.
The resulting convolution integrals contain poles in the quark
propagators, and we examined two methods for avoiding these poles.
The first method involves using an effective charm mass that is
sufficiently large that no poles occur in the physical region,
while the second involves a vertex function that cancels the
propagators in the denominator, thereby simulating the dynamics
of quark confinement.
For comparison we also considered a simplified model in which the
charm distributions are approximated by $\delta$ functions at the
fraction of the hadron mass carried by the constituent charm or
anticharm quark.
The various methods for avoiding propagator poles give rise to
somewhat disparate distributions in hadrons, although the differences
in the resulting intrinsic charm distributions in the nucleon are
mitigated by the smearing effects of the convolution.

For values of the cutoff parameter favored by the inclusive
$\Lambda_c^+$ production data, we find that intrinsic charm is
in fact dominated by the vector meson plus spin-1/2 baryon state
$\bar{D}^{*0} \Lambda_c^+$ state, which is responsible for almost
70\% of the total intrinsic charm content.  The reasons for the
singular strength of this channel are the very strong tensor
coupling to vector mesons (reminiscent of the large tensor coupling
of $\rho$ mesons in $NN$ interactions), and the additional momentum
dependence introduced by the derivative interaction.

One of the unique characteristics of the MBM is the almost unavoidable
asymmetry in the $x$ dependence of the resulting $c$ and $\bar c$
distributions in the nucleon, which reflects the different environments
in which the charm and anticharm quarks exist (charmed baryon for $c$,
$D$ and $D^*$ mesons for $\bar c$).  Although the detailed shapes and
absolute values of the $c$ and $\bar c$ PDFs depend on the models for
the hadronic form factors and quark--spectator vertex functions,
a universal feature of the MBM framework is the significantly harder
$\bar c$ distribution compared with the $c$.  The magnitude of the
corresponding asymmetry ratio exceeds 50\% at $x \gtrsim 0.5$, a result
which is largely independent of the model parameters and scale.
Comparison with the only available data on the charm structure function
$\Fcc$ at large values of $x$ from the EMC \cite{EMC}, where the
intrinsic charm is predicted to be most significant ($x \gtrsim 0.1$),
places tighter limits on the size of the momentum cutoff $\Lambda$
in the MBM.  This suggests that the amount of intrinsic charm needed
to describe the inclusive $pp$ data would lead to an overestimate of
the DIS charm cross sections, although the $\Fcc$ data themselves are
somewhat inconclusive, with some data points indicating an excess
over the perturbative QCD expectations and others consistent with
no intrinsic charm at all.
At the very least, the large-$x$ $\Fcc$ results illustrate the
potential value of such data, and it is hoped that future measurements,
such as at the proposed Electron-Ion Collider \cite{Deshpande05,
Boer11, Accardi12}, will clarify the situation.

Other possibilities for identifying intrinsic charm experimentally
include the measurement of $W$ and $Z$ cross sections at the LHC
\cite{Halzen13}, where in the case of $W$ production, up to
1/3 of the cross sections at $\sqrt{s} = 7$~TeV and as much as 40\%
at $\sqrt{s} = 14$~TeV can arise through charm production.
However, although charm plays a significant role in these reactions,
the dominant charm contributions occur at low rapidity where
perturbative charm is expected to dominate.
Alternatively, photon plus charm jet production at LHC energies,
and in particular the transverse momentum distribution of prompt
photons, has been identified \cite{Bednyakov13} as potentially
sensitive to intrinsic charm.

Another promising set of observables to study is charge asymmetries
in the production of charmed hadrons such as $D^+/D^-$, $D_s^+/D_s^-$
or $\Lambda_c/\bar{\Lambda}_c$  using proton, pion or $\Sigma^-$
beams, as well as virtual photon probes in deep-inelastic lepton
scattering. 
Particularly useful are comparisons of cross sections involving
different incident beams, which in principle are sensitive to the
so-called leading-particle effect \cite{VB96}, which involves a
strong correlation between the quantum numbers of the produced
charmed hadron and the beam hadron(s).
For instance, in $\pi^- p \to D X$ scattering there is a preference
to produce $D^-$ and $D^0$ over $D^+$ and $\bar{D}^0$, which is
attributable to the fact that the valence $(d\bar{u})$ content of the
$\pi^-$ appears also in the valence structure of $D^-$ and $D^0$, 
but not in $D^+$ and $\bar{D}^0$.
Although significant charge asymmetries are observed in some cases,
a number of possible mechanisms exist for such asymmetries (such as
recombination of a produced charm quark with a valence quark or
diquark from the beam, or coalescence between a charm and valence
quark originating in the beam \cite{BHPS, ICHT}), and it will be
necessary to disentangle these processes from any pre-existing
charm in the initial state.
However, any confirmation of an asymmetry between $c$ and $\bar c$
distributions in the nucleon at large $x$ would provide an
unambiguous signature of intrinsic charm, beyond anything that
could be generated from perturbative QCD.

Finally, we should mention some of the limitations of the MBM framework
that has been used to compute the nonperturbative charm in this analysis.
Unlike the fluctuations of the nucleon to baryons and light pseudoscalar
mesons, such as pions, which are grounded in the chiral symmetry
properties of QCD, the dissociation into the much heavier charmed
mesons and baryons is a rather more model-dependent postulate.
A reflection of this is the {\it ad hoc} restriction of the Fock state
expansion in Eq.~(\ref{eq:Fock}) to a particular set of meson--baryon
states $MB$.
While most analyses of charm in the MBM have simply assumed dominance
by the lowest-mass $\bar{D}^0 \Lambda_c^+$ state, in this work we have
attempted a more systematic treatment, including all low-lying states
with the appropriate quantum numbers.  As a guide, we have kept a close
analogy with earlier $DN$ scattering analyses \cite{Hai07, Hai08, Hai11}
which have provided constraints on the nucleon--meson--baryon couplings.
Notwithstanding, in the absence of data on transitions to individual
charmed meson and baryon states, it has been necessary to apply a
universal cutoff parameter for all hadronic form factors, as well as
construct phenomenological models for the charm quark distributions
in the charmed hadrons.
Further development of this approach will benefit from a combined
analysis of all data, using both hadronic and leptonic probes.
The future availability of inclusive charmed baryon production data
as well as measurements of the charm structure function at large $x$
will provide vital benchmarks for establishing the presence of
intrinsic charm in the nucleon.

\acknowledgments

We are grateful to J.~Haidenbauer for helpful discussions
about the J\"ulich meson--baryon scattering models.
We also thank P.~Jimenez-Delgado, D.~Murdock, F.~Steffens and R.~Vogt
for helpful discussions, and S.~Kumano for his QCD evolution programs.
TJH and JTL were supported in part by the US National Science Foundation
under grant NSF-PHY-1205019.  The work of TJH was also supported in part
by DOE grant DE-FG02-87ER40365.  WM was supported by the DOE contract
No.~DE-AC05-06OR23177, under which Jefferson Science Associates, LLC
operates Jefferson Lab.

\appendix
\section{Derivation of meson--baryon splitting functions}
\label{sec:A_MBM}

In this appendix we provide additional technical details of the
derivations of the splitting functions for the dissociation of a
nucleon with four-momentum $P$ into a meson $M$ with momentum $k$
and baryon $B$ with momentum $p$.  We consider dissociations into
the SU(4) octet isoscalar $\Lambda_c$ and isovector $\Sigma_c$
baryons, and the decuplet $\Sigma_c^*$ baryon, accompanied by the
charmed pseudoscalar $D$ and vector $D^*$ mesons.  The transitions
to specific isospin states are obtained using appropriate isospin
transition factors, as discussed in Sec.~\ref{sec:mb}.

The contribution of a specific meson--baryon component to the
nucleon hadronic tensor $W_{\mu\nu}^N$ is defined in terms of the
contributions $\delta^{[MB]} F_{1,2}^N$ to the structure functions
as \cite{Beringer:1900zz}
\bea
\delta^{[MB]} W_{\mu\nu}^N(P,q)
&=& \widetilde{g}_{\mu \nu}\, \delta^{[MB]} F_1^N\
 +\ \frac{\widetilde{P}_\mu \widetilde{P}_\nu}{P \cdot q}\,
    \delta^{[MB]} F_2^N,
\label{eq:had_tens}
\eea
where $q$ is the four-momentum of the external electromagnetic
field, and we define
  $\widetilde{g}_{\mu\nu} = -g_{\mu \nu} + q_{\mu} q_{\nu}/q^2$, and
  $\widetilde{P}_{\mu} = P_\mu - P \cdot q\, q_\mu/q^2$.
Evaluating the one-loop diagram for the scattering from the meson $M$
gives
\bea
\delta^{[MB]} W_{\mu\nu}^N
&=& \frac{g^2_{DBN}}{16\pi^2} \int_0^1 dy \int_0^\infty 
  \frac{dk^2_\perp}{y(1-y)}
  \frac{|F(s)|^2} {(M^2 - s)^2}\
  N^{MB}_{\mu\nu},
\label{eq:del_MB}
\eea
where $y$ is the longitudinal momentum fraction carried by the meson,
$F(s)$ is the $MBN$ hadronic factor, with the invariant mass squared
of the $MB$ system $s$ defined in Eq.~(\ref{eq:CoM_En}).
The tensor $N^{MB}_{\mu\nu}$ is computed from the spin trace of the
appropriate meson and baryon propagators and vertices, with explicit
forms given below.

Within the framework of time-ordered perturbation theory evaluated
in the IMF kinematics ($P_L \to \infty$), with intermediate state
particles on their mass-shells but off their ``energy-shells'', the
standard decomposition for the momentum variables is \cite{DLY70}
\begin{subequations}
\label{eq:kin-TOPT}
\bea
P_0 &=& P_L + \frac{M^2}{2P_L} + {\cal O}\left(\frac{1}{P_L^2}\right),
\\
k_0 &=& |y| P_L + \frac{k^2_\perp + m_M^2}{2 |y| P_L} 
	+ {\cal O}\left(\frac{1}{P_L^2}\right),
\\ 
p_0 &=& |1 - y| P_L + \frac{k^2_\perp + M_B^2}{2 |1 - y| P_L}
	+ {\cal O}\left(\frac{1}{P_L^2}\right),
\eea  
\end{subequations}%
for the energies, and
\begin{subequations}
\label{eq:kin-TOPT_mom}
\bea
\bm{k} &=& |y| \bm{P} + \bm{k_\perp},
\\
\bm{p} &=& |1-y| \bm{P} - \bm{k_\perp},
\eea
\end{subequations}%
for the three-momenta, with $\bm{k}_\perp \cdot \bm{P} = 0$.
In the non-vanishing forward limit one has
  $y \in [0,1]$, such that $|1-y| = (1-y)$.
The expression in Eq.~(\ref{eq:del_MB}) are evaluated in terms of
inner products $P \cdot k$, $P \cdot p$ and $P \cdot k$ computed
from Eqs.~({\ref{eq:kin-TOPT}) and (\ref{eq:kin-TOPT_mom}), and the
corrections to the structure functions $\delta^{[MB]} F_{1,2}^N$
obtained by equating the coefficients of the tensors in
Eqs.~(\ref{eq:had_tens}) and (\ref{eq:del_MB}).
The corrections to the $c$ and $\bar c$ distributions
in Eqs.~(\ref{eq:mesoncloud}) are then extracted from
$\delta^{[MB]} F_{1,2}^N$ using parton model relations
analogous to Eq.~(\ref{eq:F2c}).

\underline{\em $N \to DB$ splitting} 

The dissociation of a nucleon to a spin-1/2 charmed baryon
$B = \Lambda_c$ or $\Sigma_c$ and a pseudoscalar $D$ meson is
derived from the effective hadronic Lagrangian \cite{Hai11}
\bea
{\cal L}_{DBN}
&=& i g\, \bar{\psi}_N\, \gamma_5\, \psi_B\, \phi_D\ +\ \mathrm{h.c.},
\label{eq:Lagrangian}
\eea
where $\psi_N$ and $\psi_B$ are the nucleon and charmed baryon
fields, respectively, $\phi_D$ is the spin-0 $D$ meson field,
and the coupling constant $g \to g_{DBN}$.
Treating the propagation of the meson and baryon fields as
appropriate for point particles, the trace factor $N^{DB}_{\mu\nu}$
can be written as
\bea
N_{\mu\nu}^{DB}
&=& \frac{1}{2} \mathrm{Tr}
    \left[ i\gamma_5 (\Psl + M) i\gamma_5\, W^D_{\mu\nu}(k,q)\,
	   (\psl + M_B)
    \right]					\nonumber\\
&=& \left( 2 P \cdot p - 2 M M_B \right)\,
    \widetilde{g}_{\mu\nu} F_1^D\, +\, \dots
\label{eq:NLD-trace}
\eea
where $W_D^{\mu\nu}$ is the hadronic tensor for the $D$ meson,
with a form similar to that in Eq.~(\ref{eq:had_tens}), and
$F_1^D$ is the corresponding structure function which depends
on the $\bar c$ distribution in $D$.  Equating the coefficients
of $\widetilde{g}_{\mu\nu}$ in Eqs.~(\ref{eq:had_tens}),
(\ref{eq:del_MB}) and (\ref{eq:NLD-trace}) then yields the
convolution expression in Eq.~(\ref{eq:mesoncloud_cb}) with
the splitting function in Eq.~(\ref{eq:DLsplit}).
Performing the analogous calculation for the recoil process involving
scattering from the baryon $B$ confirms the symmetry relation
(\ref{eq:fphi}), which follows from the global charge and momentum
conservation relations in Eqs.~(\ref{eq:recip1}) and (\ref{eq:conserv}).

\underline{\em $N \to D^* B$ splitting} 

For the interaction between the nucleon and spin-1/2 baryon with a
vector $D^*$ meson, the effective Lagrangian is given by \cite{Hai11}
\bea
{\cal L}_{D^* B N}
&=& g\, \bar{\psi}_N \gamma_\mu\, \psi_B\, \theta_{D^*}^\mu\
 +\ \frac{f}{4 M}
    \bar{\psi}_N \sigma_{\mu\nu} \psi_B\, F_{D^*}^{\mu \nu}\
 +\ \mathrm{h.c.},
\label{eq:L-S1}
\eea
where $\theta_{D^*}^\mu$ is the vector meson field, with field
strength tensor
$F_{D^*}^{\mu\nu} = \partial^\mu \theta_{D^*}^\nu
		  - \partial^\nu \theta_{D^*}^\mu$,
the tensor operator $\sigma_{\mu\nu} = (i/2) [\gamma_mu,\gamma_nu]$,
and the vector and tensor couplings are $g \to g_{D^* B N}$
and $f \to f_{D^* B N}$.
In this case the trace factor $N_{\mu\nu}^{D^* B}$ is given by
\bea 
N_{\mu\nu}^{D^* B}
&=& \frac{1}{2} \mathrm{Tr}
    \bigg[
    (\Psl + M)
    \Big( g \gamma^\alpha
	+ \frac{f}{2M} (\Delta^\alpha - \gamma^\alpha \Dsl)
    \Big)
    (\psl + M_B)					\nonumber\\
& & \hspace*{2.2cm} \times
    \Big( g \gamma^\beta
	- \frac{f}{2M} (\Delta^\beta - \gamma^\beta \Dsl)
    \Big)
    \bigg]\, W^{D^*}_{\mu\nu\alpha\beta}(k,q)		\nonumber\\
&=& \bigg( g^2 G_v + \frac{gf}{M} G_{vt} + \frac{f^2}{M^2} G_t
    \bigg)\, \widetilde{g}_{\mu\nu} F_1^{D^*}\ +\ \dots
\label{eq:S1-tr2}
\eea 
where $\Delta = P - p$, and the kinematical factors $G_v$, $G_{vt}$
and $G_t$ are given in Eqs.~(\ref{eq:vectABC-app}).
The rank-4 tensor for the interacting $D^*$ meson can be expressed
in the form \cite{MT93}
\bea
W^{D^*}_{\mu\nu\alpha\beta}(k,q)
&=& \Big(
    \widetilde{g}_{\mu\nu} F_1^{D^*}
    + \frac{\widetilde{k}_\mu \widetilde{k}_\nu}{m^2_{D^*}} F_2^{D^*}
    \Big)\,
    \widetilde{g}_{\alpha\beta},
\label{eq:rank-4}
\eea
with $F_{1,2}^{D^*}$ the corresponding vector meson structure functions.

\underline{\em $N \to D B^*$ splitting} 

The interaction of a nucleon with a spin-3/2 charmed baryon
$B = \Sigma_c^*$ and a pseudoscalar meson is given by the
Lagrangian \cite{Hai11}
\bea
{\cal L}_{D B^* N}
&=& \frac{f}{m_D}
    \Big( \bar{\Psi}_{B^*}^\mu\, \psi_N\, \partial_\mu \phi_D\
       +\ \bar{\psi}_N \Psi_{B^*}^\mu \partial_\mu \phi_D
    \Big),
\label{eq:piD-Lagr}
\eea
where $\Psi_{B^*}^\mu$ is the Rarita-Schwinger spinor-vector field,
and the coupling $f \to f_{D B^* N}$.
From Eq.~(\ref{eq:piD-Lagr}) the trace factor tensor can be written
\bea 
N_{\mu\nu}^{D B^*}
&=& \frac{1}{2} \mathrm{Tr}
    \bigg[ (\Psl + M)\,
	   \Lambda^{\alpha\beta}_{B^*}(p)\,
	   \Delta_\alpha \Delta_\beta\, W_{\mu\nu}^D(k,q)
    \bigg]						\nonumber \\
&=& \frac{4}{3}
    \left( P \cdot p + M M_{B^*} \right)
    \left( \frac{(p \cdot \Delta)^2}{M_{B^*}^2} - \Delta^2 \right)
    \widetilde{g}^{\mu\nu}\, F_1^D\ +\ \dots
\label{eq:piD-trace}
\eea 
where the energy projector for the spinor-vector is
\bea
\Lambda^{\alpha\beta}_{B^*}(p)
&=& (\psl + M_{B^*})
    \left(- g^{\alpha\beta}
          + {\gamma^\alpha \gamma^\beta \over 3}
          + {\gamma^\alpha p^\beta - \gamma^\beta p^\alpha \over 3M_{B^*}}
          + {2\, p^\alpha p^\beta \over 3 M_{B^*}^2}
    \right).
\eea
The inner products in Eq.~(\ref{eq:piD-trace}) can once again be
obtained from Eqs.~(\ref{eq:kin-TOPT}) and (\ref{eq:kin-TOPT_mom})
with $B \to B^*$, which leads directly to the splitting function
in Eq.~(\ref{eq:DS*-split}).

\underline{\em $N \to D^* B^*$ splitting} 

Finally, for the nucleon splitting to a spin-3/2 charmed baryon $B^*$
coupled to a vector meson $D^*$ the effective hadronic Lagrangian is
given by \cite{Hai11}
\bea
{\cal L}_{D^* B^* N}
&=& \frac{f}{m_{D^*}}
    i \Big( \bar{\Psi}_{B^* \nu} \gamma^5 \gamma_{\mu} \psi_N
	  - \bar{\psi}_N \gamma^5 \gamma_{\mu} \Psi_{B^* \nu} 
      \Big) F_{D^*}^{\mu\nu},
\label{eq:DsSigS-Lagr}
\eea
where $f \to f_{D^* B^* N}$.
This yields the resulting trace tensor
\bea 
N_{\mu\nu}^{D^* B^*}
&=& \frac{1}{2}
    \mathrm{Tr}
    \bigg[
	(\Psl + M)
	\gamma^5 \gamma^\alpha\,
	\Lambda_{B^*}^{\alpha'\beta'}(p)\,
	\gamma^5 \gamma^\beta
    \bigg]
    \mathcal{G}_{\alpha\beta\alpha'\beta'}\,
    \widetilde{g}_{\mu\nu} F_1^{D^*}\ +\ \dots,
\label{eq:D*S*-trace1}
\eea 
where we define
\bea
\mathcal{G}_{\alpha\beta\alpha'\beta'}
&=& \Delta_{\alpha\beta}\,   \widetilde{g}_{\alpha' \beta'}
  - \Delta_{\alpha\beta'}\,  \widetilde{g}_{\alpha' \beta}
  - \Delta_{\alpha'\beta}\,  \widetilde{g}_{\alpha  \beta'}
  + \Delta_{\alpha'\beta'}\, \widetilde{g}_{\alpha  \beta}
\eea
and $\Delta_{\alpha\beta} \equiv \Delta_\alpha \Delta_\beta$,
and the other expressions are defined above.
Evaluating the trace in Eq.~(\ref{eq:D*S*-trace1}) and equating
coefficients of the $\widetilde{g}_{\mu\nu}$ terms then leads to
the convolution relation with the splitting function in
Eq.~(\ref{eq:RS-fMB}).

\section{Charm content of charmed baryons and mesons}
\label{sec:B-MBM}

This appendix details the derivations of the $c$ and $\bar c$ quark
distributions in charmed baryons and mesons within the relativistic
quark--spectator model introduced in Sec.~\ref{sec:cinc}.
For consistency with the calculation of the hadronic splitting
functions in Appendix~\ref{sec:A_MBM}, we proceed from effective
quark--hadron Lagrangians using time-ordered perturbation theory
in the IMF to compute the PDFs in terms of phenomenological
vertex functions.

\underline{\em $\bar c$ in $D$} 

To model the distribution of a $\bar c$ quark in the pseudoscalar
$D$ meson we consider the effective Lagrangian describing the
coupling of the $D$ to the $\bar c$ and a light quark $q$,
\bea
{\cal L}_{\bar{c} q D}
&=& i g\, \bar{\psi}_{\bar c}\, \gamma_5\, \psi_q\, \phi_D\
 +\ \mathrm{h.c.},
\label{eq:Lagrangian-q}
\eea
where $\psi_q$ and $\psi_{\bar c}$ are the quark $q$ and $\bar c$
fields, and the effective coupling constant is $g \to g_{\bar{c} q D}$.
The contribution to the hadronic tensor of the $D$ meson from scattering
off the $\bar c$ quark with a spectator light quark $q$ can be written
in analogy with Eq.~(\ref{eq:del_MB}) for the hadronic calculation,
\bea
\bar{c}_D (z)
&=& \frac{N_D}{16\pi^2}
    \int_0^\infty \frac{d\hat{k}^2_\perp}{z(1-z)}
    \frac{|G(\hat{s})|^2}{(m_D^2 - \hat{s})^2}\,
    \widehat{T}^{\bar{c} q},
\label{eq:del_ctau}
\eea
where $\hat{k}_\perp$ denotes the internal quark transverse momentum
in the $D$ meson and $z$ is the Bjorken scaling variable for the quark
inside the $D$ meson, with the invariant mass squared $\hat{s}$ of the
$\bar{c} q$ pair defined in Eq.~(\ref{eq:s-hat}).  The normalization
factor $N_D$ is determined from valence quark number conservation in
the $D$ meson, Eq.~(\ref{eq:cbarDNorm}).

The trace factor $\widehat{T}^{\bar{c} q}$ can be computed from the
quark-level ``handbag'' diagram, yielding
\bea
\widehat{T}^{\bar{c} q}
&=& \frac{1}{4 \hat{k}^+}
    \mathrm{Tr}
    \left[
	i \gamma_5
	(\hat{\ksl} + m_{\bar c})\,
	\gamma^+\,
	(\hat{\ksl} + m_{\bar c})\,
	(-i \gamma_5)
	(-\hat{\psl} + m_q)
    \right],						\nonumber\\
&=& 2\, \big( \hat{p} \cdot \hat{k} + m_{\bar c} m_q \big),
\label{eq:N_c-Lc1}
\eea
which follows from the on-mass-shell condition in time-ordered
perturbation theory, \mbox{$\hat{k}^2 = m^2_{\bar{c}}$}, with
$\hat{p}$ the four-momentum of the spectator quark.  The $\gamma^+$
structure arises from reducing the hard scattering amplitude
  $\gamma^{\mu} (\hat{\ksl} + \qsl + m_{\bar{c}}) \gamma^{\nu}\,
   \delta((\hat{k} + q)^2 - m_{\bar c}^2)$
to its leading twist approximation \cite{Mulders:1992za},
  $\gamma^+/(2 \hat{k}^+)\, \delta(1 - z/\hat{y})$,
where $\hat{y}$ is the parton fraction of the hadron momentum,
after equating the coefficients of $g^{\mu\nu}$ and selecting
the $+$ component of the external photon current.

The result in Eq.~(\ref{eq:cbar_Dbar}) of Sec.~\ref{sec:cinc} is then
obtained by inserting the expression for $\widehat{T}^{\bar{c} q}$
in Eq.~(\ref{eq:N_c-Lc1}) into Eq.~(\ref{eq:del_ctau}), and using the
IMF momenta as in Eq.~(\ref{eq:kin-TOPT}) but with the replacements
$y \to z$, $M \to M_D$, $m_D \to m_{\bar{c}}$, and $M_B \to m_q$.

\underline{\em $\bar c$ in $D^*$} 

For the distribution in the $D^*$ meson, the following simple vector
form is chosen for the Lagrangian describing the $\bar{c} q D^*$
interaction,
\bea
{\cal L}_{\bar{c} q D^*}
&=& g\, \bar{\psi}_{\bar c}\, \gamma_\mu\, \psi_q\, \theta_{D^*}^\mu\
 +\ \mathrm{h.c.},
\label{eq:cD*-Lagr}
\eea
where $g \to g_{D^* \bar{c} q}$.  This yields the trace factor
\bea 
\widehat{T}^{(\bar{c} q)^*}
&=& \frac{1}{4 \hat{k}^+}
    \mathrm{Tr}
    \left[
	(-\hat{\psl} + m_q)
	\gamma^\alpha
	(\hat{\ksl} + m_{\bar c})\,
	\gamma^+\,
	(\hat{\ksl} + m_{\bar c})\,
	\gamma^\beta \right] \widetilde{g}_{\alpha\beta} \nonumber\\
&=& 4
    \left(
	  \frac{(P \cdot \hat{p})(P \cdot \hat{k})}{m_{D^*}^2}
	+ \frac{3}{2} m_{\bar c}\, m_q
	+ \frac{\hat{p} \cdot \hat{k}}{2}
    \right).
\label{eq:N-cbD*}
\eea 
Applying the same procedure as for the $\bar c$ distribution inside
the pseudoscalar $D$ meson, one immediately arrives at
Eq.~(\ref{eq:cbar_D-st}).

\underline{\em $c$ in $\Lambda_c$, $\Sigma_c$} 

The charm quark distributions inside charmed baryons are obtained
from an expression analogous to that in Eq.~(\ref{eq:del_ctau}),
\bea
c_B (z)
&=& \frac{N_B}{16\pi^2}
    \int_0^\infty \frac{d\hat{k}^2_\perp}{z(1-z)}
    \frac{|G(\hat{s})|^2}{(m_B^2 - \hat{s})^2}\,
    \widehat{T}^{c [qq]},
\label{eq:c_baryon}
\eea
where $\widehat{T}^{c [qq]}$ is the corresponding trace factor for
the scattering from the $c$ quark with a spectator diquark $[qq]$
in the intermediate state.
For spin-$1/2$ baryons, we consider the scalar interaction between
the $c$ and $[qq]$ quarks and the charmed baryon, given by the
Lagrangian
\bea
{\cal L}_{c [qq] B}
&=& g\, \bar{\psi}_B\, \psi_c\, \phi_{[qq]}\ +\ \mathrm{h.c.},
\label{eq:c-Lamb-Lagr}
\eea
with $g \to g_{c [qq] B}$, and $\phi_{[qq]}$ is the field of the scalar 
diquark.  The trace factor $\widehat{T}^{c [qq]}$ can be explicitly
derived as
\bea 
\widehat{T}^{c [qq]}
&=& \frac{1}{4 \hat{k}^+}
    \mathrm{Tr}
    \left[
	(\Psl + M_B) (\hat{\ksl} + m_c) \gamma^+ (\hat{\ksl} + m_c)
    \right]  \nonumber\\
&=& 2 \left( P \cdot \hat{k} + m_c\, M_B \right),
\label{eq:N_c-Lc2}
\eea 
giving the net result for the charm quark distribution inside
a spin-$1/2$ baryon in Eq.~(\ref{eq:c_in_Lambda}).

\underline{\em $c$ in $\Sigma_c^*$} 

For the charm density of the spin-$3/2$ $B^*$ baryons the following
Lagrangian is adopted,
\bea 
{\cal L}_{c [qq]^* B^*}
&=& \frac{g}{m_{[qq]^*}} \
    i \left(
	\bar{\Psi}_{B^* \nu} \gamma_{\mu} \psi_c
      - \bar{\psi}_c \gamma_{\mu} \Psi_{B^* \nu}
    \right)
    F_{[qq]^*}^{\mu\nu},
\label{eq:cSig*-Lagr}
\eea 
with $g \to g_{c [qq]^* B^*}$, which correctly gives the parities
of the physical $B^*$ and quark fields.  Again, the field strength
tensor here has the form
  $F_{[qq]^*}^{\mu \nu}
  = \partial^{\mu} \theta_{[qq]^*}^{\nu}
  - \partial^{\nu} \theta_{[qq]^*}^{\mu}$,
where $\theta_{[qq]^*}$ denotes the (spin-1) axial-vector diquark.
The trace factor in this case is found to be
\bea 
\widehat{T}^{c [qq]^*}
&=& \frac{1}{4 m^2_{[qq]^*} \hat{k}^+}
    \mathrm{Tr}
    \left[
	\Lambda_{B^*}^{\beta'\alpha'}(P)
	\gamma^\alpha
	(\hat{\ksl} + m_c)\, \gamma^+\, (\hat{\ksl} + m_c)
	\gamma^\beta
    \right]
    \mathcal{G}_{\alpha\beta\alpha'\beta'}.
\label{eq:cSig*-trace1}
\eea 
After re-indexing, $\mathcal{G}_{\alpha\beta\alpha'\beta'}$ is
given by Eq.~(\ref{eq:D*S*-trace1}), the exchange boson carries
the four-momentum $\hat{\Delta} = P - \hat{k}$, and the metric
tensor of the massive, spin-1 diquark is
  $\widetilde{g}^{\mu\nu} = 
    -g^{\mu\nu} + P^{\mu} P^{\nu}/m^2_{[qq]^*}$.
Computing the trace and contractions of Eq.~(\ref{eq:cSig*-trace1})
and evaluating the result with the  appropriate kinematic definitions
analogous to Eq.~(\ref{eq:kin-TOPT}), one may obtain
Eq.~(\ref{eq:c_Sigma-st}) as given in Sec.~\ref{ssec:c_in_B}.

\section{Phenomenological Fit Parameters}
\label{sec:C-MBM}

Following Ref.~\cite{Pum05}, we provide here simple
three-parameter fits to the $c$ and $\bar c$ PDFs in the MBM
computed using several different models for the charm quark
distributions in the charmed mesons and baryons, including the
confining model, effective mass model, and the $\delta$ function
model.
The parametric form for the charm distributions in the nucleon
for the confining and effective mass models is taken to be
\begin{subequations}
\label{eq:conf-fit}
\bea
c(x)
&=& C^{(0)}\, A\, x^\alpha (1-x)^\beta,	\\
\bar{c}(x)
&=& C^{(0)}\, \bar{A}\, x^{\bar\alpha} (1-x)^{\bar\beta},
\eea
\end{subequations}
where the normalization constants
  $A = 1/B(\alpha+1,\beta+1)$
and
  $\bar{A} = 1/B(\bar\alpha+1,\bar\beta+1)$,
with $B$ the Euler beta function, ensure that the distributions
are normalized to $C^{(0)}$.

\begin{table}[t]
\caption{Best fit parameter values for the $c$ and $\bar c$
	distributions in the nucleon in Eqs.~(\ref{eq:conf-fit})
	and (\ref{eq:deltafn-fit}) in the MBM for a central
	cutoff mass $\Lambda=3.0$~GeV.}
\centering
\begin{tabular}{c|c c c}			\hline\hline
$c$, $\bar c$ fit
	& confining
	& effective
	& $\delta$ function			\\
parameters
	& model
	& mass model
	& model					\\ \hline
$A$	
	&\ \ \ $1.720 \times 10^2$\ \ \
	&\ \ \ $1.052 \times 10^2$\ \ \
	&\ \ \ $2.638 \times 10^5$\ \ \		\\
$\alpha$
	& 1.590
	& 3.673
	& 4.266					\\
$\beta$
	& 6.586
	& 10.16
	& 4.485					\\ \hline
$\bar A$	
	&\ \ \ $7.404 \times 10^1$\ \ \
	&\ \ \ $4.160 \times 10^0$\ \ \
	&\ \ \ $2.463 \times 10^4$\ \ \		\\
$\bar\alpha$
	& 1.479
	& 4.153
	& 5.003					\\
$\bar\beta$
	& 4.624
	& 6.800
	& 4.857					\\ \hline	
\end{tabular}
\label{table:phen-fits}
\end{table}

For the $\delta$ function model, where the $c$ and $\bar c$ PDFs
in the charmed hadrons are given by $\delta$ functions in $x$,
it is more convenient to parametrize the distributions in
Eqs.~(\ref{eq:cnaive}) as
\begin{subequations}
\label{eq:deltafn-fit}
\bea
c(x)
&=& C^{(0)}\, A\, x^\alpha (x_B - x)^\beta\,
    \theta(x_B - x),					\\
\bar{c}(x)
&=& C^{(0)}\, \bar{A}\, x^{\bar\alpha} (x_M - x)^{\bar\beta}\,
    \theta(x_M - x),
\eea
\end{subequations}
with $x_B$ and $x_M$ given by Eq.~(\ref{eq:mB}).

For convenience, we give the parameters for the PDFs normalized
to $C^{(0)}$, which can be varied within each model according to
preference.  Although the actual normalization $C^{(0)}$ in the
MBM is modified through the form factor cutoff $\Lambda$,
which affects also the shape of the resulting distribution,
in practice the change in the shape parameters over the range
$2.8 \lesssim \Lambda \lesssim 3.2$~GeV is small.
The parameters in Table~\ref{table:phen-fits} are given for
the central fit value $\Lambda = 3.0$~GeV, determined from the
inclusive $\Lambda_c^+$ production data in $pp$ scattering,
for which the normalization is $C^{(0)} = 2.4\%$.
The parametrization may be extended to smaller values of the
cutoff, $\Lambda \approx 2.5$~GeV, for qualitative comparisons,
although the changes in shape begin to become more appreciable
as one moves away from the above range.


\end{document}